\documentclass[11pt]{article}

\usepackage{a4}

\usepackage{amsmath,amsfonts,amssymb}
\usepackage{bm}

\usepackage{rotating}
\usepackage{tabularx}
\usepackage{dcolumn}

\usepackage[numbers,comma,square,sort,sort&compress]{natbib}
\usepackage[colorlinks=true,linkcolor=red,citecolor=blue]{hyperref}

\usepackage{color}
\usepackage[all]{xy}
\usepackage{graphicx}

\usepackage{comment}

\usepackage{morefloats}
\newcommand{\efloatseparator}{}

\newcommand{\captionfonts}{\small}
\makeatletter
\long\def\@makecaption#1#2{%
  \vskip\abovecaptionskip
  \sbox\@tempboxa{{\captionfonts #1: #2}}%
  \ifdim \wd\@tempboxa >\hsize
    {\captionfonts #1: #2\par}
  \else
    \hbox to\hsize{\hfil\box\@tempboxa\hfil}%
  \fi
  \vskip\belowcaptionskip}
\makeatother

\newcommand{\proj}[2]{|#1\rangle\langle#2|}
\newcommand{\ppsi}[0]{\bm\psi}
\newcommand{\tr}[0]{\mbox{tr}}
\newcommand{\one}[0]{\leavevmode\hbox{\small1\normalsize\kern-.33em1}}

\begin{document}

\title{\Large\bf \parbox{\columnwidth}{Optical properties of Semiconductor Nanostructures: \\
Decoherence versus Quantum Control}}

\author{\parbox{\columnwidth}{
  Ulrich Hohenester\\ \ \\
  Institut f\"ur Physik, Theoretische Physik\\
  Karl--Franzens--Universit\"at Graz\\
  Universit\"atsplatz 5, 8010 Graz, Austria\\ \ \\
  \parbox[t]{1.2cm}{Phone:} +43 316 380 5227\\
  \parbox[t]{1.2cm}{Fax:  } +43 316 380 9820\\
  \parbox[t]{1.2cm}{www:  } \url{http://physik.uni-graz.at/~uxh/} \\
  \parbox[t]{1.2cm}{email:} \href{mailto:ulrich.hohenester@uni-graz.at}%
  {ulrich.hohenester@uni-graz.at}\\ \ \\ \ \\
  \small to appear in {\em``Handbook of Theoretical and Computational Nanotechnology''}
  }}

\date{}

\maketitle

\newpage

\tableofcontents

\newpage

\section{Introduction}\label{sec:intro}

Although there is a lot of quantum physics at the nanoscale, one often has to work hard to observe it. In this review we shall discuss how this can be done for semiconductor quantum dots. These are small islands of lower-bandgap material embedded in a surrounding matrix of higher-bandgap material. For properly chosen dot and material parameters, carriers become confined in all three spatial directions within the low-bandgap islands on a typical length scale of tens of nanometers. This three-dimensional confinement results in atomic-like carrier states with discrete energy levels. In contrast to atoms, quantum dots are not identical but differ in size and material composition, which results in large inhomogeneous broadenings that usually spoil the direct observation of the atomic-like properties. Optics allows to overcome this deficiency by means of single-dot or coherence spectroscopy. Once this accomplished, we fully enter into the quantum world: the optical spectra are governed by sharp and ultranarrow emission peaks --- indicating a strong suppression of environment couplings. When more carriers are added to the dot, e.g. by means of charging or non-linear photoexcitation, they mutually interact through Coulomb interactions which gives rise to intriguing energy shifts of the few-particle states. This has recently attracted strong interest as it is expected to have profound impact on opto-electronic or quantum-information device applications. A detailed theoretical understanding of such Coulomb-renormalized few-particle states is therefore of great physical interest and importance, and will be provided in the first part of this paper. In a nutshell, we find that nature is gentle enough to not bother us too much with all the fine details of the semiconductor materials and the dot confinement, but rather allows for much simpler description schemes. The most simple one, which we shall frequently employ, is borrowed from quantum optics and describes the quantum-dot states in terms of generic few-level schemes. Once we understand the nature of the Coulomb-correlated few-particle states and how they couple to the light, we can start to look closer. More specifically, we shall show that the intrinsic broadenings of the emission peaks in the optical spectra give detailed information about the way the states are coupled to the environment. This will be discussed at the examples of photon and phonon scatterings. Optics can do more than just providing a highly flexible and convenient characterization tool: it can be used as a {\em control},\/ that allows to transfer coherence from an external laser to the quantum-dot states and to hereby deliberately set the wavefunction of the quantum system. This is successfully exploited in the fields of quantum control and quantum computation, as will be discussed in detail in later parts of the paper.

The field of optics and quantum optics in semiconductor quantum dots has recently attracted researchers from different communities, and has benefited from their respective scientific backgrounds. This is also reflected in this paper, where we review genuine solid-state models, such as the rigid-exciton or independent-boson ones, as well as quantum-chemistry schemes, such as configuration interactions or genetic algorithms, or quantum-optics methods, such as the unraveling of the master equation through ``quantum jumps'' or the adiabatic population transfer. The review is intended to give an introduction to the field, and to provide the interested reader with the key references for further details. Throughout, I have tried to briefly explain all concepts and to make the manuscript as self-contained as possible. The paper has been organized as follows. In sec.~\ref{sec:motivation} we give a brief overview of the field and introduce the basic concepts. Section \ref{sec:few-particle} is devoted to an analysis of the Coulomb-renormalized few-particle states and of the more simplified few-level schemes for their description. How these states can be probed optically is discussed in sec.~\ref{sec:spectroscopy}. The coherence and decoherence properties of quantum-dot states are addressed in sec.~\ref{sec:decoherence}, and we show how single-photon sources work. Finally, secs.~\ref{sec:control} and \ref{sec:computation} discuss quantum-control and quantum-computation applications. To keep the paper as simple as possible, we have postponed several of the computational details to the various appendices.

\section{Motivation and overview}\label{sec:motivation}

\subsection{Quantum confinement}\label{sec:motivation.confinement}

The hydrogen spectrum

\begin{equation}\label{eq:hydrogen}
  \epsilon_n=-\frac{E_0}{n^2},\quad n=1,2,\dots\,.
\end{equation}

\noindent provides a prototypical example for quantized motion: only certain eigenstates characterized by the quantum number $n$ (together with the angular quantum numbers $\ell$ and $m_\ell$) are accessible to the system. While the detailed form is due to the Coulomb potential exerted by the nucleus, eq.~\eqref{eq:hydrogen} exhibits two generic features: first, the spectrum $\epsilon_n$ is discrete because the electron motion is confined in all three spatial directions; second, the Rydberg energy scale $E_0=e^2/(2a_0)$ and Bohr length scale $a_0=\hbar^2/(m e^2)$ are determined by the natural constants describing the problem, i.e. the elementary charge $e$, the electron mass $m$, and Planck's constant $\hbar$. These phenomena of quantum confinement and natural units prevail for the completely different system of semiconductor quantum dots. These are semiconductor nanostructures where the carrier motion is confined in all three spatial directions \cite{woggon:97,bimberg:98,hawrylak:98}. Figure~\ref{fig:confinement} sketches two possible types of quantum confinement: in the {\em weak confinement regime}\/ of fig.~\ref{fig:confinement}a the carriers are localized at monolayer fluctuations in the thickness of a semiconductor quantum well; in the {\em strong confinement regime}\/ of fig.~\ref{fig:confinement}b the carriers are confined within small islands of lower-bandgap material embedded in a higher-bandgap semiconductor. Although the specific physical properties of these systems can differ drastically, the dominant role of the three-dimensional quantum confinement establishes a common link that will allow us to treat them on the same footing. To highlight this common perspective as well as the similarity to atoms, in the following we shall frequently refer to quantum dots as {\em artificial atoms}.\/ In the generalized expressions for Rydberg and Bohr 

\begin{eqnarray}\label{eq:aus.rydberg}
  E_s=e^2/(2\kappa_s a_s) & \sim 5\, \mbox{meV}\quad\dots 
  & \mbox{semiconductor Rydberg} \\ \label{eq:aus.bohr}
  a_s=\hbar^2\kappa_s/(m_s e^2) & \sim 10\, \mbox{nm}\quad\dots 
  & \mbox{semiconductor Bohr} 
\end{eqnarray}

\noindent we account for the strong dielectric screening in semiconductors, $\kappa_s\sim 10$, and the small electron and hole effective masses $m_s\sim 0.1\,m$ \cite{haug:93,yu:96}. Indeed, eqs.~\eqref{eq:aus.rydberg} and \eqref{eq:aus.bohr} provide useful energy and length scales for artificial atoms: the carrier localization length ranges from 100 nm in the weak to about 10 nm in the strong confinement regime, and the primary level splitting from 1 meV in the weak to several tens of meV in the strong confinement regime.

The artificial-atom picture can be further extended to optical excitations. Quite generally, when an undoped semiconductor is optically excited an electron is promoted from a valence to a conduction band. In the usual language of semiconductor physics this process is described as the creation of an {\em electron-hole pair}\/ \cite{haug:93,yu:96}: the electron describes the excitation in the conduction band, and the hole accounts for the properties of the missing electron in the valence band. Conveniently electron and hole are considered as independent particles with different effective masses, which mutually interact through the attractive Coulomb interaction. What happens when an electron-hole pair is excited inside a semiconductor quantum dot? Things strongly differ for the weak and strong confinement regime: in the first case the electron and hole form a Coulomb-bound electron-hole complex ---the so-called {\em exciton}\/ \cite{haug:93}--- whose center-of-mass motion becomes localized and quantized in presence of the quantum confinement; in the latter case confinement effects dominate over the Coulomb ones, and give rise to electron-hole states with dominant single-particle character. However, in both cases the generic feature of quantum confinement gives rise to discrete, atomic-like absorption and emission-lines --- and thus allows for the artificial-atom picture advocated above.

\subsection{Scope of the paper}

Quantum systems can usually not be measured directly. Rather one has to perturb the system and measure indirectly how it reacts to the perturbation. This is schematically shown in fig.~\ref{fig:scope} (shaded boxes): an external perturbation, e.g. a laser field, acts upon the quantum system and promotes it from the ground to an excited state; the excitation decays through environment coupling, e.g. photo emission, and finally a measurement is performed on some part of the environment, e.g. through photo detection. As we shall see, this mutual interaction between system and environment ---the environment influences the system and in turn becomes influenced by it (indicated by the arrows in fig.~\ref{fig:scope} which show the flow of information)--- plays a central role in the understanding of decoherence and the measurement process \cite{zurek:03}.

Optical spectroscopy provides one of the most flexible measurement tools since it allows for a remote excitation and detection. It gives detailed information about the system and its environment. This is seen most clearly at the example of atomic spectroscopy which played a major role in the development of quantum theory and quantum electrodynamics \cite{scully:97}, and most recently has even been invoked in the search for non-constant natural constants \cite{fritzsch:03}. In a similar, although somewhat less fundamental manner spectroscopy of artificial atoms allows for a detailed understanding of both electron-hole states (sec.~\ref{sec:spectroscopy}) and of the way these states couple to their environment (sec.~\ref{sec:decoherence}). On the other hand, such detailed understanding opens the challenging perspective to use the external fields in order to {\em control}\/ the quantum system. More specifically, the coherence properties of the exciting laser are transfered to quantum coherence in the system, which allows to deliberately set the state of the quantum system (see lower part of fig.~\ref{fig:scope}). Recent years have seen spectacular examples of such light-matter manipulations in atomic systems, e.g. Bose-Einstein condensation or freezing of light (see, e.g. \citet{chu:02} and references therein). This tremendous success also initiated great stimulus in the field of solid-state physics, as we shall discuss for artificial atoms in sec.~\ref{sec:control}. More recently, the emerging fields of quantum computation \cite{bennett:00,bouwmeester:00,nielsen:00} and quantum communication \cite{gisin:02} have become another driving force in the field. They have raised the prospect that an almost perfect quantum control would allow for computation schemes that would outperform classical computation. In turn, a tremendous quest for suited quantum systems has started, ranging from photons over molecules, trapped ions and atomic ensembles to semiconductor quantum dots. We will briefly review some proposals and experimental progress in sec.~\ref{sec:computation}.

\subsection{Quantum coherence}\label{sec:motivation.coherence}

Quantum coherence is the key ingredient and workhorse of quantum control and quantum computation. To understand its essence, let us consider a generic two-level system with ground state $|0\rangle$ and excited state $|1\rangle$, e.g. an artificial atom with one electron-hole pair absent or present. The most general wavefunction can be written in the form

\begin{equation}\label{eq:state.two-level}
  \alpha|1\rangle+\beta|0\rangle\,,
\end{equation}

\noindent with $\alpha$ and $\beta$ arbitrary complex numbers subject to the condition $|\alpha|^2+|\beta|^2=1$. Throughout this paper we shall prefer the slightly different description scheme of the {\em Bloch vector}\/ picture \cite{haug:93,mandel:95,scully:97}. Because the state \eqref{eq:state.two-level} is unambiguously defined only up to an arbitrary phase factor ---which can for instance be used to make $\alpha$ real---, it can be characterized by three real numbers. A convenient representation is provided by the Bloch vector

\begin{equation}\label{eq:bloch}
  \bm u=\left(\begin{array}{c}
    2\,\Re e(\;\,\alpha^*\beta\,) \\ 
    2\,\Im m(\,\alpha^*\beta\,) \\ 
    |\alpha|^2-|\beta|^2 \\
  \end{array}\right)\,,
\end{equation}

\noindent where the $z$-component accounts for the population inversion, which gives the probability for finding the system in either the upper or lower state, and the $x$ and $y$ components account for the phase relation between $\alpha$ and $\beta$, i.e. the {\em quantum coherence}.\/ As we shall see, this coherence is at the heart of quantum computation and is responsible for such characteristic quantum features as interference or entanglement. For an isolated system whose dynamics is entirely {\em coherent},\/ i.e. completely governed by Schr\"odinger's equation, the norm of the Bloch vector is conserved. A pictorial description is provided by the {\em Bloch sphere}\/ shown in fig.~\ref{fig:bloch}, where in case of a coherent evolution $\bm u$ always stays on the surface of the sphere.

\subsection{Decoherence}

Isolated quantum systems are idealizations that can not be realized in nature since any quantum system interacts with its environment. In general, such environment couplings corrupt the quantum coherence and the system suffers {\em decoherence}.\/ Strictly speaking, decoherence can no longer be described by Schr\"odinger's equation but calls for a more general density-matrix description, within which, as will be shown in sec.~\ref{sec:decoherence}, the $x$ and $y$ components of the Bloch vector are diminished --- $\bm u$ dips into the Bloch sphere. The most simple description for the evolution of the Bloch vector in presence of environment couplings is given by \cite{haug:93,scully:97}

\begin{equation}\label{eq:bloch.t}
  \dot u_1=-\frac{u_1}{T_2},\quad
  \dot u_2=-\frac{u_2}{T_2},\quad
  \dot u_3=-\frac{u_3+1}{T_1}\,,
\end{equation}

\noindent where the first two equations account for the above-mentioned decoherence losses, and the last one for relaxation where the system is scattered from the excited to the ground state because of environment couplings. $T_1$ and $T_2$ are the relaxation and decoherence time, sometimes referred to as {\em longitudinal}\/ and {\em transverse}\/ relaxation times. They are conveniently calculated within the framework of Fermi's golden rule, where

\begin{equation}\label{eq:golden.rule}
  (1/T)=2\pi\int D(\omega) d\omega\, g^2\, \delta(E_i-E_f-\omega)
\end{equation}

\begin{table}
\caption{Relations between $T_1$ and $T_2$ and typical scattering times for spontaneous photon emission and phonon-assisted dephasing \cite{mahan:81,borri:01,krummheuer:02}. In the last two columns we report experimental values measured in the weak and strong confinement regime, respectively.}
\label{table:decoherence}
\begin{tabularx}{\columnwidth}{XXcr}
\hline\hline
Interaction mechanism & Relation & weak & strong \\
\hline
Photon emission & $T_2=2\,T_1$ & $\sim 40$ ps \cite{bonadeo:98a} & ns \cite{borri:01}\\
Phonon dephasing & $T_1\to\infty$ & ? & $\sim 5$ ps \cite{borri:01} \\
\hline
\hline
\end{tabularx}
\end{table}

\noindent accounts for the scattering from the initial state $i$ to the final state $f$ through creation of an environment excitation with energy $\omega$, e.g. photon; $D(\omega)$ is the density of states and $g$ the matrix element associated to the interaction. In semiconductors of higher dimension $T_2$ is always much shorter than $T_1$ \cite{haug:93,rossi:02} because all elastic scatterings (i.e. processes where no energy is exchanged, such as impurity or defect scatterings) contribute to decoherence, whereas only inelastic scatterings (i.e. processes where energy is exchanged, such as phonon or mutual carrier scatterings) contribute to relaxation. On very general grounds one expects that scatterings in artificial atoms become strongly suppressed and quantum coherence substantially enhanced: this is because in higher-dimensional semiconductors carriers can be scattered from a given initial state to a continuum of final states ---and therefore couple to all environment modes $\omega$---, whereas in artificial atoms the atomic-like density of states only allows for a few selective scatterings with $\omega=E_i-E_f$. Indeed, in the coherence experiment of \citet{bonadeo:98b} the authors showed for a quantum dot in the weak confinement regime that the broadening of the optical emission peaks is completely lifetime limited, i.e. $T_2\cong 2\,T_1\sim 40$ ps --- a remarkable finding in view of the extremely short sub-picosecond decoherence times in conventional semiconductor structures. Similar results were also reported for dots in the strong confinement regime \cite{borri:01}. However, there it turned out that at higher temperatures a decoherence channel dominates which is completely ineffective in higher-dimensional systems. An excited electron-hole pair inside a semiconductor provides a perturbation to the system and causes a slight deformation of the surrounding lattice. As will be discussed in sec.~\ref{sec:phonon-dephasing}, in many cases of interest this small deformation gives rise to decoherence but not relaxation. A $T_2$-estimate and some key references are given in table~\ref{table:decoherence}.

\subsection{Quantum control}

Decoherence in artificial atoms is much slower than in semiconductors of higher dimension because of the atomic-like density of states. Yet, it is substantially faster than in atoms where environment couplings can be strongly suppressed by working at ultrahigh vacuum --- a procedure not possible for artificial atoms which are intimately incorporated in the surrounding solid-state environment. Let us consider for illustration a situation where a two-level system initially in its groundstate is excited by an external laser field tuned to the 0--1 transition. As will be shown in sec.~\ref{sec:spectroscopy}, the time evolution of the Bloch vector in presence of a driving field is of the form

\begin{equation}\label{eq:bloch.rabi}
  \dot{\bm u}=-\Omega\,\hat{\bm e}_1\times\bm u\,,
\end{equation}

\noindent with the Rabi frequency $\Omega$ determining the strength of the light-matter coupling and $\hat{\bm e}_1$ the unit vector along $x$. Figure~\ref{fig:schematics.control}a shows the trajectory of the Bloch vector that is rotated from the south pole $-\hat{\bm e}_3$ of the Bloch sphere through the north pole, until it returns after a certain time (given by the strength $\Omega$ of the laser) to the initial position $-\hat{\bm e}_3$. Because of the $2\pi$-rotation of the Bloch vector such pulses are called $2\pi$-pulses. If the Bloch vector evolves in presence of environment coupling, the two-level system becomes entangled with the environmental degrees of freedom and suffers decoherence. This is shown in fig.~\ref{fig:schematics.control}b for the phonon-assisted decoherence described above: while rotating over the Bloch sphere the length of $\bm u$ decreases, and the system does not return to its original position. Quite generally, in the process of decoherence it takes some time for the system to become entangled with its environment. If during this entanglement buildup the system is acted upon by an appropriately designed control, it becomes possible to channel back quantum coherence from the environment to the system and to suppress decoherence. This is shown in fig.~\ref{fig:schematics.control} for an optimized laser field ---for details see sec.~\ref{sec:optimal-control}--- which drives $\bm u$ from the south pole through a sequence of excited states back to the initial position without suffering any decoherence losses. Alternatively, in presence of strong laser fields the quantum dot states become renormalized, which can be exploited for efficient population transfers. Thus, quantum control allows to suppress or even overcome decoherence losses. In sec.~\ref{sec:control} we will discuss prototypical quantum-control applications and ways to combat decoherence in the solid state.

\begin{sidewaystable}
\caption{
Summary of some important parameters for semiconductor quantum dots in the weak and strong confinement regime. The primary and fine-structure splittings refer to the level splittings associated to charge and spin degrees of freedom, respectively. For details see text.
}\label{table:summary}

\begin{tabularx}{\columnwidth}{XcXXX}
\hline\hline
Property & Confinement & Group & Method & Result \\
\hline
Carrier localization length & weak & \citet{matsuda:03} & Nearfield microscopy & $100\times 100\times 5$ nm$^3$\\
Inhomogeneous broadening & weak & \citet{guest:02} & Optical spectroscopy & 1 -- 5 meV\\
Primary level splitting & weak & \citet{guest:02} & Local spectroscopy & 0.1 -- 1 meV \\
Fine-structure splitting & weak & \citet{tischler:02} & Magnetoluminescence & 10 -- 100 $\mu$eV \\
Exciton decoherence time & weak & \citet{bonadeo:98a,bonadeo:98b} &
Coherence spectroscopy & 40 ps\\
& & & & \\
Carrier localization length & strong & \citet{bimberg:98} & & $10\times 10\times 5$ nm$^3$ \\
Inhomogeneous broadening & strong & \citet{bimberg:98} & & 10 -- 50 meV \\
Primary level splitting & strong & \citet{bimberg:98} &  & 10 -- 100 meV \\
Fine-structure splitting & strong & \citet{bayer:02} & Magnetoluminescence & 10 -- 100 $\mu$eV \\
Exciton decoherence time & strong & \citet{borri:01} &
Four-wave mixing & $\sim$ ns$^{\rm a}$\\
& strong & \citet{lenihan:02} & Coherence spectroscopy & 750 ps \\
Spin relaxation time & strong & \citet{lenihan:02} & Coherence spectroscopy & $\ge 40$ ns \\
 & strong & \citet{paillard:01} & Luminescence & ``frozen carrier spins'' \\
\hline\hline
\end{tabularx}

$^a$There exists an additional phonon-assisted decoherence which is strongly temperature-dependent and can dominate at higher temperatures.
\end{sidewaystable}

\subsection{Properties of artificial atoms}

The picture we have developed so far describes artificial atoms in terms of effective few-level schemes. They can be characterized by a few parameters, which can be either obtained from ab-initio-type calculations or can be inferred from experiment. Table \ref{table:summary} reports some of the relevant parameters for artificial atoms. For both the weak and strong confinement regime the inhomogeneous broadening due to dot-size fluctuations is comparable to the primary level splittings themselves, and spectroscopy of single dots (sec.~\ref{sec:spectroscopy}) is compulsory to observe the detailed primary and fine-structure splittings. Decoherence and relaxation times for electron-hole states range from tens to hundreds of picoseconds, which is surprisingly long for the solid state. This is because of the atomic-like density of states, and the resulting inhibition of mutual carrier scatterings and the strong suppression of phonon scatterings. Yet, when it comes to more sophisticated quantum-control or quantum-computation applications (secs.~\ref{sec:control} and \ref{sec:computation}) such sub-nanoscecond relaxation and decoherence appears to be quite limiting. A possible solution may be provided by spin excitations, with their long lifetimes because of weak solid-state couplings.

\section{Few-particle states}\label{sec:few-particle}

Electron-hole states in semiconductor quantum dots can be described at different levels of sophistication, ranging from ab-intio-type approaches over effective solid-state models to generic few-level schemes. All these approaches have their respective advantages and disadvantages. For instance, ab-initio-type approaches provide results that can be quantitatively compared with experiment, but require a detailed knowledge of the confinement potential which is difficult to obtain in many cases of interest and often give only little insight into the general physical trends. On the other hand, few-level schemes grasp all the essential features of certain electron-hole states in a most simple manner, but the relevant parameters have to be obtained from either experiment or supplementary calculations. Depending on the physical problem under consideration we shall thus chose between these different approaches. A description in terms of complementary models is not at all unique to artificial atoms, but has proven to be a particularly successful concept for many-electron atoms. These are highly complicated objects whose physical properties depend on such diverse effects as spin-orbit coupling, exchange interactions, or Coulomb correlations --- and thus make first-principles calculations indispensable for quantitative predictions. On the other hand, in the understanding of the {\em aufbau}\/ principle of the periodic table it suffices to rely on just a few general rules, such as Pauli's principle, Hund's rules for open-shell atoms, and Coulomb correlation effects for transition metals. Finally, for quantum optics calculations one usually invokes generic few-level schemes, e.g. the celebrated $\Lambda$- and V-type ones, where all details of the relevant states are lumped into a few effective parameters. As we shall see, similar concepts can be successfully extended to semiconductor quantum dots. In the remainder of this section we shall discuss how this is done.

Throughout we assume that the carrier states in semiconductor quantum dots are described within a many-body framework such as density functional theory \cite{dreizler:90}, and can be described by the effective single-particle Schr\"odinger equation ($\hbar=1$ throughout)

\begin{equation}\label{eq:dft}
  \left(-\frac{\nabla^2}{2m}+U(\bm r)\right)\,\psi(\bm r)=
  \epsilon\,\psi(\bm r)\,.
\end{equation}

\noindent In the parentheses on the left-hand side the first term accounts for the kinetic energy, where $m$ is the free electron mass, and the second one for the atomic-like potential of the crystal structure. For an ideal periodic solid-state structure the eigenstates $\psi_{n\bm k}(\bm r)=u_{n\bm k}(\bm r)\exp(i\bm k\bm r)$ are given by the usual Bloch function $u$, with $n$ the band index and $\bm k$ the wavevector, and the eigenenergies $\epsilon_{n\bm k}$ provide the semiconductor bandstructure \cite{haug:93,yu:96,dreizler:90}. How are things modified for semiconductor nanostructures? In this paper we shall be concerned with quantum dots with spatial extensions of typically tens of nanometers in each direction, which consist of approximately one million atoms. This suggests that the detailed description of the atomic potential $U(\bm r)$ of eq.~\eqref{eq:dft} is not needed and can be safely replaced by a more phenomenological description scheme. A particularly simple and successful one is provided by the {\em envelope-function approach}\/ \cite{haug:93,yu:96}, which assumes that the single-particle wavefunctions $\psi(\bm r)$ are approximately given by the Bloch function $u$ of the ideal lattice modulated by an envelope part $\phi(\bm r)$ that accounts for the additional quantum confinement. In the following we consider direct {\em III--V}\/ semiconductors, e.g. GaAs or InAs, whose conduction and valence band extrema are located at $\bm k=0$ and describe the bandstructure near the minima by means of effective masses $m_{e,h}$ for electrons and holes. Then,

\begin{equation}\label{eq:sp.qd}
  \left(-\frac{\nabla^2}{2m_i}+U_i(\bm r)\right)\phi^i_\lambda(\bm r)=
  \epsilon^i_\lambda\,\phi^i_\lambda(\bm r)
\end{equation}

\noindent approximately accounts for the electron and hole states in presence of the confinement. Here, $U_{e,h}(\bm r)$ is the effective confinement potential for electrons or holes, $m_i$ is the effective mass of electrons or holes which may depend on position, e.g. to account for the different semiconductor materials in the confinement of fig.~\ref{fig:confinement}b. In the literature numerous theoretical work ---mostly based on the $k\cdot p$ \cite{jiang:97,pryor:98,pryor:99,stier:99} or empirical pseudopotential framework \cite{williamson:00,shumway:01,bester:03}--- has been concerned with more sophisticated calculation schemes for single particle states. These studies have revealed a number of interesting peculiarities associated to effects such as piezoelectric fields, strain, or valence-band mixing, but have otherwise supported the results derived within the more simple-minded envelope-function and effective-mass description scheme for few-particle states in artificial atoms.

\subsection{Excitons}\label{sec:excitons}

\subsubsection{Semiconductors of higher dimension}

What happens for optical electron-hole excitations which experience in addition to the quantum confinement also Coulomb interactions? We first recall the description of a Coulomb-correlated electron-hole pair inside a bulk semiconductor. Within the envelope-function and effective mass approximations

\begin{equation}\label{eq:exciton-bulk}
  H=-\sum_{i=e,h}\frac{\nabla_{\bm r_i}^2}{2\,m_i}-\frac {e^2}{\kappa_s |\bm r_e-\bm r_h|}
\end{equation}

\noindent is the hamiltonian for the interacting electron-hole system, with $\kappa_s$ the static dielectric constant of the bulk semiconductor. In the solution of eq.~\eqref{eq:exciton-bulk} one usually introduces the center-of-mass and relative coordinates $\bm R=(m_e\bm r_e+m_h\bm r_h)/M$ and $\bm\rho=\bm r_e-\bm r_h$ \cite{haug:93}, and decomposes $H=\mathcal{H}+h$ into the parts 

\begin{equation}\label{eq:exciton-bulk2}
  \mathcal{H}=-\frac{\nabla_{\bm R}^2}{2 M},\qquad
  h=-\frac{\nabla_{\bm\rho}^2}{2\,\mu}-\frac {e^2}{\kappa_s |\bm\rho|}\,,
\end{equation}

\noindent with $M=m_e+m_h$ and $\mu=m_e m_h/M$. Correspondingly, the total wavefunction can be decomposed into parts $\Phi(\bm R)$ and $\phi(\bm\rho)$ associated to the center-of-mass and relative motion, respectively, whose solutions are provided by the Schr\"odinger equations

\begin{equation}\label{eq:exciton-bulk3}
  \mathcal{H}\Phi(\bm R) = \mathcal{E}\Phi(\bm R),\qquad
  h\phi(\bm\rho) = \epsilon \phi(\bm\rho) \,.
\end{equation}

\noindent Here, the first equation describes the motion of a free particle with mass $M$, and the second one the motion of a particle with mass $\mu$ in a Coulomb potential $-e^2/\kappa_s|\bm\rho|$. The solutions of the latter equation are those of the hydrogen atom but for the modified Rydberg energy $E_s$ and Bohr radius $a_s$ of eqs.~\eqref{eq:aus.rydberg} and \eqref{eq:aus.bohr}. Similar results apply for the lower-dimensional quantum wells and quantum wires provided that $\phi(\bm\rho)$ is replaced by the corresponding two- and one-dimensional wavefunction, respectively. For instance \cite{haug:93},

\begin{equation}\label{eq:exciton.2d}
  \phi_0(\bm\rho)\cong\frac 4 {a_s} \,\exp\left({-\frac{2\rho}{a_s}}\right)
\end{equation}

\noindent is the approximate groundstate wavefunction for a two-dimensional quantum well, whose energy is $\epsilon_0=-4\,E_s$. For a quantum well of finite width, eq.~\eqref{eq:exciton.2d} only accounts for the in-plane part of the exciton wavefunction. If the quantum well is sufficiently narrow, the total wavefunction is approximately given by the product of \eqref{eq:exciton.2d} with the single-particle wavefunctions for electrons and holes along $z$ ---i.e. those of a ``particle in the box'' \cite{haug:93}---, and the exciton energy is the sum of $\epsilon_0$ with the single-particle energies for the $z$-motion of electrons and holes \cite{kleinman:83,bastard:89,zimmermann:97}.

\subsubsection{Semiconductor quantum dots}

How are the results of the previous section modified in presence of additional quantum confinements $U_e(\bm r_e)$ and $U_h(\bm r_h)$ for electrons and holes? In analogy to eq.~\eqref{eq:exciton-bulk} we describe the interacting electron and hole subject to the quantum-dot confinement through the hamiltonian

\begin{equation}\label{eq:exciton-qd}
  H=\sum_{i=e,h}\left(-\frac{\nabla_{\bm r_i}^2}{2\,m_i}+U_i(\bm r_i)\right)
  -\frac {e^2}{\kappa_s |\bm r_e-\bm r_h|}\,.
\end{equation}

\noindent The first term on the right-hand side accounts for the motion of the carriers in presence of $U_i$. Because of the additional terms $U_i(\bm r_i)$ a separation into center-of-mass and relative motion is no longer possible. Provided that the potentials are sufficiently strong, the carrier motion becomes confined in all three spatial directions. Suppose that $L$ is a characteristic confinement length. Then two limiting cases can be readily identified in eq.~\eqref{eq:exciton-qd}: in case of {\em weak confinement}\/ where $L\gg a_s$ the dynamics of the electron-hole pair is dominated by the Coulomb attraction, and the confinement potentials $U_i(\bm r_i)$ only provide a weak perturbation; in the opposite case of {\em strong confinement}\/ where $L\ll a_s$ confinement effects dominate, and the Coulomb part of eq.~\eqref{eq:exciton-qd} can be treated perturbatively. In the following we shall discuss both cases in slightly more detail.

\paragraph{Weak confinement regime.}

We first consider the weak-confinement regime. A typical example is provided by monolayer interface fluctuations in the width of a semiconductor quantum well, as depicted in fig.~\ref{fig:confinement}a, where the electron-hole pair becomes confined within the region of increased quantum-well thickness \cite{zrenner:94,brunner:94,hess:94,gammon:95,gammon:96,wu:00,guest:02,matsuda:03}. If the resulting confinement length $L$ is much larger than the Bohr radius $a_s$, the correlated electron-hole wavefunction factorizes into a center-of-mass and relative part, where, to a good degree of approximation, the relative part is given by the wavefunction of the quantum well (``rigid-exciton approximation'' \cite{zimmermann:97}). It then becomes possible to integrate over $\bm\rho$ and to recover an effective Schr\"odinger equation for the exciton center-of-mass motion (for details see appendix \ref{sec:rigid})

\begin{equation}\label{eq:rigid-exciton}
  \left(-\frac{\nabla_{\bm R}^2}{2M}+\bar U(\bm R)\right)\Phi_x(\bm R)=\mathcal{E}_x\Phi_x(\bm R)\,,
\end{equation}

\noindent where $\bar U(\bm R)$ is a potential obtained through convolution of $U_e(\bm r_e)$ and $U_h(\bm r_h)$ with the two-dimensional exciton wavefunction \eqref{eq:exciton.2d}. Figure \ref{fig:confinement.weak} shows for a prototypical square-like confinement the corresponding $\bar U(\bm R)$ which only depicts small deviations from the rectangular shape. The corresponding wavefunctions and energies closely resemble those of a particle in a box. Figure \ref{fig:exciton.weak} shows for the confinement potential depicted in fig.~\ref{fig:confinement.weak} the square modulus of (a) the $s$-like groundstate, (b,c) the $p$-like excited states with nodes along $x$ and $y$, and (d) the third excited state with two nodes along $x$. A word of caution is at place. Despite the single-particle character of the envelope-part $\Phi(\bm R)$ of the exciton wavefunction, that of the total wavefunction $\Phi(\bm R)\,\phi_0(\bm r)$ is dominated by Coulomb correlations. This can be easily seen by comparing the length scale of single-particle states $L_n\sim L/n$ ($n$ is the single-particle quantum number) with the excitonic Bohr radius $a_s$. To spatially resolve the variations of $\phi_0(\bm\rho)$ on the length scale of $a_s$, we have to include states up to $L_n\sim a_s$. Hence, $n\sim L/a_s$ which, because of $L\gg a_s$ in the weak-confinement regime, is a large number.

\paragraph{Strong confinement regime.}

Things are completely different in this strong-confinement regime where the confinement length is smaller than the excitonic Bohr radius $a_s$. This situation approximately corresponds to that of most types of self-assembled quantum dots \cite{woggon:97,bimberg:98,hawrylak:98} where carriers are confined in a region of typical size $10\times 10\times 5$ nm$^3$. To the lowest order of approximation, the groundstate $\Psi_0$ of the interacting electron-hole system is simply given by the product of electron and hole single-particle states of lowest energy (see also fig.~\ref{fig:exciton.strong})

\begin{eqnarray}\label{eq:exciton.strong}
  \Psi_0(\bm r_e,\bm r_h)&\cong& \phi^e_0(\bm r_e)\phi^h_0(\bm r_h)\,,\\
  E_0&\cong& \epsilon^e_0+\epsilon^h_0-
  \int d\bm r_e d\bm r_h\, \frac{|\phi^e_0(\bm r_e)|^2|\phi^h_0(\bm r_h)|^2}
  {\kappa_s|\bm r_e-\bm r_h|}\,.\label{eq:exciton.strong.b}
\end{eqnarray}

\noindent Here, the groundstate energy $E_0$ is the sum of the electron and hole single-particle energies reduced by the Coulomb attraction between the two carriers. Excited electron-hole states $\Psi_x$ can be obtained in a similar manner by promoting the carriers to excited single-particle states. In many cases the wavefunction ansatz of eq.~\eqref{eq:exciton.strong} is oversimplified. In particular when the confinement length is comparable to the exciton Bohr radius $a_s$, the electron-hole wavefunction can no longer be written as a simple product \eqref{eq:exciton.strong} of two single-particle states. We shall now briefly discuss how an improved description can be obtained. To this end we introduce the fermionic field operators $c_\mu^\dagger$ and $d_\nu^\dagger$ which, respectively, describe the creation of an electron in state $\mu$ or a hole in state $\nu$ (for details see appendix \ref{sec:second-quantization}). The electron-hole wavefunction of eq.~\eqref{eq:exciton.strong} can then be written as

\begin{equation}\label{eq:exciton.strong2}
  c_{0_e}^\dagger\,d_{0_h}^\dagger\, |0\rangle\,,
\end{equation}

\noindent where $|0\rangle$ denotes the semiconductor vacuum, i.e. no electron-hole pairs present, and $0_e$ and $0_h$ denote the electron and hole single-particle states of lowest energy. While eq.~\eqref{eq:exciton.strong2} is an eigenstate of the single-particle hamiltonian it is only an approximate eigenstate of the Coulomb hamiltonian. We now follow \citet{hawrylak:99} and consider a simplified quantum-dot confinement with cylinder symmetry. The single-particle states can then be labeled by their angular momentum quantum numbers, where the groundstate $0$ has $s$-type symmetry and the degenerate first excited states $\pm 1$ have $p$-type symmetry. Because Coulomb interactions preserve the total angular momentum \cite{hawrylak:98}, the only electron-hole states coupled by Coulomb interactions to the groundstate are those indicated in fig.~\ref{fig:CI-exciton}. We have used that the angular momentum of the hole is opposite to that of the missing electron. Within the electron-hole basis $|0\rangle=|0_e,0_h\rangle$ and $|\pm 1\rangle=|\pm 1_e,\pm 1_h\rangle$ the full hamiltonian matrix is of the form

\begin{equation}\label{eq:CI-exciton}
  E_0+
  \left(\begin{array}{ccc}
  0 & V_{sp} & V_{sp} \\
  V_{sp} & \Delta_p & V_{pp} \\
  V_{sp} & V_{pp} & \Delta_p \\
  \end{array}\right)\,,
\end{equation}

\noindent where $E_0$ is the energy \eqref{eq:exciton.strong.b} of the exciton groundstate, $\Delta_p$ the detuning of the first excited state in absence of Coulomb mixing, and $V_{sp}$ and $V_{pp}$ describe the Coulomb couplings between electrons and holes in the $s$ and $p$ shells (see fig.~\ref{fig:CI-exciton}). The Coulomb renormalized eigenstates and energies can then be obtained by diagonalizing the matrix \eqref{eq:CI-exciton}. Results of such configuration-interaction calculations will be presented in sec.~\ref{sec:spectroscopy} (see appendix \ref{sec:CI} for more details).

\subsubsection{Spin structure}\label{sec:exciton.spin}

Besides the orbital degrees of freedom described by the envelope part of the wavefunction, the atomic part additionally introduces spin degrees of freedom. For {\em III--V}\/ semiconductors an exhaustive description of the band structure near the minima (at the so-called $\Gamma$ point) is provided by an eight-band model \cite{kane:66,haug:93} containing the $s$-like conduction band states $|s,\pm \frac 1 2\rangle$ and the $p$-like valence band states $|\frac 3 2,\pm\frac 3 2\rangle$, $|\frac 3 2,\pm\frac 1 2\rangle$, and $|\frac 1 2,\pm\frac 1 2\rangle$ [note that these $s$- and $p$-states refer to the atomic orbitals and have nothing to do with those introduced in eq.~\eqref{eq:CI-exciton}]. In the problem of our present concern four of the six valence band states can be approximately neglected: first, the $|\frac 1 2,\pm\frac 1 2\rangle$ ones which are energetically split off by a few hundred meV because of spin-orbit interactions \cite{kane:66,haug:93,yu:96,zutic:04}; second, the states $|\frac 3 2,\pm\frac 1 2\rangle$ associated to the {\em light-hole}\/ band which are energetically split off in case of a strong quantum confinement along the growth direction $z$ --- e.g. those shown in fig.~\ref{fig:confinement}. Thus, the atomic part of the electron and hole states of lowest energy is approximately given by the $s$-type conduction band states $|s,\pm \frac 1 2\rangle$ and the $p$-type states $|\frac 3 2,\pm \frac 3 2\rangle$ associated to the {\em heavy-hole}\/ band. From these two electron and hole states we can form four possible electron-hole states $|\pm \frac 1 2,\pm \frac 3 2\rangle$ and $|\pm \frac 1 2,\mp \frac 3 2\rangle$, where the entries account for the $z$-projection of the total angular momentum $m_j$ for the electron and hole, respectively. A word of caution is at place: since the hole describes the properties of the {\em missing}\/ electron in the valence band, its $m_j$ value is opposite of that of the corresponding valence band state. For that reason, the usual optical selection rules $\Delta j=0$ and $\Delta m_j=\pm 1$ \cite{haug:93} for the optical transitions under consideration translate to the matrix elements

\begin{equation}\label{sec:selection-rules}
  \Bigl< 0\Bigr|e\bm r\Bigl| \frac 1 2,-\frac 3 2 \Bigr>=\mu_0\hat{\bm e}_+,\quad
  \Bigl< 0\Bigr|e\bm r\Bigl|-\frac 1 2, \frac 3 2 \Bigr>=\mu_0\hat{\bm e}_-,\quad
\end{equation}

\begin{table}
\caption{Spin structure of electron-hole states. The first column reports the $z$-components of the angular momenta for the electron and hole, the second column indicates whether the exciton can be optically excited (bright) or not (dark), the third column shows the polarization vector of the transition, and the last column gives the short-hand notation used in this paper; the upper triangles indicate whether the electron spin points upwards ($\blacktriangle\!\triangledown$) or downwards ($\vartriangle\!\!\!\blacktriangledown$) and the lower triangles give the corresponding information about the hole spin.
}
\label{table:exciton.strong}
\begin{tabularx}{\columnwidth}{XXcc}
\hline\hline
Electron-hole state & Optical coupling & Polarization & notation \\
\hline
$|+\frac 1 2,-\frac 3 2\rangle$ & bright & $\hat{\bm e}_+$ & 
$\blacktriangle\!\triangledown\atop\blacktriangledown\!\vartriangle$\\
$|-\frac 1 2,+\frac 3 2\rangle$ & bright & $\hat{\bm e}_-$ & 
$\vartriangle\!\blacktriangledown\atop\triangledown\!\blacktriangle$\\
$|+\frac 1 2,+\frac 3 2\rangle$ & dark &  & 
$\blacktriangle\!\triangledown\atop\triangledown\!\blacktriangle$\\
$|-\frac 1 2,-\frac 3 2\rangle$ & dark &  & 
$\vartriangle\!\blacktriangledown\atop\blacktriangledown\!\vartriangle$\\
\hline
\hline
\end{tabularx}
\end{table}

\noindent with $\mu_0$ the optical dipole matrix element, $|0\rangle$ the semiconductor vacuum, and $\hat{\bm e}_\pm$ the polarization vector for left- or right-handed circularly polarized light. Below we shall refer to hole states with $m_j=\pm\frac 3 2$ as holes with spin-up or spin-down orientation, and to exciton states $|\pm\frac 1 2,\mp\frac 3 2\rangle$ as excitons with spin-up or spin-down orientation. Thus, for optically allowed excitons the spins of the electron and hole point into opposite directions, as indicated in fig.~\ref{fig:exciton.strong} and table~\ref{table:exciton.strong}. The degeneracy of the four exciton states of table \ref{table:exciton.strong} is usually split. First, the bright and dark excitons are separated by a small amount $\delta\sim 10$--100 $\mu$eV because of the {\em electron-hole exchange interaction}\/ \cite{bayer:99,bayer:02,tischler:02,bester:03}. This is a genuine solid-state effect which accounts for the fact that an electron promoted from the valence to the conduction band no longer experiences the exchange interaction with itself, and we thus have to correct for this missing interaction in the bandstructure description. It is a repulsive interaction which is only present for electrons and holes with opposite spin orientations. Additionally, in case of an asymmetric dot confinement the exciton eigenstates can be computed from the phenomenological hamiltonian \cite{bayer:02}

\begin{equation}\label{eq:h-exchange}
  H_{\rm exchange}=\frac 1 2\left(\begin{array}{cccc}
    \phantom{-}\delta & \phantom{-}\delta' & 0 & 0 \\
    \phantom{-}\delta' & \phantom{-}\delta & 0 & 0 \\
    \phantom{-}0 & \phantom{-}0 & -\delta & \phantom{-}\delta'' \\
    \phantom{-}0 & \phantom{-}0 & \phantom{-}\delta'' & -\delta \\
  \end{array}\right)\,,
\end{equation}

\noindent where $\delta'$ and $\delta''$ are small constants accounting for the asymmetry of the dot confinement. The corresponding eigenstates are linear combinations of the exciton states of table~\ref{table:exciton.strong}, e.g. $(|\frac 1 2,-\frac 3 2\rangle\pm |-\frac 1 2,\frac 3 2\rangle)/\sqrt 2$ for the optically allowed excitons which are linearly polarized along $x$ and $y$. If a magnetic field is applied along the growth direction $z$ the two bright exciton states become energetically further split. Alternatively, if in the Voigt geometry a magnetic field $B_x$ is applied along $x$ the two bright exciton states become mixed \cite{bayer:02}; we will use this fact later in the discussion of possible exciton-based quantum computation schemes.

\subsection{Biexcitons}\label{sec:biexcitons}

In semiconductors of higher dimension a few other Coulomb-bound electron-hole complexes exist: for instance, the negatively charged exciton, which consists of one hole and two electrons with opposite spin orientations, and the biexciton, which consists of two electron-hole pairs with opposite spin orientations. In both cases the binding energy is of the order of a few meV and is attributed to genuine Coulomb correlations: in the negatively charged exciton the carriers arrange such that the hole is preferentially located in-between the two electrons and thus effectively screens the repulsive electron-electron interaction; similarly, in the biexciton the four carriers arrange in a configuration reminiscent of the $H_2$ molecule, where the two heavier particles ---the holes--- are located at a fixed distance, and the lighter electrons are delocalized over the whole few-particle complex and are responsible for the binding (see insets of fig.~\ref{fig:confinement.weak}). In the literature a number of variational wavefunction {\em ans\"atze}\/ are known for the biexciton description, e.g. that of Kleinman \cite{kleinman:83}

\begin{equation}\label{eq:kleinman}
  \bar\phi_0(\bm r_e,\bm r_h,\bm r_{e'},\bm r_{h'})=
  \exp[-(s_e+s_{e'})/2]\,\cosh[\beta(t_e-t_{e'})]\,\chi(r_{hh'})\,,
\end{equation}

\noindent with $s_e=r_{eh}+r_{eh'}$, $t_e=r_{eh}-r_{eh'}$, and $r_{ij}$ the distance between particles $i$ and $j$. The first two terms on the right-hand side account for the attractive electron-hole interactions, and $\chi(r_{hh'})$ for the repulsive hole-hole one ($\beta$ is a variational parameter). In the inset of figure \ref{fig:confinement.weak} we plot the probability distribution for the electron and hole as computed from eq.~\eqref{eq:kleinman}: in comparison to the exciton the biexciton is much more delocalized, and correspondingly the biexciton binding is much weaker \cite{kleinman:83,filinov:03}.

\subsubsection{Weak confinement regime}

Suppose that the biexciton is subject to an additional quantum confinement, e.g. induced by the interface fluctuations depicted in fig.~\ref{fig:confinement}a. If the characteristic confinement length $L$ is larger than the excitonic Bohr radius $a_s$ and the extension of the biexciton, one can, in analogy to excitons, introduce a ``rigid-biexciton'' approximation: here, the biexciton wavefunction \eqref{eq:kleinman} of the ideal quantum well is modulated by an envelope function which depends on the center-of-mass coordinate of the biexciton. An effective confinement for the biexciton can be obtained through appropriate convolution of $U_i(\bm r_i)$ (for details see appendix \ref{sec:rigid}), which is shown in fig.~\ref{fig:confinement.weak} for a representative interface fluctuation potential. Because of the larger extension of the biexciton wavefunction, the effective potential exhibits a larger degree of confinement and correspondingly the biexciton wavefunction of fig.~\ref{fig:biexciton.weak} is more localized. We will return to this point in the discussion of local optical spectroscopy in sec.~\ref{sec:snom}.

\subsubsection{Strong confinement regime}

In the strong confinement regime the ``binding'' of few-particle complexes is not due to Coulomb correlations but to the quantum confinement, whereas Coulomb interactions only introduce minor energy renormalizations. It thus becomes possible to confine various few-particle electron-hole complexes which are unstable in semiconductors of higher dimension. We start our discussion with the few-particle complex consisting of two electrons and holes. In analogy to higher-dimensional semiconductors, we shall refer to this complex as a {\em biexciton}\/ keeping in mind that the binding is due to the strong quantum confinement rather than Coulomb correlations. To the lowest order of approximation, the biexciton groundstate $\bar\Psi_0$ in the strong confinement regime is given by the product of two excitons \eqref{eq:exciton.strong} with opposite spin orientations

\begin{eqnarray}\label{eq:biexciton.strong}
  &&\bar\Psi_0(\bm r_e,\bm r_h,\bm r_{e'},\bm r_{h'})\cong 
  \Psi_0(\bm r_e,\bm r_h)\Psi_0(\bm r_{e'},\bm r_{h'})\\
  \label{eq:biexciton.energy}
  &&\bar E_0\cong 2\,E_0+
  \langle\bar\Psi_0|\, H_{ee'}+H_{hh'}+H_{eh'}+H_{e'h}\,|\bar\Psi_0\rangle\,.
\end{eqnarray}

\noindent The second term on the right-hand side of eq.~\eqref{eq:biexciton.energy} accounts for the repulsive and attractive Coulomb interactions not included in the exciton ground state energy $E_0$. If electron and hole single-particle states have the same spatial extension, the repulsive contributions $H_{ee'}$ and $H_{hh'}$ are exactly canceled by the attractive contributions $H_{eh'}$ and $H_{e'h}$ and the biexciton energy is just twice the exciton energy, i.e. there is no binding energy for the two neutral excitons. In general, this description is too simplified. If the electrons and holes arrange in a more favorable configuration, such as the $H_2$ one in the weak confinement regime, the Coulomb energy can be reduced. Within the framework of configuration interactions outlined in appendix~\ref{sec:CI}, such correlation effects imply that the biexciton wavefunction no longer is a single product of two states but acquires additional components from excited states. A rough estimate for the magnitude of such correlation effects is given in first order perturbation theory by $\langle V\rangle^2/(\Delta\epsilon)$, with $\langle V\rangle$ the average gain of Coulomb energy (typically a few meV) and $\Delta\epsilon$ the splitting of single-particle states (typically a few tens of meV). In general, it turns out to be convenient to parameterize the biexciton energy through

\begin{equation}\label{eq:biexciton.binding}
  \bar E_0=2\,E_0-\Delta\,,
\end{equation}

\noindent where $\Delta$ is the biexciton binding energy. Its value is usually positive and somewhat smaller than the corresponding quantum-well value, but can sometimes even acquire negative values (``biexciton anti-binding'' \cite{rodt:03}). We shall find that the Coulomb renormalization $\Delta$ has the important consequence that the biexciton transition is at a different frequency than the exciton one, which will allow us to distinguish the two states in incoherent and coherent spectroscopy.

\subsection{Other few-particle complexes}

Besides the exciton and biexciton states, quantum dots in the strong confinement regime can host a number of other few-particle complexes. Depending on whether they are neutral, i.e. consist of an equal number of electrons and holes, or charged, we shall refer to them as {\em multi excitons}\/ or {\em multi-charged excitons}.\/ Since electron-hole pairs are neutral objects, quantum dots can be populated by a relatively large number of pairs ranging from six \cite{bayer:00a} to several tens \cite{rinaldi.prb:00,raymond:04} dependent on the dot confinement. In experiments such multi-exciton population is usually achieved as follows: a pump pulse creates electron-hole pairs in continuum states (e.g., wetting layer) in the vicinity of the quantum
dot, and some of the carriers become captured in the dot; because of the fast subsequent carrier relaxation (sec.~\ref{sec:decoherence}) the few-particle system relaxes to its state of lowest energy, and finally the electron and hole recombine by emitting a photon. Thus, in a steady-state experiment information about the few-particle carrier states can be obtained by varying the pump intensity and monitoring the luminescence from the quantum dot \cite{landin:98,dekel:98,bayer:00a,zrenner:00}. Results of such multi-exciton spectroscopy experiments will be briefly presented in the next section. Experimentally it is also possible to create electron-hole complexes with an unequal number of electrons and holes. Figure~\ref{fig:charged-exciton} shows how this can be done \cite{warburton:97,warburton:00,findeis.prb:01}: a quantum dot is placed within a $n$-$i$ field-effect structure; when an external gate voltage is applied, the energy of the electron groundstate drops below the Fermi energy of the $n$-type reservoir and an electron tunnels from the reservoir to the dot, where further charging is prohibited because of the {\em Coulomb blockade},\/ i.e. because of the strong Coulomb repulsion between electrons in the dot; when the dot is optically excited, e.g. by the same mechanism of off-resonant excitation and carrier capture described above, one can create charged excitons. A further increase of the gate voltage allows to promote more electrons from the reservoir to the dot, and to hereby create multi-charged excitons with up to two surplus electrons. In \citet{regelman:01} a quantum dot was placed in a $n$-$i$-$p$ structure which allowed to create in the same sample either negatively (more electrons than holes) or positively (more holes than electrons) charged excitons by varying the applied gate voltage. A different approach was pursued by \citet{hartmann.prl:00}, where charging was achieved by unintentional background doping and the mechanism of photo depletion, which allowed to charge quantum dots with up to five surplus electrons. Luminescence spectra of such multi-charged excitons will be presented in sec.~\ref{sec:luminescence}.

\subsection{Coupled dots}

We conclude this section with a brief discussion of coupled quantum dots. In analogy to {\em artificial atoms},\/ we may refer to coupled dots as {\em artificial molecules}.\/ Coupling is an inherent feature of any high-density quantum dot ensemble, as, e.g. needed for most optoelectronic applications \cite{bimberg:98}. On the other hand, it is essential to
the design of (quantum) information devices, for example quantum dot cellular automata \cite{snider:99} or quantum-dot implementations of quantum computation (sec.~\ref{sec:computation}). Artificial molecules formed by two or more coupled dots are extremely interesting also from the fundamental point of view, since the interdot coupling can be tuned far out of the regimes accessible in natural molecules, and the relative importance of single-particle tunneling and Coulomb interactions can be varied in a controlled way. The interacting few-electron states in a double dot were studied theoretically \cite{rontani.ssc:01,partoens:00,martin-moreno:00} and experimentally by tunneling and capacitance experiments
\cite{oosterkamp:98a,oosterkamp:98b,fujisawa:98,schmidt:97,blick:98a,blick:98b,brodsky:00,amaha:01}, and correlations were found to induce coherence effects and novel ground-state phases depending on the interdot coupling regime. For self-organized dots stacking was demonstrated \cite{fafard:00}, and the exciton splitting in a single artificial molecule was observed and explained in terms of single-particle level filling of delocalized bonding and anti-bonding electron and hole states
\cite{schedelbeck:97,bayer:01a,borri:03}. When a few photoexcited particles are present, Coulomb coupling between electrons and holes adds to the homopolar electron-electron and hole-hole couplings. In addition, single-particle tunneling and kinetic energies are affected by the different spatial extension of electrons and holes, and the correlated ground and excited states are governed by the competition of these effects \cite{rontani.ssc:01,troiani.prb:02,janssens:02a,janssens:02b}. A particularly simple parameterization of single-exciton and biexciton states in coupled dots is given by the Hubbard-type hamiltonian \cite{koskinen.ssc:03}

\begin{equation}\label{eq:double-dot}
    H=E_0\sum_\sigma\left(\hat n_{L\sigma}+\hat n_{R\sigma}\right)-
    t\sum_\sigma\left(b_{L\sigma}^\dagger b_{R\sigma}^{\phantom{\dagger}}+
                      b_{R\sigma}^\dagger b_{L\sigma}^{\phantom{\dagger}}
		      \right)-
    \Delta\sum_{\ell=L,R}\hat n_{\ell\uparrow}\hat n_{\ell\downarrow}
    \,,
\end{equation}

\noindent with $b_{\ell\sigma}^\dagger$ the creation operator for excitons with spin orientation $\sigma=\pm$ in the right or left dot, $\hat n_{\ell\sigma}=b_{\ell\sigma}^\dagger b_{\ell\sigma}^{\phantom\dagger}$ the exciton number operator, $t$ the tunneling matrix element, and $\Delta$ the biexciton binding. Indeed, the hamiltonian \eqref{eq:double-dot} accounts properly for the formation of bonding and anti-bonding exciton states, and the fact that in a biexciton state the two electron-hole pairs preferentially stay together to benefit from the biexciton binding $\Delta$ \cite{rontani.ssc:01,troiani.prb:02}. We will return to coupled dots in the discussion of quantum control (sec.~\ref{sec:control}) and quantum computation (sec.~\ref{sec:computation}).

\section{Optical spectroscopy}\label{sec:spectroscopy}

In the last section we have discussed the properties of electron-hole states in semiconductor quantum dots. We shall now show how these states couple to the light and can be probed optically. Our starting point is given by eq.~\eqref{eq:dft} which describes the propagation of one electron subject to the additional quantum confinement $U$. Quite generally, the light field is described by the vector potential $\bm{\mathcal{A}}$ and the light-matter coupling is obtained by replacing the momentum operator $\bm p=-i\nabla$ with $\bm p-(q/c)\bm{\mathcal{A}}$ \cite{scully:97,yu:96}, where $q=-e$ is the charge of the electron and $c$ the speed of light. The light-matter coupling then follows from

\begin{equation}\label{eq:light-matter.dft}
  \frac{\left(\bm p+ \frac e c\,\bm{\mathcal{A}}\right)^2}{2 m}+U(\bm r)=
  H_0+\frac e {mc}\, \bm{\mathcal{A}}\,\bm p+
  \frac{e^2}{2mc^2}\,\bm{\mathcal{A}}^2\,,
\end{equation}

\noindent where we have used the Coulomb gauge $\nabla\bm{\mathcal{A}}=0$ \cite{jackson:62} to arrive at the $\bm{\mathcal{A}}\,\bm p$ term. In many cases of interest the spatial dependence of $\bm{\mathcal{A}}$ can be neglected on the length scale of the quantum states, i.e. in the {\em far-field}\/ limit ---recall that the length scales of light and matter are given by microns and nanometers, respectively---, and we can perform a gauge transformation to replace the $\bm{\mathcal{A}}\,\bm p$ term by the well-known dipole coupling \cite{scully:97}

\begin{equation}\label{eq:light-matter}
  H_{\rm op}=\frac e {mc}\, \bm{\mathcal{A}}\,\bm p\cong e\bm r\,\bm{\mathcal{E}}\,.
\end{equation}

\noindent The relation between the vector potential and the electric field is given by $\bm{\mathcal{E}}=(1/c)\partial\bm{\mathcal{A}}/\partial t$. In this far-field limit we can also safely neglect the $\bm A^2$ term. This is because the matrix elements $\langle 0|\pm\frac 1 2,\mp\frac 3 2\rangle$ between the atomic states introduced in sec.~\ref{sec:exciton.spin} vanish owing to the orthogonality of conduction and valence band states. Equation~\eqref{eq:light-matter} is suited for both classical and quantum light fields. In the first case $\bm{\mathcal{E}}$ is treated as a $c$-number, in the latter case the electric field of photons reads \cite{scully:97,mandel:95,walls:95,andreani:99}

\begin{equation}\label{eq:efield.photon}
  \bm{\mathcal{E}}\cong i\sum_{\bm k\lambda}
  \left(\frac{2\pi\omega_k}{\kappa_s}\right)^{\frac 1 2}
  \left(\hat{\bm e}_{\bm k\lambda}\,  a_{\bm k\lambda}-
        \hat{\bm e}_{\bm k\lambda}^*\,a_{\bm k\lambda}^\dagger\right)\,.
\end{equation}

\noindent Here, $\bm k$ and $\lambda$ are the photon wavevector and polarization, respectively, $\omega_k=ck/n_s$ is the light frequency, $n_s=\sqrt\kappa_s$ the semiconductor refractive index, $\hat{\bm e}_{\bm k\lambda}$ the photon polarization vector, and $a_{\bm k\lambda}$ denotes the usual bosonic field operator. The free photon field is described by the hamiltonian $H_0^\gamma=\sum_{\bm k\lambda}\omega_k\,a_{\bm k\lambda}^\dagger a_{\bm k\lambda}$.

\paragraph{Optical dipole moments.}

Optical selection rules were already introduced in sec.~\ref{sec:exciton.spin} where we showed that light with appropriate polarization $\lambda$ ---i.e. for light propagation along $z$ either circular polarization for symmetric dots or linear polarization for asymmetric ones--- can induce electron-hole transitions. We shall now show how things are modified when additionally the envelope part of the carrier wavefunctions is considered. In second quantization (see appendix \ref{sec:CI}) the light-matter coupling of eq.~\eqref{eq:light-matter} reads
 
\begin{equation}\label{eq:light-matter2}
  H_{\rm op}\cong \sum_\lambda\int d\bm r\,
  \left(\mu_0\hat{\bm e}_\lambda\,
  \ppsi_{\bar\lambda}^h(\bm r)\ppsi_\lambda^e(\bm r)+\mbox{h.c.}\right)
  \,\bm{\mathcal{E}}\,,
\end{equation}

\noindent where $\bar\lambda$ is the polarization mode orthogonal to $\lambda$. Because of the envelope-function approximation the dipole operator $e\bm r$ has been completely absorbed in the bulk moment $\mu_0$ \cite{haug:93,yu:96}. The first term in parentheses of eq.~\eqref{eq:light-matter2} accounts for the destruction of an electron-hole pair, and the second one for its creation. Similarly, in an all-electron picture the two terms can be described as the transfer of an electron from the conduction to the valence band or vice versa. This single-particle nature of optical excitations translates to the requirement that electron and hole are destroyed or created at the {\em same}\/ position $\bm r$. Equation~\eqref{eq:light-matter2} usually comes together with the so-called {\em rotating-wave approximation}\/ \cite{haug:93,scully:97}. Consider the light-matter coupling \eqref{eq:light-matter2} in the interaction picture according to the hamiltonian of the unperturbed system: the first term, which accounts for the annihilation of an electron-hole pair, then approximately oscillates with $e^{-i\omega_0 t}$ and the second term with $e^{i\omega_0 t}$, where $\omega_0$ is a frequency of the order of the semiconductor band gap. Importantly, $\omega_0$ sets the largest energy scale (eV) of the problem whereas all exciton or few-particle level splittings are substantially smaller. If we accordingly separate $\bm{\mathcal{E}}$ into terms oscillating with approximately $e^{\pm i\omega_0 t}$, we encounter in the light-matter coupling of eq.~\eqref{eq:light-matter2} two possible combinations of exponentials: first, those with $e^{\pm i(\omega_0-\omega_0)t}$ which have a slow time dependence and have to be retained; second, those with $e^{\pm i(\omega_0+\omega_0)t}$ which oscillate with twice the frequency of the band gap. In the spirit of the random-phase approximation, the latter off-resonant terms do not induce transitions and can thus be neglected. Then, the light-matter coupling of eq.~\eqref{eq:light-matter2} becomes

\begin{equation}\label{eq:light-matter3}
  H_{\rm op}\cong \frac 1 2
  \left(\bm{\mathcal{P}}\,\bm{\mathcal{E}}^{(-)}+ 
         \bm{\mathcal{P}}^\dagger\,\bm{\mathcal{E}}^{(+)}\right)\,,
\end{equation}

\noindent where $\bm{\mathcal{E}}^{(\pm)}\propto e^{\mp i\omega_0 t}$ solely evolves with positive or negative frequency components, and the complex conjugate of $\bm{\mathcal{E}}^{(-)}$ is given by $\bm{\mathcal{E}}^{(+)}$ \cite{mandel:95}. In eq.~\eqref{eq:light-matter3} $\bm{\mathcal{P}}=\sum_\lambda\mu_0\hat{\bm e}_\lambda\int d\bm r\,\ppsi_{\bar\lambda}^h(\bm r)\ppsi_\lambda^e(\bm r)$ is the usual interband polarization operator \cite{haug:93,rossi:98}. Let us finally briefly discuss the optical dipole elements for excitonic and biexcitonic transitions. Within the framework of second quantization the exciton and biexction states $|x_\lambda\rangle$ and $|b\rangle$ can be expressed as

\begin{eqnarray}\label{eq:exciton-field}
  |x_\lambda\rangle &=& \int d\bm\tau\,\Psi_x(\bm r_e,\bm r_h)\,
  \ppsi_\lambda^{e\,\dagger}(\bm r_e)\ppsi_{\bar\lambda}^{h\,\dagger}(\bm r_h)\,|0\rangle\\
  \label{eq:biexciton-field}
  |b\rangle&=&\int d\bar{\bm\tau}\,
  \bar\Psi_b(\bm r_e,\bm r_h,\bm r_e',\bm r_h')\,
  \ppsi_\lambda^{e\,\dagger}(\bm r_e)\ppsi_{\bar\lambda}^{h\,\dagger}(\bm r_h)
  \ppsi_{\bar\lambda}^{e\,\dagger}(\bm r_e')\ppsi_\lambda^{h\,\dagger}(\bm r_h')\,
  |0\rangle\,,\quad
\end{eqnarray}

\noindent with $d\bm\tau$ and $d\bar{\bm\tau}$ denoting the phase space for excitons and biexcitons, respectively. In eq.~\eqref{eq:exciton-field} the exciton state consists of one electron and hole with opposite spin orientations, and in eq.~\eqref{eq:biexciton-field} the biexciton of two electron-hole pairs with opposite spin orientations. From the light-matter coupling \eqref{eq:light-matter2} we then find for the optical dipole elements

\begin{alignat}{4}\label{eq:dipole-exciton}
  \langle 0|\bm{\mathcal{P}}|x_\lambda\rangle &=& M_{0x}\hat{\bm e}_\lambda &=
  \mu_0\,\hat{\bm e}_\lambda\int d\bm r\,\Psi_x(\bm r,\bm r)\\
  \label{eq:dipole-biexciton}
  \langle x_{\bar\lambda}|\bm{\mathcal{P}}|b\rangle &=&
  M_{xb}\,\hat{\bm e}_\lambda &=
  \mu_0\,\hat{\bm e}_\lambda\int d\bm rd\bm r_ed\bm r_h\,
  \Psi_x^*(\bm r_e,\bm r_h)\hat\Psi_b(\bm r,\bm r,\bm r_e,\bm r_h)\,.
\end{alignat}

\noindent In eq.~\eqref{eq:dipole-exciton} the dipole moment is given by the spatial average of the exciton wavefunction $\Psi_x(\bm r,\bm r)$ where the electron and hole are at the same position $\bm r$. Similarly, in eq.~\eqref{eq:dipole-biexciton} the dipole moment is given by the overlap of exciton and biexciton wavefunctions subject to the condition that the electron with spin $\bar\lambda$ and the hole with spin $\lambda$ are at the same position $\bm r$, whereas the other electron and hole remain at the same position. In appendix \ref{sec:dipole.rigid} we show that in the weak confinement regime the oscillator strength for optical transitions scales with the confinement length $L$ according to $|M_{0x}|^2\propto L^2$, i.e. it is proportional to the confinement area. For that reason excitons in the weak-confinement regime couple much stronger to the light than those in the strong confinement regime, which makes them ideal candidates for various kinds of optical coherence experiments \cite{bonadeo:98a,bonadeo:98b,bonadeo:99,andreani:99,chen:00,guest:01,li:03,matsuda:03}.

\paragraph{Fluctuation-dissipation theorem.}

We next discuss how to compute optical spectra in linear response. As a preliminary task we consider the general situation where a generic quantum system is coupled to an external perturbation $X(t)$, e.g. an exciting laser light, via the system operator $A$ through $A\,X(t)$ \cite{kubo:85}. Let $\langle B\rangle$ be the expectation value of the operator $B$ in the perturbed system and $\langle B\rangle_0$ that in the unperturbed one. In linear-response theory the change $\langle\Delta B\rangle=\langle B\rangle-\langle B\rangle_0$ is assumed to be {\em linear}\/ in the perturbation $X(t)$ --- an approximation valid under quite broad conditions provided that the external perturbation is sufficiently weak. We can then derive within lowest-order time-dependent perturbation theory the famous {\em fluctuation-dissipation theorem}\/ \cite{kubo:85}

\begin{equation}\label{eq:fluctuation-dissipation}
  \langle\Delta B(0)\rangle=i\int_{-\infty}^0 dt'\,
  \langle[A(t'),B(0)]\rangle_0\,X(t')\,,
\end{equation}

\noindent where operators $A$ and $B$ are given in the interaction picture according to the unperturbed system hamiltonian $H_0$. In eq.~\eqref{eq:fluctuation-dissipation} we have assumed that the external perturbation has been turned on at sufficiently early times such that the system has reached equilibrium. The important feature of eq.~\eqref{eq:fluctuation-dissipation} is that it relates the expectation value of $B$ in the {\em perturbed}\/ system to the correlation $[A,B]$ ---or equivalently to the fluctuation $[\Delta A,\Delta B]$ because commutators with $c$-numbers always vanish--- of the {\em unperturbed system}.\/ Usually, the expression on the right-hand side of eq.~\eqref{eq:fluctuation-dissipation} is much easier to compute than that on the left-hand side. We will next show how the fluctuation-dissipation theorem \eqref{eq:fluctuation-dissipation} can be used for the calculation of linear optical absorption and luminescence.

\subsection{Optical absorption}\label{sec:absorption}

Absorption describes the process where energy is transferred from the light field to the quantum dot, i.e. light becomes absorbed. Absorption is proportional to the loss of energy of the light field, or equivalently to the gain of energy of the system

\begin{equation}\label{eq:absorption.energy}
  \alpha(\omega)\propto \frac d {dt} \left< H_0+H_{\rm op} \right> =
  \left< \frac{\partial H_{\rm op}}{\partial t}\right> \,.
\end{equation}

\noindent Consider a mono-frequent excitation $\mathcal{E}_0\hat{\bm e}_\lambda\cos\omega t$, where $\mathcal{E}_0$ is the amplitude of the light field. Inserting this expression into eq.~\eqref{eq:absorption.energy} gives after some straightforward calculation $\alpha(\omega)\propto -\omega\mathcal{E}_0\,\Im m\left(e^{i\omega t}\langle\hat{\bm e}_\lambda^*\bm{\mathcal{P}}\rangle\right)$. From the fluctuation-dissipation theorem we then find

\begin{equation}\label{eq:absorption.fluc-diss}
  \alpha_\lambda(\omega)\propto\Im m\left(i
  \int_{-\infty}^0 dt'\, \left<\left[\hat{\bm e}_\lambda^*\bm{\mathcal{P}}(0),
  \hat{\bm e}_\lambda\bm{\mathcal{P}}^\dagger(t') 
  \right]\right>_0\,e^{-i\omega t'}\right)\,,
\end{equation}

\noindent i.e. optical absorption is proportional to the spectrum of interband polarization fluctuations. We emphasize that this is a very general and important result that holds true for systems at finite temperatures, and is used for ab-initio type calculations of optically excited semiconductors \cite{onida:02}. We next show how to evaluate eq.~\eqref{eq:absorption.fluc-diss}. Suppose that the quantum dot is initially in its groundstate. Then only the term $\langle 0|\bm{\mathcal{P}}(0)\,\bm{\mathcal{P}}^\dagger(t')|0\rangle$ contributes in eq.~\eqref{eq:absorption.fluc-diss} because no electron-hole pair can be destroyed in the vacuum, i.e. $\bm{\mathcal{P}}|0\rangle=0$. Through $\bm{\mathcal{P}}^\dagger(t')|0\rangle$ an electron-hole pair is created in the quantum dot which propagates in presence of the quantum confinement and the Coulomb attraction between the electron and hole. Thus, the propagation of the interband polarization can be computed by use of the exciton eigenstates $|x\rangle$ through $\langle x|\bm{\mathcal{P}}^\dagger(t')|0\rangle=e^{iE_xt}\,\langle x|\bm{\mathcal{P}}^\dagger|0\rangle$. Inserting the complete set of exciton eigenstates in eq.~\eqref{eq:absorption.fluc-diss} gives 

\begin{equation}\label{eq:absorption.fluc-diss2}
  \alpha(\omega)\propto \Im m\left(i\sum_{x\lambda}
  \int_{-\infty}^0 dt'\,\left|M_{0x_\lambda}\right|^2\,e^{-i(\omega-E_{x_\lambda})t'}\right)\,.
\end{equation}

\noindent To evaluate the integral in eq.~\eqref{eq:absorption.fluc-diss2} we have to assume that the exciton energy has a small imaginary part $E_x-i\gamma$ associated to the finite exciton lifetime because of environment couplings (sec.~\ref{sec:decoherence}). Then, $\int_{-\infty}^0 dt'e^{-i(\omega-E_x+i\gamma)}=i/(\omega-E_x+i\gamma)$ and we obtain for the optical absorption the final result

\begin{equation}\label{eq:absorption-spectrum}
  \alpha(\omega)\propto \sum_{x\lambda}
  \left|M_{0x_\lambda}\right|^2\,\delta_\gamma(\omega-E_{x_\lambda})\,.
\end{equation}

\noindent Here $\delta_\gamma(\omega)=\gamma/(\omega^2+\gamma^2)$ is a Lorentzian which in the limit $\gamma\to 0$ gives Dirac's delta function. According to eq.~\eqref{eq:absorption-spectrum} the absorption spectrum of a single quantum dot is given by a comb of delta-like peaks at the energies of the exciton states, whose intensities ---sometimes referred to as the {\em oscillator strengths}--- are given by the square modulus of the dipole moments \eqref{eq:dipole-exciton}.

\subsubsection{Weak confinement}

In the weak confinement regime the absorption spectrum is given by

\begin{equation}\label{eq:absorption.weak}
  \alpha(\omega)\propto\sum_x\left|\int d\bm R\,\Phi_x(\bm R)\right|^2\,
  \delta_\gamma(\omega-\mathcal{E}_x)\,,
\end{equation}

\noindent where $\Phi_x(\bm R)$ is the center-of-mass wavefunction introduced in sec.~\ref{sec:excitons}. Let us consider the somewhat simplified example of a rectangular confinement with infinite barriers whose solutions are $\Phi(X,Y)=2/(L_1L_2)^{\frac 1 2}\,\sin(n_1\pi X/L_1)\sin(n_2\pi\, Y/L_2)$. Here $X$ and $Y$ are the center-of-mass coordinates along $x$ and $y$, $L_1$ and $L_2$ are the confinement lengths in $x$- and $y$-direction, and $n_1$ and $n_2$ the corresponding quantum numbers. The energy associated to this wavefunction is $\mathcal{E}=\pi^2/(2M)\,[(n_1/L_1)^2+(n_2/L_2)^2]$. Inserting these expressions into eq.~\eqref{eq:absorption.weak} shows that the oscillator strength is zero when $n_1$ or $n_2$ is an even number, and proportional to $L_1L_2/(n_1n_2)$ otherwise. Figure \ref{fig:absorption-weak} shows absorption spectra computed within this framework for (a) an inhomogeneously broadened ensemble of quantum dots and (b) a single dot. The first situation corresponds to typical optical experiments performed on ensembles of quantum dots. Single dots can be measured by different types of local spectroscopy such as submicron apertures \cite{brunner:94,zrenner:94,zrenner:00,guest:02,batteh:04}, solid immersion microscopy \cite{wu:99,wu:00}, or scanning near-field microscopy \cite{guest:01,guenther:02,matsuda:03}. Note that such single-dot spectroscopy is indispensable for the observation of the atomic-like optical density of states depicted in fig.~\ref{fig:absorption-weak}b, which is completely hidden in presence of the inhomogeneous broadening of fig.~\ref{fig:absorption-weak}a.

\subsubsection{Strong confinement}

In the strong confinement regime the optical response is governed by the single-particle properties. However, Coulomb interactions are responsible for renormalization effects which leave a clear fingerprint in the optical response. In the context of quantum-dot based quantum computation schemes (sec.~\ref{sec:computation}) it is precisely this fingerprint that allows the optical manipulation of individual few-particle states. Similarly to the absorption \eqref{eq:absorption.weak} in the weak-confinement regime, the linear optical absorption in the strong-confinement regime reads

\begin{equation}\label{eq:absorption.strong}
  \alpha(\omega)\propto\sum_x\left|\int d\bm r\, \Psi_x(\bm r,\bm r)\right|^2\,
  \delta(\omega-E_x)\,.
\end{equation}

\noindent Since the electron and hole are confined within a small space region, the oscillator strength is much smaller as compared to the weak confinement regime. The approximately product-type structure \eqref{eq:exciton.strong} of the exciton wavefunction and the similar shape of electron and hole wavefunctions gives rise to optical selection rules where only transitions between electron and hole states with corresponding quantum numbers, e.g. $s$--$s$ or $p$--$p$, are allowed. Indeed, such behavior is observed in fig.~\ref{fig:absorption-strong} showing absorption spectra representative for In$_x$Ga$_{1-x}$As dots: the three major peaks can be associated to transitions between the respective electron and hole ground states, and the first and second excited states \cite{hawrylak:98,hohenester.pss-b:00,simserides.prb:00}. In our calculations we assume parabolic confinement potentials for electrons and holes, with a 2:1 ratio between the electron and hole single-particle splittings \cite{sosnowskii:98}, and compute the spectra within a full configuration-interaction approach (appendix \ref{sec:CI}) for the respective six electron and hole single-particle states of lowest energy. The single-dot spectrum of fig.~\ref{fig:absorption-strong}b shows that Coulomb interactions result in a shift of oscillator strength to the transitions of lower energy (in a pure single-particle framework the ratio would be simply 1:2:3, reflecting the degeneracy of single-particle states), and the appearance of additional peaks \cite{simserides.prb:00,hawrylak:00}. For the dot ensemble, fig.~\ref{fig:absorption-strong}a, we observe that the broadening of the groundstate transition is much narrower than that of the excited ones. This is because the excited states are less confined and are accordingly stronger affected by Coulomb interactions. Note that for the level broadening considered in the figure the second and third exciton transitions even strongly overlap.

\subsection{Luminescence}\label{sec:luminescence}

Luminescence is the process where in a carrier complex one electron-hole pair recombines by emitting a photon. To account for the creation of photons we have to adopt the framework of second quantization of the light field \cite{mandel:95,walls:95,scully:97} and use expression \eqref{eq:efield.photon} for the electric field of photons. Then,

\begin{equation}\label{eq:light-matter.photon}
  H_{op}\cong-\frac i 2\sum_{\bm k\lambda}
  \left(\frac{2\pi\omega_k}{\kappa_s}\right)^{\frac 1 2}
  \left(\hat{\bm e}_{\bm k\lambda}^*\,a_{\bm k\lambda}^\dagger\,\bm{\mathcal{P}}-
  \hat{\bm e}_{\bm k\lambda}\,a_{\bm k\lambda}\,\bm{\mathcal{P}}^\dagger
  \right)
\end{equation}

\noindent is the hamilton operator which describes the photon-matter coupling within the envelope-function and rotating-wave approximations. The first term on the right-hand side describes the destruction of an electron-hole pair through photon emission, and the second one the reversed process. We shall now show how to compute from eq.~\eqref{eq:light-matter.photon} the luminescence spectrum $L(\omega)$. It is proportional to the increase in the number of photons $\langle a_{\bm k\lambda}^\dagger a_{\bm k\lambda}^{\phantom\dagger}\rangle$ emitted at a given energy $\omega$. With this approximation we obtain

\begin{equation}\label{eq:luminescence0}
  L(\omega)\propto \frac d {dt}\left(\sum_{\bm k\lambda}
  \left< a_{\bm k\lambda}^\dagger a_{\bm k\lambda}\right>
  \,\delta(\omega-\omega_{\bm k\lambda})\right)\,.
\end{equation}

\noindent We next use Heisenberg's equation of motion $\dot{\mathcal{O}}=-i[\mathcal{O},H]$, with $\mathcal{O}$ an arbitrary time-independent operator, and $\langle [a_{\bm k\lambda}^\dagger a_{\bm k\lambda}^{\phantom\dagger},H]\rangle=2i\,\Im m\langle a_{\bm k\lambda}^\dagger [a_{\bm k\lambda},H_{\rm op}]\rangle$. By computing the commutator with $H_{\rm op}$, eq.~\eqref{eq:light-matter.photon}, we obtain

\begin{equation}\label{eq:luminescence1}
  L(\omega)\propto\sum_{\bm k\lambda}
  \left(\frac{2\pi\omega_k}{\kappa_s}\right)^{\frac 1 2}
  \Im m\left< a_{\bm k\lambda}^\dagger\,\hat{\bm e}_{\bm k\lambda}^* 
  \bm{\mathcal{P}} \right>\,\delta(\omega-\omega_{\bm k\lambda})\,.
\end{equation}

\noindent The expression $\langle a^\dagger \bm{\mathcal{P}}\rangle$ is known as the {\em photon-assisted density matrix}\/ \cite{kira:98,rossi:02}. It describes the correlations between the photon and the electron-hole excitations in the quantum dot. Again, we can use the fluctuation-dissipation theorem \eqref{eq:fluctuation-dissipation} to compute eq.~\eqref{eq:luminescence1} and to provide a relation between the photon-assisted density matrix and the correlation function $\langle [a_{\bm k\lambda}^\dagger(0)\bm{\mathcal{P}}(0),a_{\bm k'\lambda'}(t)\bm{\mathcal{P}}^\dagger(t)] \rangle$. The calculation can be considerably simplified if we make the reasonable assumption that before photon emission no other photons are present. Then, only $\langle a_{\bm k'\lambda'}^{\phantom\dagger} a_{\bm k\lambda}^\dagger \rangle=\delta_{\bm k\bm k'}\delta_{\lambda\lambda'}$ does not vanish. It is diagonal in $\bm k$ and $\lambda$ because the only photon that can be destroyed by the annihilation operator $a$ is the one created by $a^\dagger$. Thus,

\begin{equation}
  L(\omega)\propto\pi\sum_{\bm k\lambda}
  \frac{2\pi\omega_k\mu_0^2}{\kappa_s}\,
  \Im m\left( i
  \int_{-\infty}^0 dt'\;\left<   
    \hat{\bm e}_{\bm k\lambda}   \bm{\mathcal{P}}^\dagger(t')\,
    \hat{\bm e}_{\bm k\lambda}^* \bm{\mathcal{P}}(0)
  \right> e^{i\omega t}\right)\,\delta(\omega-\omega_{\bm k\lambda})\,.
\end{equation}

\noindent Suppose that the system is initially in the eigenstate $|i\rangle$ of the unperturbed system which has energy $E_i$. We next insert a complete set of eigenstates $\sum_f\proj f f$ in the above equation (note that because the interband polarization operator $\bm{\mathcal{P}}$ can only remove one electron-hole pair the states $f$ and $i$ differ by one electron and hole), and assume that only photons propagating along $z$ with polarization $\lambda$ are detected. Then,

\begin{equation}\label{eq:luminescence-spectrum}
  L(\omega)\propto \sum_{f}
  \left|\langle f|\hat{\bm e}_\lambda^*\bm{\mathcal{P}}|i\rangle\right|^2\,
  \delta_\gamma(E_f+\omega-E_i)
\end{equation}

\noindent gives the expression for calculating luminescence spectra. Here, the photon energies  at the peak positions equal the energy differences of initial and final states, and the oscillator strengths are given by the overlap between the two wavefunctions subject to the condition that one electron-hole pair is removed through the interband polarization operator $\bm{\mathcal{P}}$.

\subsection{Multi and multi-charged excitons}

Let us first discuss the luminescence of multi-excitons, i.e. carrier complexes with an equal number of electron-hole pairs. To observe Coulomb renormalization effects in the optical spectra (such as, e.g. the biexciton shift $\Delta$) it is compulsory to measure single dots. For dot ensembles all line splittings would be completely hidden by the inhomogenenous broadening. The challenge to detect luminescence from single quantum dots (the density of typical self-assembled dots in the strong confinement regime is of the order of $5\times 10^{10}$ cm$^{-2}$ \cite{bimberg:98}) is accomplished by means of various experimental techniques, such as shadow masks or mesas \cite{zrenner:02}. Such {\em single-dot spectroscopy}\/ \cite{landin:98,motohisa:98,dekel:98,bayer:98,bayer:00a,dekel:00,rinaldi.prb:00,findeis:00} has revealed a surprisingly rich fine-structure in the optical spectra, with the main characteristic that whenever additional carriers are added to the dot the optical spectra change because of the resulting additional Coulomb interactions. This has the consequence that each quantum-dot spectrum uniquely reflects its electron-hole configuration. In the following we adopt the model of a quantum dot with cylinder symmetry (sec.~\ref{sec:excitons}), and compute the luminescence spectra for an increasing number of electron-hole pairs within a full configuration-interaction approach. Results are shown in fig.~\ref{fig:luminescence}. For the single-exciton decay the luminescence spectra exhibit a single peak at the exciton energy $E_0$ whose intensity is given by $|M_{0x}|^2$. In the biexciton decay one electron-hole pair in the Coulomb-renormalized carrier complex recombines by emitting a photon, whose energy is reduced by $\Delta$ because of Coulomb correlation effects. Things become more complicated when the number of electron-hole pairs is further increased. Here one pair has to be placed in the excited $p$-shell, which opens up the possibility for different decay channels. Because of wavefunction symmetry only electrons and holes in corresponding shells can recombine and emit a photon \cite{hawrylak:98,hawrylak:99}. For recombination in the $p$ shell, the energy $\omega\sim E_0+\Delta\epsilon_e+\Delta\epsilon_h$ of the emitted photon is blue-shifted by the energy splitting $\Delta\epsilon_e+\Delta\epsilon_h$ of single-particle states. On the other hand, recombination in the $s$-shell brings the system to an excited biexciton state with one electron-hole pair in the $s$ and one in the $p$ shell. Such excited states are subject to pronounced Coulomb renormalizations, which can be directly monitored in the luminescence spectra of fig.~\ref{fig:luminescence}. The two main features of luminescence from the different single-particle shells and the unambiguous spectroscopic fingerprint for each few-particle state because of Coulomb correlations prevail for the other multi-exciton complexes. Similar conclusions also apply for multi-charged excitons \cite{warburton:97,hartmann.prl:00,warburton:00,findeis.prb:01,finley:01,regelman:01,urbaszek:03,besombes:03}. Because of the strong single-particle character, the aufbau principle for negatively and positively charged excitons is dominated by successive filling of single-particle states, whereas Coulomb interactions only give rise to minor energy renormalizations. The only marked difference in comparison to multi excitons is the additional Coulomb repulsion due to the imbalance of electrons and holes, which manifests itself in the carrier-capture characteristics \cite{hartmann.prl:00,lomascolo:02} and in the instability of highly charged carrier complexes \cite{warburton:00,findeis.prb:01}. Typical multi-charged exciton spectra are shown in the right panel of fig.~\ref{fig:luminescence}. For negatively charged dots, the main peaks red-shift with increasing doping because of exchange and correlation effects, and each few-particle state has its own specific fingerprint in the optical response. When the dot is positively charged, the emission-peaks preferentially shift to the blue \cite{regelman:01}. This unique assignment of peaks or peak multiplets to given few-particle configurations allows in optical experiments to unambiguously determine the configuration of carrier complexes.

\subsection{Near-field scanning microscopy}\label{sec:snom}

Up to now we have been concerned with optical excitation and detection in the far-field regime, where the spatial dependence of the electric field $\bm{\mathcal{E}}(\bm r)$ can be safely neglected. However, the diffraction limit $\lambda/2$ of light can be significantly overcome through near-field optical microscopy \cite{paesler:96,hecht:00}. This is a technique based on scanning tunneling microscopy, where an optical fiber is used as the tip and light is quenched through it. Most importantly, close to the tip the electric field contribution is completely different from that in the far-field \cite{paesler:96,hanewinkel:97,jackson:62}. For the quantum dots of our present concern the carrier wavefunctions are always much stronger confined in the $z$-direction than in the lateral ones, which allows us to replace the generally quite complicated electro-magnetic field distribution in the vicinity of the tip \cite{paesler:96,bryant:99,liu:99} by a more simple shape, e.g. a Gaussian with a given full width of half maximum $\sigma_\mathcal{E}$. Up to now most of the local-spectroscopy experiments were performed with spatial resolutions $\sigma_\mathcal{E}$ larger than the extension of the semiconductor nanostructures themselves \cite{guest:02}. This allowed to locate them but not not spatially resolve their electron-hole wavefunctions. Only very recently, \citet{matsuda:02,matsuda:03} succeeded in a beautiful experiment to spatially map the exciton and biexciton wavefunctions of a quantum dot in the weak confinement regime. In the following we briefly discuss within the framework developed in refs.~\cite{mauritz:99,mauritz:00,simserides.prb:00} the main features of such local-spectroscopy experiments, and point to the difficulties inherent to their theoretical interpretation. Our starting point is given by the light-matter coupling \eqref{eq:light-matter2}. We assume, however, that the electric field $\bm{\mathcal{E}}$ has an explicit space dependence through

\begin{equation}\label{eq:electric.near-field}
  \bm{\mathcal{E}}^{(+)}(\bm r)=\mathcal{E}_0\,e^{-i\omega t}\,\bm\xi(\bm R-\bm r)\,.
\end{equation}

\noindent Here, $\mathcal{E}_0$ is the amplitude of the exciting laser with frequency $\omega$, and $\bm\xi(\bm R-\bm r)$ is the profile of the electric field in the vicinity of the fiber tip. The tip is assumed to be located at position $\bm R$. Scanning the tip over the sample thus allows to measure the local absorption (or luminescence \cite{savasta:02,pistone:04}) properties at different positions, and to acquire information about the electron-hole wavefunction on the nanoscale. The light-matter coupling for the electric field profile \eqref{eq:electric.near-field} is of the form

\begin{equation}\label{eq:light-matter.near}
  H_{\rm op}\cong \mathcal{E}_0\,\mu_0\sum_\lambda\int d\bm r\,
  \left(\hat{\bm e}_\lambda\,\ppsi_{\bar\lambda}^h(\bm r)\ppsi_\lambda^e(\bm r)\,
  e^{i\omega t}\bm\xi^*(\bm R-\bm r)+\mbox{h.c.}\right)\,.
\end{equation}

\noindent The remaining calculation to obtain the optical near-field absorption spectra is completely analogous to that of far-field, sec.~\ref{sec:absorption}, with the only difference that the light-matter coupling \eqref{eq:light-matter.near} has to be used instead of eq.~\eqref{eq:light-matter2}. We finally arrive at

\begin{equation}\label{eq:absorption.near}
  \alpha(\omega)\propto\sum_{x\lambda}\left|\int d\bm r\,
  \bm\xi^*(\bm R-\bm r)\,\hat{\bm e}_\lambda\Psi_{x_\lambda}(\bm r,\bm r)\right|^2\,
  \delta_\gamma(\omega-E_{x_\lambda})\,.
\end{equation}

\noindent In comparison to eq.~\eqref{eq:absorption-spectrum} the optical matrix element is given by the convolution of the electro-magnetic profile $\bm\xi^*(\bm R-\bm r)\,\hat{\bm e}_\lambda$ with the exciton wavefunction, rather than the simple spatial average of $\Psi(\bm r,\bm r)$. Two limiting cases can be readily identified in eq.~\eqref{eq:absorption.near}. First, for a far-field excitation $\bm\xi_0$ which does not depend on $\bm r$ one recovers precisely the far-field absorption \eqref{eq:absorption-spectrum}. In the opposite limit of infinite resolution, where $\bm\xi$ resembles a $\delta$-function, the oscillator strength is given by the square modulus of the exciton wavefunction $\Psi(\bm R,\bm R)$ at the tip position. Finally, within the intermediate regime of a narrow but finite probe, $\Psi(\bm r,\bm r)$ is averaged over a region which is determined by the spatial extension of the light beam. Therefore, excitonic transitions which are optically forbidden in the far-field may become visible in the near-field. Figure~\ref{fig:snom} shows near-field spectra as computed from eq.~\eqref{eq:absorption.near} for a quantum dot in the weak-confinement regime. The confinement for excitons and biexcitons is according to fig.~\ref{fig:confinement.weak}. In the second and third rows we report our calculated optical near-field spectra for spatial resolutions of 25 and 50 nm. Note that the first (fig.~b) and second excited state (not shown) are dipole forbidden, but have large oscillator strengths for both resolutions. As a result of interference effects, the spatial maps at finite spatial resolutions differ somewhat from the wavefunction maps, particularly for the excited states: the apparent localization is weaker, and in (c) the central lobe is very weak for both resolutions \cite{simserides.prb:00,hohenester.apl:04}. For the nearfield mapping of the biexciton we have to be more specific of how the system is excited. We assume that the dot is initially populated by the ground state exciton and that the near-field tip probes the transition to the biexciton ground state. This situation approximately corresponds to that
of Ref.~\cite{matsuda:02,matsuda:03} with non-resonant excitation in the non-linear power regime. Similarly to eq.~\eqref{eq:absorption.near}, the local spectra for biexcitons are given by

\begin{equation}\label{eq:absorption.near2}
  \alpha(\omega)\propto \left|
  \int d\bm rd\bm r_ed\bm r_h\, \bm\xi(\bm R-\bm r)\,\hat{\bm e}_\lambda^*\,
  \Psi_0^*(\bm r_e,\bm r_h)\bar\Psi_0(\bm r,\bm r,\bm r_e,\bm r_h)\right|^2\,
  \delta_\gamma(\omega+E_0-\bar E_0)\,.
\end{equation}

\noindent The corresponding spectra are shown in column (d) of figure \ref{fig:snom}. We observe that for the smaller spatial resolution the biexciton ground state depicts a stronger degree of localization than the exciton one, in nice agreement with the
recent experiment of \citet{matsuda:02,matsuda:03}.

\subsection{Coherent optical spectroscopy}

Absorption in a single quantum dot is the absorption of a single photon, which can usually not be measured. Other experimental techniques exist which allow to overcome this problem. In photo-luminescence-excitation spectroscopy an exciton is created in an excited state; through phonon scattering it relaxes to its state of lowest energy (sec.~\ref{sec:phonon-dephasing}), and finally recombines by emitting a photon which is detected. Other techniques are more sensitive to the coherence properties to be discussed below. For instance, in four-wave mixing spectroscopy \cite{shah:96,rossi:02}, the system is first excited by a sufficiently strong pump pulse, which creates a polarization. When at a later time a probe pulse arrives at the sample, a polarization grating is formed and light is emitted into a direction determined by those of the pump and probe pulse \cite{shah:96}. This signal carries direct information about how much of the polarization introduced by the first pulse is left at a later time, i.e. it is a direct measure of the coherence properties. Another technique is coherent nonlinear optical spectroscopy \cite{bonadeo:00} which is often used for quantum dots in the weak confinement regime. It offers a much better signal-to-noise ratio, and gives detailed information about the coherence properties of excitons and biexcitons.

\section{Quantum coherence and decoherence}\label{sec:decoherence}

Quantum coherence and decoherence are the two key players in the fields of quantum optics and semiconductor quantum optics. The light-matter coupling \eqref{eq:light-matter3} is mediated by the interband polarization which, in a microscopic description, corresponds to a coherent superposition of quantum states. For isolated systems this provides a unique means for quantum control, where the system wavefunction can be brought to any desired state \cite{rabitz:04}. Decoherence is the process that spoils such ideal performance. It is due to the fact that any quantum system interacts with its environment, e.g. photons or phonons, and hereby acquires an uncontrollable phase. This introduces a kind of ``random noise'' and diminishes the quantum-coherence properties. While from a pure quantum-control or quantum-computation perspective decoherence is often regarded as ``the enemy'' \cite{viola:98}, from a more physics-oriented perspective it is the grain of salt: not only it provides a means to monitor the state of the system, but also allows for deep insights to the detailed interplay of quantum systems with their environment. This section is devoted to a more careful analysis of these two key players. We first briefly review the basic concepts of light-induced quantum coherence and its loss due to environment couplings. Based on this discussion, we then show how decoherence can be directly monitored in optical spectroscopy and how it can be successfully exploited for single-photon devices.

\subsection{Quantum coherence}

In most cases we do not have to consider the full spectrum of quantum-dot states.  For instance, if the laser frequencies are tuned to the exciton groundstate it completely suffices to know the energy of the exciton together with the optical matrix element connecting the states. It is physical intuition together with the proper choice of the excitation scenario which allows to reduce a complicated few-particle problem to a relatively simple few-level scheme. This situation is quite different as compared to the description of carrier dynamics in higher-dimensional semiconductors, where such a clear-cut separation is not possible because of the scattering-type nature of carrier-carrier interactions \cite{haug:93,shah:96,rossi:02}. It is, however, quite similar to quantum optics \cite{mandel:95,walls:95,scully:97} which relies on phenomenological level schemes, e.g. $\Lambda$- or V-type schemes, with a few effective parameters --- a highly successful approach despite the tremendously complicated nature of atomic states. Let us denote the generic few-level scheme with $|i\rangle$, where $i$ labels the different states under consideration. If the artificial atom would be isolated from its environment we could describe it in terms of the wavefunction

\begin{equation}\label{eq:state.few-level}
  |\Psi\rangle=\sum_i C_i|i\rangle\,,
\end{equation}

\noindent with $C_i$ the coefficients subject to the normalization condition $\sum_i |C_i|^2=1$. Such wavefunction description is no longer possible for a system in contact with its environment. Because the coefficients $C_i$ acquire random phases through environment couplings and the system can suffer scatterings, we can only state with a certain probability that the system is in a given state. In statistical physics this lack of information is accounted for by the {\em density operator}\/ \cite{reichl:98}

\begin{equation}\label{eq:density-operator}
  \bm\rho=\sum_k p_k \proj{\Psi_k}{\Psi_k}=\overline{\proj\Psi\Psi}\,,
\end{equation}

\noindent where the sum of $k$ extends over an ensemble of systems which are with probability $p_k$ in the state $|\Psi_k\rangle$. The last term on the right-hand side provides the usual short-hand notation of this ensemble average. By construction, $\bm\rho$ is a hermitian operator whose time evolution is given by the Liouville von-Neumann equation $i\dot{\bm\rho}=[H,\bm\rho]$. This can be easily proven by differentiating eq.~\eqref{eq:density-operator} with respect to time and using Schr\"odinger's equation for the hamiltonian $H$ \cite{haug:93,scully:97,reichl:98}. If we insert the few-level wavefunctions \eqref{eq:state.few-level} into \eqref{eq:density-operator} we obtain

\begin{equation}\label{eq:density.few-level}
  \bm\rho=\sum_{ij}\overline{C_iC_j^*}\,\proj ij=\sum_{ij}\rho_{ij}\,\proj ij\,.
\end{equation}

\noindent Here, $\rho_{ij}=\overline{C_iC_j^*}$ is the {\em density matrix}\/ of the few-level system which contains the maximum information we possess about the system. The diagonal elements $\rho_{ii}$ account for the probability of finding the system in state $i$, and the off-diagonal elements $\rho_{ij}$ for the quantum coherence between states $i$ and $j$. As consequence $\bm\rho$ fulfills the trace relation $\tr\,\bm\rho=\sum_i \rho_{ii}=1$ which states that the system has to be in one of its states.

\subsubsection{Two-level system}

A particularly simple and illustrative example is given by a generic two-level system. This may correspond to an artificial atom that is either in its groundstate $0$ or in the single-exciton state $1$. In optical experiments the population of excited exciton states can be strongly suppressed through appropriate frequency filtering (recall that the principal level splittings are of the order of several tens of meV), and that of biexcitons through appropriate light polarization. Thus, systems with dominant two-level character can indeed be identified in artificial atoms. The density matrix is of dimension two. It has four complex matrix elements corresponding to eight real numbers. Because $\bm\rho$ is a hermitian matrix only four of them are independent, which additionally have to fulfill the normalization condition $\tr\,\bm\rho=1$. A convenient representation of $\bm\rho$ is through the Pauli matrices 

\begin{equation}\label{eq:pauli}
  \sigma_1=\proj 1 0 + \proj 0 1,\quad
  \sigma_2=-i\,\left(\proj 1 0 - \proj 0 1\right),\quad
  \sigma_3=\proj 1 1 - \proj 0 0\,,
\end{equation}

\noindent which together with the unit matrix $\one=\proj 1 1+\proj 0 0$ provide a complete basis within the two-level subspace. The Pauli matrices are hermitian and have trace zero (appendix~\ref{sec:two-level}). Thus, the density matrix can be expressed as

\begin{equation}\label{eq:density.pauli}
  \bm\rho=\frac 1 2\left(\one+\sum_i u_i\,\sigma_i\right)=
  \frac 1 2\left(\one+\bm u\,\bm\sigma\right)\,,
\end{equation}

\noindent with $\bm\sigma=(\sigma_1,\sigma_2,\sigma_3)$ and the first term guarantees the trace relation $\tr\,\bm\rho=1$. The system is thus fully characterized by the three-dimensional {\em Bloch vector}\/ $\bm u=(u_1,u_2,u_3)$ which was already introduced in sec.~\ref{sec:motivation.coherence}: its $x$- and $y$-components $u_1$ and $u_2$ account for the real and imaginary part of the quantum coherence ---or interband polarization---, respectively, and the $z$-component $u_3$ gives the population inversion between the excited and ground state. Equation~\eqref{eq:density.pauli} demonstrates that the Bloch-vector picture advocated in sec.~\ref{sec:motivation} prevails for density matrices. What happens when the system is excited by an external laser? As discussed in appendix \ref{sec:two-level}, within a rotating frame the system subject to an exciting laser can be described by the hamiltonian \cite{mandel:95}

\begin{equation}\label{eq:hamiltonian.two-level}
  H= \frac 1 2\biggl(\Delta\,\sigma_3-\Omega^*\,\proj 0 1 - \Omega\,\proj 1 0\biggr)\,,
\end{equation}

\noindent where $\Delta$ is the detuning between the laser and the two-level transition, and $\Omega$ is the {\em Rabi frequency}\/ which determines the strength of the light-matter coupling. Note that $\Omega$ describes the envelope part of the laser pulse which is constant for a constant laser and has an only small time dependence for typical pulses. From the Liouville von-Neumann equation $\dot{\bm\rho}=-i[H,\bm\rho]$ we then obtain the equation of motion for the Bloch vector

\begin{equation}\label{eq:bloch-equations}
  \dot{\bm u}=\bm\Omega\times\bm u,\quad
  \bm\Omega=(-\Re e(\Omega),\Im m(\Omega),\Delta)\,.
\end{equation}

\noindent We are now in the position to quantitatively describe the buildup of quantum coherence. Suppose that the system is initially in its groundstate $0$ where the Bloch vector points into the negative $z$-direction. When the laser is turned on, the Bloch vector is rotated perpendicularly to $\bm\Omega$. Upon expansion of the solutions of the Bloch equations \eqref{eq:bloch-equations} in powers of the driving field $\Omega$, we observe that to the lowest order the population, i.e. the $z$-component of the Bloch vector, remains unchanged and only a quantum coherence ---described by $u_1$ and $u_2$--- is created. This is due to the fact that the light couples indirectly, i.e. through the interband polarizations, to the quantum-state populations. When we consider in the solutions of eq.~\eqref{eq:bloch-equations} higher orders of $\Omega$ we find that this induced polarization acts back on the system and modifies the populations. A particularly simple and striking example of such nonlinear light-matter interactions is given by a constant driving field where the solutions of \eqref{eq:bloch-equations} can be found analytically \cite{rabi:37,mandel:95}

\begin{eqnarray}\label{eq:bloch-solutions}
  u_1(t) &=& \phantom{-} (\Delta\,\Omega)/(\Omega_{\rm eff}^2)\,
  \left(1-\cos\Omega_{\rm eff}t\right) \nonumber\\
  u_2(t) &=& -\Omega/(\Omega_{\rm eff})\,\sin\Omega_{\rm eff}t\nonumber\\
  u_3(t) &=& - \left(\Delta^2+\Omega^2\,\cos\Omega_{\rm eff}t\right)/(\Omega_{\rm eff}^2)\,.
\end{eqnarray}

\noindent Here $\Omega_{\rm eff}^2=\Omega^2+\Delta$ and we have assumed that $\Omega$ is entirely real. From eq.~\eqref{eq:bloch-solutions} we readily observe that after a time $T$ given by $\Omega_{\rm eff}\,T=2\pi$ the system returns into the initial state. For that reason pulses of duration $T$ are called $2\pi$-pulses. The phenomenon of a $2\pi$-rotation of the Bloch vector has been given the name {\em Rabi rotation}\/ \cite{rabi:37}. Figure \ref{fig:bloch.detuning} shows the trajectories of the Bloch vector for a pulse with $\Omega\,T=2\pi$ and for different detunings: only on resonance, i.e. for $\Delta=0$, the Bloch vector returns at time $T$ to its intial position, whereas off resonance $\bm u$ ends up in an excited state. We shall return to this point later in the discussion of self-induced transparency (sec.~\ref{sec:sit}). Rabi oscillations are a striking and impressive example of the nonlinear light-matter interaction. Indeed, the solutions \eqref{eq:bloch-solutions} clearly show that terms up to infinite order in $\Omega$ are required to account for the return of $\bm u$ to its initial position. For semiconductor quantum dots Rabi oscillations have been measured both in the time \cite{zrenner:02,htoon:02,borri:02,li:03} and frequency \cite{kamada:01} domain. A particularly beautiful experimental setup is due to \citet{zrenner:02}, where the authors used a quantum dot embedded in a field effect structure to convert the final exciton population to a photocurrent which could be directly measured. A somewhat different approach was pursued by \citet{kamada:01} where the appearance of additional peaks in the resonance-luminescence spectra at frequencies $\omega_0\pm\Omega_{\rm eff}$ centered around the laser frequency $\omega_0$ were observed --- a clear signature of Rabi-type oscillations \cite{mandel:95,scully:97}. We conclude this section with a short comment on the coherence properties of the Bloch-vector propagation. In fact, the time evolution \eqref{eq:bloch-solutions} of the isolated quantum system can be described in terms of the projector $\bm\rho=\proj\Psi\Psi$, where $|\Psi\rangle$ is the system wavefunction.  Such projector-like density operators have the unique property $\bm\rho^2=\bm\rho$. For the two-level system under consideration this has the consequence that the length of the Bloch vector remains one throughout. In other words, the trajectory of $\bm u$ is located on the surface of the {\em Bloch sphere}\/ (figs.~\ref{fig:bloch} and \ref{fig:bloch.detuning}), i.e. the unit sphere in the Bloch space.

\subsection{Decoherence}

Decoherence describes the process where a quantum system in contact with its environment loses its quantum-coherence properties. We shall assume that the environment ---sometimes referred to as a {\em reservoir}--- has an infinite number of degrees of freedom, and we are not able to precisely specify the corresponding statevector. To account for this lack of information, in the following we adopt the framework of statistical physics. Suppose that the problem under consideration is described by the hamiltonian $H_S+H_R+V$, where $H_S$ and $H_R$ account for the system and reservoir, respectively, and $V$ for their coupling. Let $\bm w(t)$ be the density operator of the total system in the interaction representation according to $H_S+H_R$. The quantity of interest is the {\em reduced density operator}\/ $\bm\rho$ of the system alone. It is obtained from the total density operator by tracing over the reservoir degrees of freedom through $\bm\rho=\tr_R\,\bm w$. We shall now derive the equation of motion for $\bm\rho$ when it is in contact with the environment. As a starting point we trace in the Lioville von-Neumann equation for $\bm w$ over the reservoir degrees of freedom and obtain

\begin{equation}\label{eq:liouville1}
  \dot{\bm\rho}(t)=\tr_R\,\dot{\bm w}(t)=-i\,\tr_R\,\left[V(t),\bm w(t)\right]\,.
\end{equation}

\noindent The important feature of this equation is that we are only able to trace out the reservoir on the left-hand side but not on the right-hand side which still depends on the full density operator $\bm w$. The simple and physical deep reason for this is that through $V$ the system and environment become entangled and can no longer be described independently. Suppose that at an early time $t_0$ system and reservoir were uncorrelated such that $\bm w(t_0)=\bm\rho(t_0)\otimes\bm\rho_R$, with $\bm\rho_R$ is the density operator of the reservoir. We can then use the usual time evolution operator $U(t,t_0)$ to establish a relation between $\bm w(t)$ and $\bm w(t_0)$. To the lowest order in the system-environment coupling $V$ we find \cite{walls:95}

\begin{equation}\label{eq:liouville2}
  \bm w(t)=\bm w(t_0)-\int_{t_0}^t dt'\,\left[V(t'),\bm\rho(t_0)\otimes\bm\rho_R\right]+
  \mathcal{O}(V^2)\,.
\end{equation}

\noindent This expression no longer depends on the density operator $\bm w(t)$ of the interacting system and environment. We can thus insert eq.~\eqref{eq:liouville2} into \eqref{eq:liouville1} to obtain an equation of motion for $\bm\rho$ which depends on the system degrees of freedom only. One additional approximation proves to be useful. To the same order in the system-environment interaction $V$ one can replace $\bm\rho(t_0)$ by $\bm\rho(t)$ \cite{walls:95} --- a well-defined procedure which can be justified for any given order of $V$ \cite{breuer:99,breuer:02}. This replacement is known as the {\em Markov approximation}\/ and has the advantage that the time evolution of the density operator $\bm\rho(t)$ only depends on its value at the same instant of time. Then,

\begin{equation}\label{eq:liouville3}
  \dot{\bm\rho}(t)\cong -\int_{t_0}^t dt'\,\tr_R\,\left[ V(t),\left[ V(t'),
  \bm\rho_R\otimes\bm\rho(t)\right]\right]\,,
\end{equation}

\noindent and we have made the assumption $\tr_R\,[V,\bm\rho(t_0)\otimes\bm\rho_R]=0$ which holds true in most cases of interest \cite{walls:95,scully:97}. Equation \eqref{eq:liouville3} is our final result. It has the evident structure that at time $t'$ the system becomes entangled with the environment, the entangled system and reservoir propagate for a while ---note that $V$ is given in the interaction representation according to $H_S+H_R$---, and finally a back-action on the system occurs at time $t$. Because of the finite interaction time, within the processes of decoherence and relaxation the system can acquire an uncontrollable phase or can exchange energy with the environment. The general structure of eq.~\eqref{eq:liouville3} prevails if higher orders of the interaction $V$ are considered \cite{nakajima:58,zwanzig:60,zwanzig:61,fick:90}, as will be also discussed at the example of phonon-assisted dephasing in sec.~\ref{sec:phonon-dephasing}.

\subsubsection{Caldeira-Leggett-type model}\label{sec:leggett}

To be more specific, in the following we consider the important case where a generic two-level system is coupled linearly to a bath of harmonic oscillators \cite{caldeira:81,caldeira:83,leggett:87}

\begin{equation}\label{eq:leggett}
  H=\frac{E_0}2\,\sigma_3\,+\sum_i\omega_i\,a_i^\dagger a_i
  +i\sum_i g_i\left(a_i^\dagger\,\sigma_--a_i\,\sigma_+\right)\,.
\end{equation}

\noindent Here, $E_0$ is the energy splitting between ground and excited state, $\omega_i$ the energies of the harmonic oscillators which are described by the bosonic field operator $a_i$, and $g_i$ the system-oscillator coupling constant which is assumed to be real. The last term on the right-hand side defines the system-environment interaction $V$ where we have introduced the lowering and raising operators $\sigma_-=\proj 0 1$ and $\sigma_+=\proj 1 0$ for the two-level system. If we insert $V$ in the interaction representation according to $H_S+H_R$ into eq.~\eqref{eq:liouville3} we obtain

\begin{eqnarray}\label{eq:leggett2}
  \dot{\bm\rho}(t)&\cong& -\int_{t_0}^t dt' \sum_{ij} g_ig_j\,\biggl(\nonumber\\
  & &\left< a_i(t)a_j^\dagger(t')\right>\sigma_+(t)\,\sigma_-(t')\,\bm\rho(t)
   + \left< a_j(t')a_i^\dagger(t)\right>\bm\rho(t)\,\sigma_+(t')\,\sigma_-(t)\nonumber\\
  &-&\left< a_i(t)a_j^\dagger(t')\right>\sigma_-(t')\,\bm\rho(t)\,\sigma_+(t)
   - \left< a_j(t')a_i^\dagger(t)\right>\sigma_-(t)\,\bm\rho(t)\,\sigma_+(t')\nonumber\\
  &+&\left< a_i^\dagger(t)a_j(t')\right>\sigma_-(t)\,\sigma_+(t')\,\bm\rho(t)
   + \left< a_j^\dagger(t')a_i(t)\right>\bm\rho(t)\,\sigma_-(t')\,\sigma_+(t)\nonumber\\
  &-&\left< a_i^\dagger(t)a_j(t')\right>\sigma_+(t')\,\bm\rho(t)\,\sigma_-(t)
   - \left< a_j^\dagger(t')a_i(t)\right>\sigma_+(t)\,\bm\rho(t)\,\sigma_-(t')\,
  \biggr)\,.\nonumber\\
\end{eqnarray}

\noindent Here, the terms in brackets have been derived by use of cyclic permutation under the trace and describe the propagation of excitations in the environment. The remaining terms with $\bm\rho$ and $\sigma$ account for the effects of environment coupling on the system. The expression in the second and third line describe {\em emission} processes where an environment excitation is created prior to its destruction, and those in the fourth and fifth line {\em absorption}\/ processes where the destruction is prior to the creation. The latter processes usually only occur at finite temperatures. In the following we suppose that $\bm\rho_R$ describes the reservoir in thermal equilibrium such that $\langle a_i^\dagger a_j\rangle\cong \delta_{ij}\,n(\omega_i)$, with $n(\omega)$ the usual Bose-Einstein distribution function. Within this spirit we have also neglected in eq.~\eqref{eq:leggett2} terms with $\langle aa\rangle$ and $\langle a^\dagger a^\dagger\rangle$ which would only play a role in specially prepared environments such as squeezed reservoirs \cite{walls:95}. Let us next consider one specific term in \eqref{eq:leggett2}, which can be simplified according to

\begin{equation}\label{eq:memory}
  \int_0^{t-t_0} d\tau\,\sum_{ij} g_ig_j\,\left< a_i(\tau)a_j^\dagger(0)\right>\,e^{iE_0\tau}
  \cong\sum_i g_i^2\,\left(1+n(\omega_i)\right)\,\gamma(E_0-\omega,t-t_0)\,,
\end{equation}

\noindent with $\gamma(\Omega,t)=\sin\Omega t/\Omega$ (and we have neglected terms which only contribute to energy renormalizations but not to decoherence and relaxation \cite{walls:95}). $\gamma(\Omega,t)$ has the important feature that in the adiabatic limit $\lim_{t\to\infty}\gamma(\Omega,t)=\pi\delta(\Omega)$ it gives Dirac's delta function. Thus, for sufficiently long times the various terms in eq.~\eqref{eq:leggett2} account for energy-conserving scattering processes (as discussed below the strict adiabatic $t\to\infty$ limit is usually not needed and it suffices to assume that the reservoir memory is sufficiently short-lived). Within the adiabatic limit the environment couplings can then be described by the scattering rates $\Gamma_1=2\pi \sum_i g_i^2 \,\left(1+n(\omega_i)\right)\,\delta(\omega_i-E_0)$ and $\Gamma_2=2\pi \sum_i g_i^2 \, n(\omega_i)\,\delta(\omega_i-E_0)$ for emission and absorption, respectively, and the Lindblad operators $L_1=\sqrt\Gamma_1\,\sigma_-$ and $L_2=\sqrt\Gamma_2\,\sigma_+$ associated to emission and absorption. We can bring eq.~\eqref{eq:leggett2} to the compact form 

\begin{equation}\label{eq:lindblad}
  \dot{\bm\rho}=-i(H_{\rm eff}\bm\rho-\bm\rho H_{\rm eff}^\dagger)+\sum_i L_i\bm\rho L_i^\dagger\,,
\end{equation}

\noindent where $H_{\rm eff}=H_S-(i/2)\sum_i L_i^\dagger L_i$ is an effective, non-hermitian hamiltonian. Equation \eqref{eq:lindblad} is known as a {\em master equation of Lindblad form}\/ \cite{lindblad:76,walls:95}. It has the intriguing feature that it is guaranteed that during the time evolution the trace over $\bm\rho$ remains one throughout.

\subsubsection{Unraveling of the master equation}

In many cases of interest the master equation \eqref{eq:lindblad} of Lindblad form can be solved by a simple and particularly transparent scheme. It is known as the {\em unraveling of the master equation} \cite{dum:92,dalibard:92,plenio:98}. Recall that the density operator is a statistical mixture of state vectors, $\bm\rho=\sum_k p_k\,\proj{\Psi_k}{\Psi_k}$, where the summation over $k$ results from the statistical average of the various pure states $|\Psi_k\rangle$. For simplicity we restrict ourselves to a single state vector $|\Psi\rangle$. The general case \eqref{eq:density-operator} then follows from a straightforward generalization. On insertion of the projector $\proj\Psi\Psi$ into the master equation \eqref{eq:lindblad} we obtain

\begin{equation}\label{eq:unraveling}
  \frac d {dt}\,\proj\Psi\Psi=
  -i\left(H_{\rm eff}\,\proj\Psi\Psi-\proj\Psi\Psi\,H_{\rm eff}^\dagger\right)+
  \sum_i L_i\,\proj\Psi\Psi\,L_i^\dagger\,.
\end{equation}

\noindent The first term on the right-hand side can be interpreted as a non-hermitian, Schr\"odinger-like evolution $i|\dot\Psi\rangle=H_{\rm eff}|\Psi\rangle$ under the influence of $H_{\rm eff}$. In contrast, the second term describes a time evolution where $|\Psi\rangle$ is projected ---or jumps--- to one of the possible states $L_i|\Psi\rangle$. For sufficiently small time intervals $\delta t$ the time evolution according to $H_{\rm eff}$ is given by $|\Psi(t+\delta t)\rangle=(1-iH_{\rm eff}\,\delta t)|\Psi(t)\rangle$. Note that $H_{\rm eff}$ is non-hermitian and consequently the wavefunction at later time is not normalized. To the lowest order in $\delta t$ the decrease of norm $\delta p$ is given by 

\begin{equation}\label{eq:unraveling2}
  \delta p=i\delta t\,\Bigl<\Psi(t)\Bigr|H_{\rm eff}-H_{\rm eff}^\dagger\Bigl|\Psi(t)\Bigr>
  =\delta t\sum_i\,\Bigl<\Psi(t)\Bigr|L_i^\dagger L_i\Bigl|\Psi(t)\Bigr>=
  \sum_i \delta p_i\,.
\end{equation}

\noindent The full master equation evolution has to preserve the norm. This missing norm $\delta p$ is brought in by the states $L_i|\Psi(t)\rangle$ to which the system is scattered with probability $\delta p_i$. The time evolution of the density operator can thus be decomposed into

\begin{equation}\label{eq:unraveling3}
  \bm\rho(t+\delta t)\cong 
  (1-\delta p)\,\bm\rho_0(t+\delta t)+\delta p\,\bm\rho_1(t+\delta t)\,,
\end{equation}

\noindent where $\bm\rho_0(t+\delta t)=U_{\rm eff}\bm\rho(t)U_{\rm eff}^\dagger$ with the non-unitary time evolution $U_{\rm eff}\cong 1-iH_{\rm eff}\delta t$ accounts for the unscattered part of the density operator, and $\bm\rho_1(t+\delta t)=\sum_i\delta p_i\,L_i\,\bm\rho(t)\,L_i^\dagger$ for the remainder where a scattering has occurred with probability $\delta p$. For that reason, the different parts of $\bm\rho$ are often refered to as {\em conditional density operators}\/ \cite{plenio:98}. This allows for a simple interpretation of the master equation \eqref{eq:lindblad}: the first term on the right-hand describes the propagation of the system in presence of $H_S$ and out-scatterings, which are responsible for decoherence, and the second one for in-scatterings which result in relaxation. This decomposition will prove particularly useful in the discussion of single-photon sources (sec.~\ref{sec:single-photon}).

\subsection{Photon scatterings}\label{sec:photons}

The light-photon coupling of equation \eqref{eq:light-matter.photon} can be described within the framework of the Caldeira-Leggett-type model. For a generic two-level system it is of the form \eqref{eq:leggett} where the summation index $i$ includes the photon wavevector $\bm k$ and polarization $\lambda$. The coupling constant reads

\begin{equation}\label{eq:coupling-photon}
  g_{\bm k\lambda}=\left(\frac{2\pi\omega_k}{\kappa_s}\right)^{\frac 1 2}
  (\hat{\bm e}_{\bm k\lambda}^*\,\hat{\bm e})\,M_{0x}\,,
\end{equation}

\noindent with $\hat{\bm e}$ the exciton polarization defined in sec.~\ref{sec:exciton.spin} and $M_{0x}$ the optical dipole matrix element of eq.~\eqref{eq:dipole-exciton}. Inserting eq.~\eqref{eq:coupling-photon} into \eqref{eq:memory} gives for the memory kernel

\begin{equation}\label{eq:memory2}
  \sum_{\bm k\lambda}\frac{2\pi\omega_k}{\kappa_s}\,
  (\hat{\bm e}_{\bm k\lambda}^*\,\hat{\bm e})^2\,\left|M_{0x}\right|^2
  \gamma(E_0-\omega_k,t)=
  \frac{4n}{3\pi c^2}\,\left|M_{0x}\right|^2\int_0^\infty\omega^3d\omega\,
  \gamma(E_0-\omega,t)\,.
\end{equation}

\noindent To arrive at the right-hand side we have replaced the summation over $\bm k$ by $\sum_{\bm k}\to(2\pi)^{-3}\int_0^\infty k^2 dk\,\int d\Omega$, and $\int d\Omega$ is the integration over all angles which has been performed analytically \cite{mandel:95,scully:97,hohenester.prb:03}. The integral over $\omega$ accounts for the temporal buildup of photon scatterings. It is shown in figure \ref{fig:gamma} as a function of time. Most remarkably, the asymptotic value is reached on a timescale of femtoseconds. Thus, in the description of scatterings one does not have to invoke the strict adiabatic limit $t\to\infty$ but the asymptotic scattering behavior is rather due to the extremely short-lived memory kernel of the reservoir. This is because the system is coupled to an infinite number of photon modes which interfere destructively in the scattering process. If we replace the integral in eq.~\eqref{eq:memory2} by its asymptotic value $\pi\omega_0^3$ we find for the scattering rate of spontaneous photon emission

\begin{equation}\label{eq:wigner-weisskopf}
  \Gamma=\frac{4n_s\mu_0^2\omega_0^3}{3c^3}\,\int d\bm r\,\left|\Psi_x(\bm r,\bm r)\right|^2
  \sim \int d\bm r\,\left|\Psi_x(\bm r,\bm r)\right|^2 \,\times\mbox{1 ns}^{-1}\,.
\end{equation}

\noindent This is the generalized Wigner-Weisskopf decay rate for a dipole radiator embedded in a medium with refractive index $n_s$. The nanosecond timescale given on the right-hand side of eq.~\eqref{eq:wigner-weisskopf} represents a typical value for GaAs- or InGaAs-based quantum dots. The values for $\int d\bm r\,\left|\Psi_x(\bm r,\bm r)\right|^2$ range from one in the strong-confinement regime to several tens in the weak-confinement regime (appendix.~\ref{sec:dipole.rigid}), and the corresponding scattering times $1/\Gamma$ from nanoseconds to a few tens of picoseconds \cite{bonadeo:98b,borri:01,lenihan:02}. Such finite lifetime of excited exciton states affects both the coherence properties and the lineshape of optical transitions. We first consider the case of Rabi-type oscillations in presence of a constant laser, which were already discussed at the beginning of this section. The equation of motion for the coherent time evolution follows from eq.~\eqref{eq:bloch-equations} and that for the incoherent part by eq.~\eqref{eq:bloch.t}, where the transverse and longitudinal scattering times $T_2/2=T_1=1/\Gamma$ are given by the Wigner-Weisskopf decay time (for details see appendix \ref{sec:two-level}). For simplicity we consider a resonant excitation $\Delta=0$ and assume $\Omega$ to be real. The motion of the Bloch vector then only takes place in the $(y,z)$-plane and can be computed from

\begin{equation}\label{eq:bloch-damped}
  \dot u_2 =  \Omega\,u_3-\frac\Gamma 2\,u_2\,,\quad
  \dot u_3 = -\Omega\,u_2-\Gamma(u_3+1)\,.
\end{equation}

\noindent Typical results for the propagation are shown in fig.~\ref{fig:bloch-damped}. We observe that Rabi-flopping occurs but is damped because of the finite exciton lifetime. In the limit $t\to\infty$ all oscillations become completely damped and the Bloch vector approaches $\bm u\cong\bm 0$. Such loss of coherence properties is a general property of decoherence and we will encounter similar results for the phonon-assisted dephasing (sec.~\ref{sec:phonon-dephasing}). We next discuss the influence of a finite exciton lifetime on the lineshape of optical transitions measured in absorption experiments (analogous conclusions hold for luminescence). Our starting point is given by eq.~\eqref{eq:absorption.fluc-diss} which, for the two-level system under consideration, states that the absorption spectrum is given by the spectrum of polarization fluctuations $\langle\sigma_-(0)\sigma_+(t)\rangle$. The objective to calculate from the equation of motion for the Bloch vector, which depends on only one time argument, the two-time correlation functions can be accomplished by different means. A popular one is based on the {\em quantum regression theorem}\/ which relates for a system initially decoupled from its environment the density-matrix to the two-time correlation functions \cite{lax:63,lax:68,mandel:95,walls:95}. The primary idea of this approach is as follows. Let $\bm\rho(t_0)$ denote the density operator at time $t_0$. The density operator at later time can be obtained by use of the time evolution operator $U(t,t_0)$ through $\bm\rho(t)=U(t,t_0)\,\bm\rho(t_0)\,U^\dagger(t,t_0)$. Upon insertion of a complete set of eigenstates $|i\rangle$ this equation can be transformed to matrix form

\begin{eqnarray}\label{eq:regression}
  \rho_{ij}(t)&=&\sum_{kl}
    \langle i|U(t,t_0)|k\rangle
    \langle k|\bm\rho(t_0)|l\rangle
    \langle l|U^\dagger(t,t_0)|j\rangle\nonumber\\
  &=& \sum_{kl} U_{ik}(t,t_0)\,U_{jl}^*(t,t_0)\,\rho_{kl}(t_0)\nonumber\\
  &=& \sum_{kl} G_{ij,kl}(t,t_0)\,\rho_{kl}(t_0)\,,
\end{eqnarray}

\noindent where the last equality defines the Green function $G_{ij,kl}(t,t_0)$. Once we know the Green function we can compute the expectation value for any operator $A$ according to $\langle A\rangle=\sum_{ij}\rho_{ij}(t)\,A_{ji}=\sum_{ij,kl} G_{ij,kl}(t,t_0)\,\rho_{kl}(t_0)\,A_{ji}$. However, it also allows the calculation of multi-time expectation values. Consider the correlation function for two operators $A$ and $B$ in the Heisenberg picture

\begin{eqnarray}\label{eq:regression2}
  \langle A(t)B(t_0)\rangle &=& \tr\,\bm\rho\,(t_0)\,U^\dagger(t,t_0)\,A\,U(t,t_0)\,B
  \nonumber\\
  &=& \langle i|\bm\rho(t_0)\proj j j U^\dagger(t,t_0)\proj l l A \proj m m
  U(t,t_0)\proj n n B |i\rangle\nonumber\\
  &=& U_{mn}(t,t_0)\,U_{lj}^*(t,t_0)\,B_{ni}\,\rho_{ij}(t_0)\,A_{lm}\nonumber\\
  &=& G_{ml,nj}(t,t_0)\,B_{ni}\,\rho_{ij}(t_0)\,A_{lm}\,,
\end{eqnarray}

\noindent where we have made use of the usual Einstein summation convention. Importantly, eq.~\eqref{eq:regression2} shows that the two-time correlation function can be computed by replacing the density-matrix $\rho_{ij}(t_0)$ at time $t_0$ by the modified expression $\sum_k B_{ik}\,\rho_{kj}(t_0)$. The result, which is know as the quantum regression theorem, implies that the fluctuations regress in time like the macroscopic averages. Equation \eqref{eq:regression2} holds exactly but the factorization of the density operator at time $t_0$ plays an essential role in the derivation \cite{mandel:95}. We shall now show how this result can be used to compute the polarization fluctuations $\langle\sigma_-(t)\sigma_+(t_0)\rangle$. According to the regression theorem we have to use instead of the initial density operator $\bm\rho(t_0)=\proj 0 0$ the modified $\sigma_+\,\bm\rho(t_0)=\proj 1 0=\sigma_+$ one. Inserting $\sigma_+$ into the Lindblad equation \eqref{eq:lindblad} gives 

\begin{equation}\label{eq:regression.photon}
  \dot\sigma_+=-i\left(E_0-i\frac\Gamma 2\right)\sigma_+\,,
\end{equation}

\noindent which shows that the excitation $\sigma_+$ propagates with the transition energy $E_0$ but is damped because of spontaneous photon emissions. For the correlation function we obtain $\langle\sigma_-(t)\sigma_+(t_0)\rangle=\exp-i[E_0-i(\Gamma/2)](t-t_0)$ which, upon insertion into eq.~\eqref{eq:absorption.fluc-diss}, gives the final result

\begin{equation}\label{eq:lineshape.photon}
  \alpha(\omega)\propto \frac{\Gamma/2}{(\omega-E_0)^2+\left(\Gamma/2\right)^2}\,.
\end{equation}

\noindent The lineshape for optical transitions of a two-level system subject to spontaneous photon emissions is a Lorentzian centered at $E_0$ and with a full-width of half maximum of $\Gamma/2$. In a nonlinear coherent-spectroscopy experiment \citet{bonadeo:98a} made the important observation that for excitons in the weak-confinement regime energy relaxation and dephasing rates are comparable and predominantly due to photon emissions, thus reflecting the absence of significant pure dephasing. Such behavior is quite surprising for the solid state since all interaction mechanisms can contribute to $T_2$ but only a few to $T_1$. Similar results were also found in the strong confinement regime where, however, things turn out to be more complicated (sec.~\ref{sec:phonon-dephasing}) \cite{borri:01,lenihan:02,borri:02b,borri:03,zwiller:04}.

\subsection{Single-photon sources}\label{sec:single-photon}

Single-photon sources are one of the most promising quantum-dot based quantum devices. The creation of a single photon on demand ---first a trigger is pushed and one single photon is emitted after a given time interval--- plays an important role in quantum cryptography, e.g. for secure key  distributions \cite{bouwmeester:00,gisin:02}. \citet{gerard:99} were the first to propose a turnstile single-photon source based on artificial atoms. Their proposal exploits two peculiarities of artificial atoms: first, because of Coulomb renormalizations of the few-particle states in the decay of a multi-exciton state each photon is emitted at a different frequency (see fig.~\ref{fig:luminescence}); second, because of environment couplings ---for details see below--- photons are always emitted from the few-particle state of lowest energy. Thus, in the cascade decay of a multi-exciton complex the last photon will always be that of the single-exciton decay, and this photon can be distinguished from the others through spectral filtering. This is how the quantum-dot based single-photon source works: a short pump laser excites electron-hole pairs in the continuum states in the vicinity of the quantum dot, where some become captured in the dot; the resulting multi-exciton complex decays by emitting photons --- because of Coulomb renormalizations each photon has a different frequency and because of environment couplings emission only takes place from the respective few-particle states of lowest energy; finally, the ``single photon on demand'' comes from the last single-exciton decay. Spectral filtering of the last photon is usually accomplished by placing the quantum dot in an optical resonator such as a microcavity \cite{yamamoto:00,michler:00}. The theoretical description of single scatterings is a quite non-trivial task. The framework of environment couplings developed in this section is based on statistical physics and thus applies to ensembles of (identical) systems only. How do things have to be modified for the description of a single system? Surprisingly enough not too much. The question of how to theoretically describe such problems first arose almost two decades ago when it became possible to store single ions in a Paul trap and to continuously monitor their resonance fluorescence, and led to the development of the celebrated {\em quantum-jump approach}\/ \cite{dum:92,dalibard:92,plenio:98}. This approach combines the usual master-equation approach with the rules of demolition quantum measurements \cite{hegerfeldt:93,plenio:98} and provides a flexible tool for the description of single-system dynamics subject to continuous monitoring. Suppose that the artificial atom and the photon environment at time $t_0$ are described by the density operator $\bm w(t_0)=\bm\rho(t_0)\otimes\bm\rho_R$. We shall now let the system evolve  for a short time $\delta t$ in presence of the light-matter coupling. This time $\delta t$ is supposed to be long enough to allow the photon to become separated from the dot ---see fig.~\ref{fig:gamma} for the buildup time of scatterings---, and short enough that only a single photon is emitted. What is the probability that a photon is detected within $\delta t$? Let $\mathbb{P}_0=\proj {0_R}{0_R}$ denote the projector on the photon vacuum $0_R$. Then

\begin{equation}\label{eq:measurement}
  P_0=\tr\,\mathbb{P}_0\,U(t_0+\delta t,t_0)\,\bm w(t_0)\,
  U^\dagger(t_0,t_0+\delta t)\,\mathbb{P}_0
\end{equation}

\noindent gives the probability that within $\delta t$ no photon is emitted. The term $U\bm wU^\dagger$ describes the propagation of the quantum dot coupled to photons, and the projection operators $\mathbb{P}_0$ the photon detection. With probability $P_0$ no photon is detected. If we correspondingly project in eq.~\eqref{eq:measurement} on the single-photon subspace $\mathbb{P}_1$ we get the probability $P_1$ that one photon is detected within $\delta t$. Obviously $P_0+P_1=1$ must be fulfilled. There is one important conclusion to be drawn from eq.~\eqref{eq:measurement}. If we compute according to eq.~\eqref{eq:liouville3}  the time evolution of the density operator to the lowest order in $V$, but replace the trace over the reservoir by the projection operators $\mathbb{P}_0$ and $\mathbb{P}_1$ ---which are associated to the outcome of the measurement--- we encounter expressions which are completely similar to those of the emission processes in the second and third line of eq.~\eqref{eq:leggett2}. However, the terms in the second line only show up for projection on $\mathbb{P}_0$ and those in the third line only for projection on $\mathbb{P}_1$. This dependence can be understood as follows. In the quantum-mechanical time evolution \eqref{eq:liouville3} of the master equation the density operator splits up into two terms associated to the situations where a photon is emitted or not. Through the measurement ---described by the projection operators $\mathbb{P}_0$ and $\mathbb{P}_1$--- we acquire additional information whether a photon has been emitted or not, and we corrispondingly have to modify the density operator. A particularly transparent description scheme for this propagation subject to quantum measurements is given by the master equation \eqref{eq:lindblad} of Lindblad form and its unraveling \eqref{eq:unraveling3} \cite{hegerfeldt:93,plenio:98}. In the time evolution of $\bm\rho(t)$ we assume that after each time interval $\delta t$ a gedanken measurement is performed \cite{hegerfeldt:93,plenio:98}, where either no photon or a single photon is detected. These two situations correspond to the two terms of eq.~\eqref{eq:unraveling3} with probabilities $P_0=1-\delta p$ and $P_1=\delta p$. The decomposition of the time evolution of the density operator into no-scattering and scattering contributions provides an elegant means to calculate probabilities for finite time intervals $[t_0,t]$. Let us introduce the conditional density operator $\bm\rho_0$, associated to no-photon detection, whose time evolution is given by

\begin{equation}\label{eq:conditional.dt}
  \dot{\bm\rho}_0=-i\left(H_{\rm eff}\,\bm\rho_0-\bm\rho_0\,H_{\rm eff}^\dagger\right)\,,
\end{equation}

\noindent subject to the initial condition $\bm\rho_0(t_0)=\bm\rho(t_0)$. Because $H_{\rm eff}$ is a non-hermitian operator, the trace $P_0(t)=\tr\,\bm\rho_0(t)$ decreases and gives the probability that the system has not emitted a photon within $[t_0,t]$. The probability that a photon is emitted at $t$ is given by $-\delta t\,\dot P_0(t)$. Once a photon has been detected, we acquire additional information about the system, and accordingly have to change its density operator. This is the point where the second term of the unraveled master equation \eqref{eq:unraveling3} comes into play. For the photon emission described by the Lindblad operator $L_i$ ($i$ may correspond to the photon polarization),

\begin{equation}\label{eq:conditional.new}
  \rho(t+\delta t)\longrightarrow \frac{L_i\bm\rho_0(t)L_i^\dagger}{\tr[L_i\bm\rho_0(t)L_i^\dagger]}
\end{equation}

\begin{sidewaystable}
\caption{The primary quantities and equations of interest of the quantum-jump approach \cite{plenio:98}. For discussion see text.
}\label{table:quantum-jump}
\begin{tabularx}{\columnwidth}{XX}
\hline\hline
Description & Expression \\
\hline
Full density operator & $\bm w(t)$ \\
Time evolution \eqref{eq:liouville3} of $\bm w(t)$ & $\dot{\bm w}(t)=-i[V(t),\bm w(t)]$ \\
& \\
Conditional density operator for no-photon emission & $\bm\rho_0(t)$ \\
Time evolution of $\bm\rho_0(t)$ & 
$\dot{\bm\rho}_0=-i(H_{\rm eff}\bm\rho_0-\bm\rho_0 H_{\rm eff}^\dagger)$ \\
Conditional time evolution operator for no-photon emission &
$U_{\rm eff}(t,t_0)=\exp[-iH_{\rm eff}(t-t_0)]$ for time independent $H_{\rm eff}$ \\
Conditional time evolution for no-photon emission in $[t_0,t]$ &
$U_{\rm eff}(t,t_0)\,\bm\rho(t_0)\,U_{\rm eff}^\dagger(t,t_0)$ \\
& \\
Projection on photon vacuum; no photon detected & 
$\mathbb{P}_0$ \\
Projection on single-photon subspace; photon detected & 
$\mathbb{P}_1$ \\
& \\
Probability that photon is detected in $[t,t+\delta t]\,^{a}$ &
$\tr\,\mathbb{P}_1 U(t,t+\delta t)\bm w(t)U^\dagger(t,t+\delta t)\mathbb{P}_1\to
\delta t\sum_i\tr[\bm\rho(t)\,L_i^\dagger L_i]$ \\
Probability that no photon is detected in $[t_0,t]\,^{a}$ &
$P_0(t)=\tr\,\mathbb{P}_0 U(t,t_0)\bm w(t_0)U^\dagger(t,t_0)\to
\tr[U_{\rm eff}(t,t_0)\bm\rho(t_0)U_{\rm eff}^\dagger(t,t_0)]$ \\
& \\
Full density operator after photon detection at time $t$ &
$\bm w(t)=\tr_R[\mathbb{P}_1\bm w(t)\mathbb{P}_1]\,\mathbb{P}_0$\\
System density operator after detection of photon $i$ &
$\bm\rho(t)=L_i\bm\rho_0(t)L_i^\dagger/\tr[.]$ \\
\hline\hline
\end{tabularx}
$^a$The two expressions on the right-hand side correspond, respectively, to the full density operator $\bm w(t)$ and the reduced density operator $\bm\rho(t)$ of the system. The latter is computed within the approximation of a master equation \eqref{eq:lindblad} in Lindblad form.
\end{sidewaystable}

\noindent gives the density operator right after the scattering (the denominator guarantees that the density operator after the scattering fulfills the trace relation). Table \ref{table:quantum-jump} lists some of the basic quantities of this scheme which is known as the {\em quantum-jump approach}\/ \cite{plenio:98}. Let us consider as a first example the situation where a quantum dot is initially in the single-exciton state. With the Lindblad operator $L=\sqrt\Gamma\,\sigma_-$ associated to photon emission, we obtain for the effective hamiltonian $H_{\rm eff}=-i(\Gamma/2)\proj 1 1$. Thus the decay of $P_0(t)=\exp-\Gamma t$ is mono-exponentially and the probability to detect a photon at time $t$ is given by $-\delta t\,\dot P_0(t)=\delta t\,\Gamma\,\exp-\Gamma t$.

\paragraph{Photon antibunching.}

We next consider the situation where a single quantum dot is driven by a constant laser with Rabi frequency $\Omega$, and the resonance luminescence ---sometime refered to as resonance fluorescence--- is measured. The quantity we are interested in is the probability that once a photon has been detected at time zero the {\em next}\/ photon is detected at time $t$. It is similar to the two-photon correlation function $g^{(2)}(t)$ \cite{mandel:95,scully:97,plenio:98}, with the only difference that we ask for the next instead of any subsequent photon. With the framework of the quantum-jump approach we are in the position to readily compute things. This is how it goes. Suppose that the system is initially in state $|0\rangle$. The effective hamiltonian in presence of the driving laser and of photon emissions is $H_{\rm eff}=(1/2)[\Delta\,\sigma_3-\Omega\,\sigma_1-i\Gamma(\one+\sigma_3)/2]$, with $\Delta$ the detuning between the laser and the two-level transition which we assume to be zero. As shown in appendix \ref{sec:two-level}, the probability $P_0(t)$ for no-photon emission can then be computed analytically 

\begin{equation}\label{eq:prob.no-photon}
  P_0(t)=e^{-\sinh\theta\,\Omega_{\rm eff}t}
  \Bigl(1+\sinh\theta\,\sin\Omega_{\rm eff}t\Bigr)\,,
\end{equation}

\noindent where $\Omega_{\rm eff}^2=\Omega^2-(\Gamma/2)^2$ for $\Omega>\Gamma/2$ and we have defined the angle $\theta$ through $\tanh\theta=(\Gamma/2\Omega)$. Equation \eqref{eq:prob.no-photon} gives the probability that a system ---which is initially in its groundstate and is subject to a constant laser field--- has emitted no photon within $[0,t]$. Small angles $\theta$ refer to the case that photon scatterings occur seldomly on the time scale of $1/\Omega$, and large angles to the case where photon emissions and Rabi flopping take place on the same time scale. Figure \ref{fig:photon-count}a shows $P_0(t)$ as computed from eq.~\eqref{eq:prob.no-photon} for different values of $\theta$. We observe that in all cases $P_0(t)$ decays exponentially ---the decay constant is given by $\sinh\theta\,\Omega_{\rm eff}$--- and is modulated by the Rabi-type oscillations of $\sin\Omega_{\rm eff}t$. The latter oscillations reflect the fact that it requires the driving field $\Omega$ to bring the system from the ground to the excited state and eventually back to the ground state, and that photons can only be emitted from the excited state. This is also clearly shown in fig.~\ref{fig:photon-count}b which shows the probability distribution $-\dot P_0(t)$ for the emission of the next photon: all three curves start at zero and it requires a finite time to bring the system to the excited state where it can emit a photon. Finally, fig.~\ref{fig:photon-count}c shows the probability distribution that after a photon count at time zero any other photon is detected at $t$. It is computed from the quantum regression theorem \eqref{eq:regression2} for an initial density operator $\proj 0 0$. While at later times no correlation between the first and the subsequent photon count exist, at early times there is a strong anti-correlation because of the above-mentioned laser-mediated excitation of the upper state. This is a genuine single-system effect ---for an ensemble of two-level systems the two-photon correlation would be a Poissonian distribution \cite{mandel:95}---, and is known as {\em photon antibunching}.\/ Indeed, such behavior has been clearly observed in the two-photon correlations of single quantum dots. Using pulsed laser excitation single-photon turnstile devices that generate trains of single-photon pulses were demonstrated \cite{michler:00,santori:01,pelton:02,santori:02,waks:02,baier:04}. In a somewhat different scheme, electroluminescence from a single quantum dot within the intrinsic region of a $p$-$i$-$n$ junction was shown to act as an electrically driven single-photon source \cite{yuan:02,benson:00}. Also the decay of multi-exciton states has attracted great interest, since it provides a source for multicolor photons with tunable correlation properties \cite{regelman:01b}. In the quantum cascade decay of the biexciton it was demonstrated that the first photon emitted from the biexciton-to-exciton decay is always followed by the photon of the single-exciton decay, i.e. {\em photon bunching}\/ \cite{moreau:01}. When both photons are emitted along $z$ ---which could be achieved e.g. by an appropriate design of the microcavity--- the two photons not only differ in energy but also in their polarizations, which could be used for the creation of entangled photons \cite{benson:00,gywat:02,stace:03,sifel.apl:03,hohenester.prb:03}. 

We conclude this section with a more conceptual problem. It is known that under quite broad conditions quantum measurements lead to a wavefunction collapse. How does this collapse show up in the quantum-jump approach under consideration? Suppose that a photon is detected within the time interval $[t+\delta t]$. According to the von-Neumann-L\"uders rule the total density operator $\bm w$ has to be changed to \cite{hegerfeldt:93,plenio:98}

\begin{eqnarray}\label{eq:reduction.w}
  \bm w(t+\delta t) \longrightarrow \tr_R\left(\mathbb{P}_1\,U(t+\delta t,t)\,\bm w(t)\,
  U^\dagger(t+\delta t,t)\,\mathbb{P}_1\right)\mathbb{P}_0\,.
\end{eqnarray}

\noindent Here, the projector $\mathbb{P}_1$ accounts for the photon detection, and the reservoir trace together with $\mathbb{P}_0$ for the demolition measurement where the photo detector absorbs the photon. The latter procedure leads to the collapse of the photon wavefunction. In eq.~\eqref{eq:conditional.new} the influence of the photon measurement on $\bm\rho$ is less obvious. In comparison to the full master equation \eqref{eq:lindblad} the main effect of the measurement is that we acquire additional information about the photon environment, and we corrispondingly modify the system density operator. Is this equivalent to a wavefunction collapse? The solution to this problem is quite subtle. In the derivation of the master equation \eqref{eq:lindblad} of Lindblad form we made the adiabatic approximation $t\to\infty$, which, as shown at the example of photons in fig.~\ref{fig:gamma}, is equivalent to the assumption of a sufficiently short-lived reservoir memory. In other words, in the process of photon emission described by the Lindblad operators $L_i$ the photon becomes fully decoupled from the system. Thus, when we measure the photon no back-action on the system occurs. The environment is used ``as a witness'' \cite{zurek:03} which provides information about the system (i.e. whether it has emitted a photon or not). For that reason, we are neither forced to explicitly introduce a wavefunction collapse in the purification process \eqref{eq:conditional.new}, nor does it matter whether the photon is detected directly after emission or travels some distance before detection (since the time evolution of the system is described identically in both cases). On the other hand, we promise that we will use the photon only to perform photon counting but will not try to accurately measure its frequency --- which would require a sufficiently long interaction time between the quantum dot and the photon, within which the two objects would become entangled.

\subsection{Phonon scatterings}\label{sec:phonon-dephasing}

In addition to the photon coupling, carriers in artificial atoms experience interactions with genuine solid-state excitations such as, e.g. phonons. For sufficiently small interlevel splittings phonon scatterings can be described within the framework of the Caldeira-Leggett model of sec.~\ref{sec:leggett}. Theoretical estimates for the corresponding relaxation times are of the order of several tens of picoseconds \cite{bockelmann:90,benisty:91,bockelmann:93,bockelmann:94}. This finally justifies our assumption made in single-dot spectroscopy that photon emission always occurs from the few-particle states of lowest energy. However, things are considerably more difficult when the interlevel splitting is larger than the phonon energies. This is the case for most types of self-assembled dots where the level splitting is of the order of 50--100 meV, to be compared with the energy of longitudinal optical phonons of 36 meV in GaAs. According to Fermi's golden rule \eqref{eq:golden.rule}, scatterings should here become completely inhibited because of the lack of energy conservation. This led to the prediction of the so-called {\em phonon bottleneck}\/ \cite{bockelmann:90,benisty:91}. Most experimental studies revealed, however, a fast intradot relaxation of optically excited carriers \cite{ohnesorge:96,heitz:97,grosse:97,sosnowskii:98}. Furthermore, \citet{borri:01,borri:02b,borri:03} observed in optical coherence spectroscopy experiments that phonon-induced decoherence can even occur in complete absence of relaxation. Such decoherence is due to the lattice deformation induced by the optical excitation and the resulting formation of a {\em polaron},\/ i.e. a composite exciton-phonon excitation. In the following we first briefly review the theoretical description of polarons and phonon-assisted dephasing. Based on this, we then reexamine relaxation processes beyond the framework of Fermi's golden rule.

\subsubsection{Spin-boson model}\label{sec:spin-boson}

Consider the model where a generic two-level system is coupled linearly to a reservoir of harmonic oscillators such that the interaction only occurs when the system is in the upper state \cite{mahan:81}

\begin{equation}\label{eq:spin-boson}
  H=E_0\,\proj 1 1+\sum_i\omega_i\,a_i^\dagger a_i+
  \sum_i g_i\left(a_i^\dagger+a_i\right)\proj 1 1\,.
\end{equation}

\noindent Here, $\omega_i$ is the phonon energy, $a_i$ the bosonic field operator for phonons, and $g_i$ the coupling constant (details will be presented below). In comparison to the Caldeira-Leggett-type model \eqref{eq:leggett}, the so-called {\em spin-boson model}\/ of eq.~\eqref{eq:spin-boson} does not induce transitions between the two levels. Yet it leads to decoherence. This can be easily seen by writing eq.~\eqref{eq:spin-boson} in the interaction picture according to the hamiltonian for the uncoupled quantum dot and phonons

\begin{equation}\label{eq:spin-boson.v}
  V(t)=\sum_i g_i\left(e^{i\omega_i t}\,a_i^\dagger+e^{-i\omega_i t}\,a_i\right)\proj 1 1\,.
\end{equation}

\noindent Through the phonon coupling the two-level system becomes entangled with the phonons, where each phonon mode evolves with a different frequency $\omega_i$. If we trace out the phonon degrees of freedom ---similarly to the procedure employed in the derivation of the Lindblad equation \eqref{eq:lindblad}---, the different exponentials $e^{\pm i\omega t}$ interfere destructively, which leads to decoherence. Because this decoherence is not accompanied by relaxation, the process has been given the name {\em pure dephasing}.\/ We shall now study things more thoroughly. We first note that the dot-phonon coupling term can be removed through the transformation \cite{duke:65,mahan:81,jacak:03}

\begin{equation}
  e^s\,H\,e^{-s}=\bar E_0\,\proj 1 1+
  \sum_i\omega_i\,a_i^\dagger a_i\,,
\end{equation}

\noindent with $s=\proj 1 1\sum_i\xi_i(a_i^\dagger-a_i)$ an anti-hermitian operator, $\bar E_0=E_0-\sum_i (g_i)^2/\omega_i$ the renormalized two-level energy, and $\xi_i=g_i/\omega_i$. The simple physical reason is that for the hamiltonian \eqref{eq:spin-boson} the oscillator equilibrium positions are different for the ground and excited state of the two-level system, and $e^s$ ---which is closely related to the usual displacement operator $D(\xi)=e^{\xi a^\dagger-\xi^* a}$ of the harmonic oscillator \cite{mandel:95,walls:95,scully:97}--- accounts for this displacement of positions. Let us first study the lineshape of optical transitions resulting from the phonon coupling \eqref{eq:spin-boson}. As shown in appendix~\ref{sec:independent-boson}, within the spin-boson model the polarization fluctuations governing the absorption spectra \eqref{eq:absorption.fluc-diss} can be computed analytically \cite{mahan:81}

\begin{equation}\label{eq:polaron.fluc}
  \langle\sigma_-(0)\sigma_+(t)\rangle=e^{i\bar E_0t}\,
  \exp{\sum_i\left(\frac{g_i}{\omega_i}\right)^2\left[
    i\,\sin\omega_i t-\bigl(1-\cos\omega_i t\bigr)\,\coth\frac{\beta\omega_i}2\right]}\,,
\end{equation}

\noindent with $\beta$ the inverse temperature. Because the final result \eqref{eq:polaron.fluc} is exact (within the limits of our model hamiltonian), it can be employed for arbitrarily strong phonon couplings $g_i$. In addition, it provides a prototypical model for decoherence which has found widespread applications in various fields of research \cite{viola:98,breuer:02,uchiyama:02,smirnov:03,vagov:03}. For semiconductor quantum dots eq.~\eqref{eq:polaron.fluc} and related expressions have been widely used for the description of optical properties \cite{kral:98,stauber:00,krummheuer:02,vagov:02,vagov:03,jacak:03}. We now follow \citet{krummheuer:02} and derive explicit results for GaAs-based quantum dots. For simplicity we assume a spherical dot model and acoustic deformation potential interactions \cite{krummheuer:02,foerstner:03a}

\begin{equation}\label{eq:acoustic-deformation}
  g_q=\left(\frac q{2\rho c_\ell}\right)^{\frac 12}\,\bigl(D_e-D_h\bigr)\,e^{-q^2 L^2/4}
\end{equation}

\noindent as the only coupling mechanism. Here, $q$ is the phonon wavevector, $\rho$ the mass density, $c_\ell$ the longitudinal sound velocity, $D_e$ and $D_h$ the deformation potentials for electrons and holes, respectively, and the electron and hole wavefunctions have been approximated by Gaussians with the same carrier localization length $L$. Because the exponential in \eqref{eq:acoustic-deformation} introduces an effective cutoff for the wavevectors $q$, we do not have to explicitly account for the cutoff at the Debye frequency. It turns out to be convenient to measure length in units of $L$, wavevectors in units of $1/L$, energy in units of $c_\ell/L$, and time in units of $L/c_\ell$. With material parameters representative for GaAs \cite{krummheuer:02,foerstner:03a} and a carrier localization length $L=5$ nm, we obtain, respectively, 1 ps, 0.7 meV, and 7.8 K for the time, energy, and temperature scale. The dot-phonon coupling strength can be expressed in dimensionless form

\begin{equation}\label{eq:dot-phonon.coupling}
  \alpha_p=\frac{(D_e-D_h)^2}{4\pi^2\rho\,c_\ell^3\,L^2}\cong 0.033\,,
\end{equation}

\noindent where the estimate on the right-hand side corresponds to the material and dot parameters listed above. With the natural units for time, energy, and temperature, the coupling strength $\alpha_p$ of eq.~\eqref{eq:dot-phonon.coupling} becomes the only parameter of the spin-boson model. We then get $\sum_i(g_i)^2/\omega_i=\alpha_p(\pi/2)^\frac 12$ for the renormalization of the two-level energy, and $\exp\,\alpha_p\,\int_0^\infty xdx\,e^{-x^2/4}\,[i\,\sin xt-(1-\cos xt)\,\coth(\beta x/2)]$ for the polarization correlation function \eqref{eq:polaron.fluc}. This function is shown in fig.~\ref{fig:polaron}a for different temperatures. We observe a decay at early times associated to phonon dephasing ---i.e. part of the quantum coherence is transfered from the two-level system to the phonons---, and the curves approach a constant value at later times. The asymptotic value decreases with increasing temperature. In a sense, this finding is reminiscent of the Franck-Condon principle of optically excited molecules: because the equilibrium positions of the ions in the ground and excited state are different, after photo excitation the molecule ends up in an excited vibrational state. However, in contrast to the molecule, which just couples to a few vibrational modes, optical excitations in artificial atoms couple to a continuum of phonon modes which all evolve with a different frequency. In the spirit of the random-phase approximation this introduces decoherence. Figure~\ref{fig:polaron}b shows the imaginary part of the fourier transform of $\langle\sigma_-(0)\sigma_+(t)\rangle$, which is proportional to absorption. In addition to the delta-peak at energy $\bar E_0$ ---which would acquire a Lorentzian shape \eqref{eq:lineshape.photon}  in presence of photon emissions---, the spin-boson coupling \eqref{eq:spin-boson} gives rise to a broad continuum in the optical spectra that increases with increasing temperature. Such behavior has been observed experimentally and attributed to phonon dephasing \cite{borri:01}.

\paragraph{Rabi-type oscillations.}

We next discuss the influence of the dot-phonon coupling \eqref{eq:spin-boson} on the coherent optical response of artificial atoms. In contrast to the previous section, where all results could be obtained analytically, in presence of a laser pulse $\Omega$ with arbitrary strength the solution of the equations of motion is more cumbersome \cite{viola:98,uchiyama:02,vagov:03} and one is forced to introduce an approximate description scheme \cite{foerstner:03a,foerstner:03b,hohenester.prl:04}. To this end, in the following we adopt a density-matrix description. Our starting point is given by the Heisenberg equations of motion for $\bm\sigma$ and $a_i$, according to the hamiltonian \eqref{eq:hamiltonian.two-level} and the spin-boson coupling \eqref{eq:spin-boson},

\begin{equation}\label{eq:polaron-heisenberg}
  \dot{\bm\sigma} = \bm\Omega\times\bm\sigma+\sum_i g_i\left(a_i^\dagger+a_i\right)\,
  \hat{\bm e}_3\times\bm\sigma\,,\quad
  \dot a_i = -i\Bigl(\omega_i\,a_i+g_i\,\proj 1 1\Bigr)\,.
\end{equation}

\noindent The vector $\bm\Omega$ is defined in eq.~\eqref{eq:bloch-equations}. Our objective now is to derive from \eqref{eq:polaron-heisenberg} an approximate equation of motion for the Bloch vector $\bm u=\langle\bm\sigma\rangle$. Multiplying in eq.~\eqref{eq:polaron-heisenberg} the total density operator $\bm w$ from the left-hand side and tracing over the system and phonon degrees of freedom, shows that the Bloch vector couples to $\langle a_i\bm\sigma\rangle$. If we would derive similarly to eq.~\eqref{eq:polaron-heisenberg} the equation of motion for $\langle a_i\bm\sigma\rangle$, we would find that it couples to higher-order density matrices such as $\langle a_ia_j\bm\sigma\rangle$. This is because through the spin-boson coupling \eqref{eq:spin-boson} the two-level system becomes entangled with the phonons, and each density matrix couples to a matrix of higher order. The resulting infinite hierarchy of density matrices has been given the name {\em density matrix hierarchy}\/ \cite{rossi:02,bonitz:98}. We shall now introduce a suitable truncation scheme. Consider a generic expectation value $\langle AB\rangle$ with two arbitrary operators $A$ and $B$. If there was no correlation between the two operators, the expectation value would be simply the product $\langle A\rangle\,\langle B\rangle$. We shall now lump all correlations between $A$ and $B$ into the correlation function $\langle\!\langle AB\rangle\!\rangle$, and express the expectation value of the two operators through $\langle AB\rangle=\langle A\rangle\,\langle B\rangle+\langle\!\langle AB\rangle\!\rangle$. Corresponding factorization schemes ---which are known as {\em cumulant expansions}\/--- apply to expectation values with more operators such as, e.g. $\langle ABC\rangle$ \cite{balescu:63,reichl:98,rossi:02}. In the common truncation scheme of the density-matrix hierarchy one selects a few cumulants, which are expected to be of importance, and neglects all remaining ones. To the lowest order of approximation, within the spin-boson model we keep the Bloch vector $\langle\bm\sigma\rangle$, the coherent phonon amplitude $s_i=\langle a_i\rangle$, and the phonon-assisted density matrix $\bm u_i=\langle\!\langle a_i\bm\sigma\rangle\!\rangle$ as dynamic variables. Their equations of motion can be readily obtained from eq.~\eqref{eq:polaron-heisenberg}, and we obtain

\begin{eqnarray}\label{eq:polaron.density}
\dot{\bm u}&=&\bm\Omega_{\rm eff}\times \bm u+2\sum_i g_i\,\hat{\bm e}_3\times\Re e(\bm u_i)\nonumber\\
\dot s_i &=& -i\,\omega_i\,s_i - \frac i 2\, g_i\, (1+u_3)\nonumber\\
\dot{\bm u}_i &=&\bm\Omega_{\rm eff}\times \bm u_i-i\,\omega_i\,\bm u_\lambda 
 +g_i\, \left(n_i+\frac 1 2\right)\, \hat{\bm e}_3\times \bm u
 +\frac i 2\, g_i\left(u_3\,\bm u-\hat{\bm e}_3\right)\,,\qquad
\end{eqnarray}

\noindent with $\bm\Omega_{\rm eff}=\bm\Omega+2\,\Re e\sum_i g_i\, s_i\,\hat{\bm e}_3$, and $n_i=\langle\!\langle a_i^\dagger a_i\rangle\!\rangle$ the phonon distribution function which we approximate by the thermal distribution $n(\omega_i)$. We can corrispondingly keep higher-order cumulants such as $\langle\!\langle a_i^\dagger a_j\bm\sigma\rangle\!\rangle$ whose equations of motion are considerably more complicated \cite{foerstner:03a,foerstner:03b}. On general grounds, we expect that the neglect of higher-order cumulants is appropriate for sufficiently low temperatures and weak dot-phonon couplings $g_i$ ---or equivalently $\alpha_p$ defined in eq.~\eqref{eq:dot-phonon.coupling}---, which is a valid assumption for GaAs-based quantum dots ($\alpha_p\cong 0.033$) at low temperatures. Figure \ref{fig:rabi-polaron} shows results of calculations based on eq.~\eqref{eq:polaron.density} for a $2\pi$-pulse and for temperatures of $T=1$ and $T=10$. We observe that Rabi flopping occurs but is damped because of the phonon-assisted dephasing. In particular at higher temperatures phonon dephasing is of strong importance, and gives rise to decoherence on a picosecond time scale. We shall return to this point in section~\ref{sec:control}.

\paragraph{Beyond the spin-boson model.}

Although in many cases of interest the spin-boson model \eqref{eq:spin-boson} provides a sufficiently sophisticated description of optical excitations in artificial atoms, there are situations where it is expected to break down. For instance, at higher temperatures anharmonic decay of phonons ---a phonon decays in an energy-conserving scattering into two phonons of lower energy--- or higher-order phonon processes \cite{uskov:00} could play a decisive role. If the artificial atom can no longer be described as a genuine two-level system, one has to additionally consider phonon-mediated scattering channels. Quite generally, the strong polar-optical coupling to longitudinal optical phonons introduces a marked deformation of the surrounding lattice and the formation of a {\em polaron}\/ \cite{hameau:99,oshiro:99,bissiri:00}. Because optical phonons have a very small dispersion, this interaction channel has no significant impact on decoherence. However, when the system is in an excited exciton or multi-exciton state, the anharmonic decay of phonons contributing to the polaron allows for relaxation processes even in absence of energy matching between the unrenormalized dot transition and the phonons \cite{kral:98,ferreira:99,verzelen:00,verzelen:02,verzelen:02b,jacak:02}. This demonstrates that phonon relaxation and decoherence in artificial atoms is more efficient than one would expect in a simple-minded Fermi's-golden-rule picture. Future work will show to what extent carrier-phonon interactions can be tailored in artificial atoms \cite{heitz:01,urayama:01}, and whether phonon-assisted dephasing can be eventually strongly suppressed or fully overcome \cite{zanardi:98,zanardi:99,hohenester.prl:04}.

\subsection{Spin scatterings}

So far we have seen that photon and phonon scatterings occur on a timescale ranging from several tens of picoseconds to nanoseconds. Such decoherence times are remarkably long for the solid state, but are rather short when it comes to more sophisticated quantum control applications (secs. \ref{sec:control} and \ref{sec:computation}). Optical excitations in quantum dots possess another degree of freedom which has recently attracted enormous interest: {\em spin}\/ \cite{zutic:04}. Spin couples weakly to the solid state environment, and is therefore expected to be long lived. Optics provides a simple means to modify spin degrees of freedom through coupling to the charge degrees. This dual nature of optical excitations is exploited in quantum-computation proposals, to be discussed in sec.~\ref{sec:computation}. What are the typical spin relaxation and decoherence times in artificial atoms? Things are quite unclear. Experimentally, it was found that at low temperature spin relaxation is almost completely quenched \cite{kamada:99,paillard:01,lenihan:02,cortez:02}. Theoretical estimates indicate relaxation times of the order of microseconds or above \cite{khaetskii:00}, whereas almost no conclusive results exist for the pertinent decoherence times. Thus, spin keeps its secret in the game and holds a lot of promise and hope.

\section{Quantum control}\label{sec:control}

Recent years have witnessed enormous interest in controlling quantum phenomena in a variety of nanoscale systems \cite{poetz:99}. Quite generally, such control allows to modify the system's wavefunction at will through appropriate tailoring of external fields, e.g., laser pulses: while in {\em quantum optics}\/ the primary interest of this wavefunction engineering lies on the exploitation of quantum coherence among a few atomic levels \cite{mandel:95,walls:95,scully:97}, in {\em quantum chemistry}\/ optical control of molecular states has even led to the demonstration of optically driven chemical reactions of complex molecules \cite{rabitz:00}; furthermore, starting with the seminal work of \citet{heberle:95} coherent-carrier control in semiconductors and semiconductor nanostructures has been established as a mature field of research on its own. This research arenas have recently received further impetus from the emerging fields of quantum computation and quantum communication \cite{bouwmeester:00}, aiming at quantum devices where the wavefunction can be manipulated with highest possible precision. It is worth emphasizing that hitherto there exists no clear consensus of how to optimally tailor the system's control, and it appears that each field of research has come up with its own strategies: for instance, quantum-optical implementations in atoms benefit from the long atomic coherence times of meta-stable states, and it usually suffices to rely on the solutions of effective models (e.g., adiabatic population transfer in an effective three-level system \cite{bergmann:98}); in contrast, in quantum chemistry the complexity of molecular states usually does not permit schemes which are solely backed from the underlying level schemes, and learning algorithms, which receive direct feedback from experiment, appear to be the method of choice. Finally, coherent control in semiconductor nanostructures has hitherto been primarily inspired by quantum-optical techniques; however, it is clear that control in future quantum devices will require more sophisticated techniques to account for the enhanced decoherence in the solid state. 

This section is devoted to an introduction into the field. Throughout we shall use the laser-induced quantum coherence as the workhorse, which allows to bring the system to almost any desired state \cite{rabitz:04}. On the other hand, such ideal performance is spoiled by the various decoherence channels at play, e.g. the ones discussed in the previous section. From the field of quantum optics a number of control strategies are known which allow to suppress or even overcome decoherence losses. In sec.~\ref{sec:stirap} we shall discuss one of the most prominent one: stimulated Raman adiabatic passage \cite{bergmann:98}, a technique which exploits the renormalized states in presence of strong laser fields for a robust and high-fidelity population transfer. Because of its simplicity, there has recently been strong interest in possible solid-state implementations \cite{lindberg:95,poetz:97,binder:98,artoni:00,brandes:00,hohenester.apl:00,brandes:01,troiani.prl:03}. There exist other control strategies, which are either based on schemes developed in the fields of nuclear magnetic resonance \cite{slichter:96,levitt:03,chen:01,piermarocchi:02,chen:04} or rely on general optimization approaches such as optimal control \cite{rabitz:00,peirce:88,borzi.pra:02} or genetic algorithms \cite{rabitz:00,zeidler:01}. Below we shall review the latter two approaches. Similar to the last section it will prove useful to rely on effective level schemes, which grasp the main features of the excitonic and multi-exciton quantum dot states (fig.~\ref{fig:level-scheme}). Finally, for a more extensive discussion of coherent optical spectroscopy and coherent carrier control in quantum dots the reader is referred to the literature \cite{bonadeo:98b,chen:00,chen:02,guenther:02,lenihan:02,li:03,zrenner:02}.

\subsection{Stimulated Raman adiabatic passage}\label{sec:stirap}

Let us first consider the $\Lambda$-type level scheme depicted in fig.~\ref{fig:level-scheme}b. It consists of two long-lived states $|0\rangle$ and $|1\rangle$ which are optically connected through a third short-lived state $|2\rangle$. Such a level scheme may correspond to a coupled dot charged with one surplus carrier, where states $|0\rangle$ and $|1\rangle$ are associated to the carrier localization in one of the dots and $|2\rangle$ to the charged exciton which allows optical coupling between states $|0\rangle$ and $|1\rangle$ \cite{hohenester.apl:00}; alternatively, we may associate the two lower states $|0\rangle$ and $|1\rangle$ to the spin orientation of one surplus electron in a single quantum dot, where in presence of a magnetic field along $x$ (i.e. Voigt geometry, sec.~\ref{sec:exciton.spin}) the two states can be optically coupled through the charged exciton $|2\rangle$ (see refs.~\cite{imamoglu:99,troiani.prl:03,pazy:03} and sec.~\ref{sec:computation}). Quite generally, for the level scheme of fig.~\ref{fig:level-scheme}b and assuming that the system is initially prepared in state $|0\rangle$, in the following we shall ask the question: what is the most efficient way to bring the system from $|0\rangle$ to $|1\rangle$? Suppose that the frequencies of two laser pulses are tuned to the 0--2 and 1--2 transitions, respectively. For reasons to become clear in a moment, we shall refer to the pulses as {\em pump}\/ and {\em Stokes}.\/ Since direct optical transitions between $|0\rangle$ and $|1\rangle$ are forbidden, we have to use $|2\rangle$ as an auxiliary state; however, intermediate population of $|2\rangle$ introduces losses through environment coupling, e.g. spontaneous photon emissions or phonon-assisted dephasing. We can use the master equation \eqref{eq:lindblad} of Lindblad form to describe the problem. Within a rotating frame we obtain the effective hamiltonian \cite{bergmann:98,hohenester:02}

\begin{equation}\label{eq:stirap-ham}
  H_{\rm eff}=-\frac 1 2\left(
    \begin{array}{ccc}
      0 & 0 & \Omega_p^* \\
      0 & 0 & \Omega_s^* \\
      \Omega_p & \Omega_s & \epsilon\\
    \end{array}\right)\,,
\end{equation}

\noindent with $\epsilon=2\,\Delta+i\Gamma$, $\Delta$ the detuning of the lasers with respect to the 0--2 and 1--2 transitions, and $\Gamma$ the inverse lifetime of the upper state. $\Omega_p$ and $\Omega_s$ are the Rabi frequencies for the pump and Stokes pulse, respectively. Throughout we assume that $\Omega_p$ only affects the 0--2 transition and $\Omega_s$ only the 1--2 one, which can e.g. be achieved through appropriate polarization filtering, or for sufficiently large 0--1 splittings through appropriate choice of laser frequencies. Figure~\ref{fig:stirap}b shows results of simulations for different time delays between the Stokes and pump pulse, and for different pulse areas $A_s=\int_{-\infty}^\infty dt\;\Omega_s(t)$ (we assume $A_s=A_p$): black corresponds to successful and white to no population transfer. In the case of the ``intuitive'' ordering of laser pulses where the pump pulse excites the system {\em before}\/ the Stokes pulse, i.e. negative time delays in fig.~\ref{fig:stirap}b, one observes enhanced population transfer for odd multiples of $\pi$. This is associated to processes where the pump pulse first excites the system from $|0\rangle$ to $|2\rangle$, and the subsequent Stokes pulse brings the system from $|2\rangle$ to $|1\rangle$. However, the large black area at positive time delays in fig.~\ref{fig:stirap}b suggests that there is a more efficient way for a population transfer. Here, the two pulses are applied in the ``counter-intuitive'' order, i.e. the pump pulse is turned on {\em after}\/ the Stokes pulse. Because of the resemblance of this scheme with a Raman-type process it has become convenient to introduce the expression of a {\em Stokes} pulse, and the whole process has been given the name {\em stimulated Raman adiabatic passage}\/ \cite{bergmann:98}. This process fully exploits the quantum coherence introduced by the intense laser fields. In presence of the Stokes pulse the dot-states become renormalized, and these renormalized states are used by the pump pulse for a robust and high-fidelity population transfer. While fig.~\ref{fig:stirap} presents solutions of the full master equation \eqref{eq:lindblad}, in the following we shall only consider the time evolution of $\bm\rho$ due to the effective hamiltonian \eqref{eq:stirap-ham}. If $\Omega_p$ and $\Omega_s$ have a sufficiently slow time dependence ---as will be specified in more detail further below---, at each instant of time the system is characterized by the eigen values and vectors of $H_{\rm eff}$. Straightforward algebra yields for $\Gamma\ll\Omega_s$, $\Gamma\ll\Omega_p$ the eigenvalues 

\begin{equation}\label{eq:stirap.eigenvalues}
  \varpi_0=0\,,\quad
  \varpi_\pm=\frac 1 2\left(\Delta\pm\Omega_{\rm eff}\right)+
  i\left(1-\Delta/\Omega_{\rm eff}\right)\,\Gamma\,,
\end{equation}

\begin{table}[b]
\caption{Time evolution of the quantities characterizing the stimulated Raman adiabatic passage \cite{bergmann:98}. For discussion see text.}
\label{table:stirap}
{\small
\begin{tabularx}{\columnwidth}{Xccc}
\hline\hline
Quantity & $\Omega_s$ on & $\Omega_s$ and $\Omega_p$ on & $\Omega_p$ on \\
\hline
Angle $\theta$ & 0 & $0\to\pi/2$ & $\pi/2$ \\
$\sim\varpi_\pm-\mbox{\small$\Delta/2$}$ & 
{\small$\pm\frac 1 2(\Delta^2+\Omega_s^2)^\frac 1 2$} & 
{\small$\pm\frac 1 2(\Delta^2+\Omega_s^2+\Omega_p^2)^{\frac 1 2}$} & 
{\small$\pm\frac 1 2(\Delta^2+\Omega_p^2)^{\frac 1 2}$} \\
$\varpi_0$ & 0 & 0 & 0 \\
Trapped state $|a_0\rangle$ & $|0\rangle$ & $|0\rangle\to|1\rangle$ & $|1\rangle$ \\
\hline
\hline
\end{tabularx}
}
\end{table}

\noindent with $\Omega_{\rm eff}^2=\Delta^2+\Omega_p^2+\Omega_s^2$. Most importantly, eigenvalue $\varpi_0$ has no imaginary part and consequently does not suffer radiative losses (this holds even true for large values of $\Gamma$). Indeed, introducing the time-dependent angle $\theta$ through $\tan\theta=\Omega_p/\Omega_s$ we observe that the corresponding eigenvector

\begin{equation}\label{eq:stirap.trapped}
  |a_0\rangle=\cos\theta|1\rangle-\sin\theta|2\rangle
\end{equation}

\noindent has {\em no}\/ component of the ``leaky'' state $|2\rangle$ --- in contrast to the eigenvectors $|a_\pm\rangle$ which are composed of all three states $|0\rangle$, $|1\rangle$, and $|2\rangle$. In $|a_0\rangle$ the amplitudes of the 0--2 and 1--2 transitions interfere destructively, such that the state is completely stable against absorption and emission from the radiation fields. For that reason state $|a_0\rangle$ has been given the name {\em trapped state}.\/ The population transfer process exploits this trapped state as a vehicle in order to transfer population between states $|0\rangle$ and $|1\rangle$. It is achieved by using overlap in time between the two laser pulses (fig.~\ref{fig:stirap}a, table~\ref{table:stirap}). Initially, the system is prepared in state $|0\rangle$. When the first (Stokes) laser is smoothly turned on, the system is excited at the Stokes frequency. At this frequency no transitions can be induced; what the pulse does, however, is to align the time-dependent state vector $|\Psi(t)\rangle$ with $|a_0(t)\rangle=|0\rangle$ (since $\theta=0$ in the sector $\Omega_s\neq 0$, $\Omega_p=0$), and to split the degeneracy of the eigenvalues $\varpi_0$ and $\varpi_\pm$. Thus, if the pump laser is smoothly turned on ---such that throughout $\Omega_{\rm eff}(t)$ remains large enough to avoid non-adiabatic transitions between $|a_0(t)\rangle$ and $|a_\pm(t)\rangle$ \cite{bergmann:98}--- all population is transferred between states $|0\rangle$ and $|1\rangle$ within an adiabatic process where $|\Psi(t)\rangle$ directly follows the time-dependent trapped state $|a_0(t)\rangle$. Stimulated Raman adiabatic passage is a process important for a number of reasons. First, it is a prototypical example of how intense laser fields can cause drastic renormalizations of carrier states; quite generally, these ``dressed'' states exhibit novel features in case of quantum interference, i.e. if three or more states are optically coupled. Second, quantum control in quantum dots has recently attracted increasing interest in view of possible quantum computation applications \cite{troiani.prb:00,chen:01,pazy:01,troiani.prl:03} aiming at an all-optical control of carrier states; in this respect, the adiabatic transfer scheme might be of some importance because of its robustness and its high fidelity (sec.~\ref{sec:computation}). More specifically, from fig.~\ref{fig:stirap}b it becomes apparent that the transfer works successfully within a relatively large parameter regime --- in contrast to the ``intuitive'' order of pulses, where a detailed knowledge of the dipole matrix elements and a precise control over the laser parameters is required. Thus, it provides a robust scheme which only relies on sufficiently smooth and strong laser pulses.

\subsection{Optimal control}\label{sec:optimal-control}

In many cases of interest it is more difficult ---or even impossible--- to guess control strategies solely based on physical intuition, and one is forced to rely on more general control schemes. Suppose that the system under investigation is described by the $n$-component vector $x=(x_1,x_2,\dots x_n)$ of dynamic variables, whose equations of motion are given by the differential equations $\dot x_i=F_i(x,\Omega)$ with $F$ a functional which depends on all state variables $x$ and the control fields $\Omega$. Here, $x$ may refer to the different components of a wavefunction $\psi$ which obeys Schr\"odinger's equation, or to the different components of the density operator $\bm\rho$, or to the cumulants of a density-matrix approach. The main assumption we shall make is that of a Markovian time evolution, i.e. we assume that for the time evolution of $x_i(t)$ the functional $F$ only depends on the variables $x(t)$ at the same instant of time. The components $x_i$ are supposed to be real, which can always be achieved by separating $\psi_i$ or $\rho_{ij}$ into their respective real and imaginary parts. Initially the system is in the state $x_0$. Quite generally, in the field of quantum control we are seeking for control fields $\Omega$ which bring the system from $x_0$ at time zero to the desired final state $x_d$ at time $T$, or promote the system within the time period $[0,T]$ through a sequence of desired states. To evaluate a given control $\Omega$, we have to quantify its success through the {\em cost functional}

\begin{equation}\label{eq:optimal.cost}
  J(x,\Omega)=J_T(x)+\tilde J(x,t)+\frac\gamma 2\int_0^T dt\,\sum_i\|\Omega_i(t)\|^2\,.
\end{equation}

\noindent Here $J_T(x)$, which only depends on $x(T)$, accounts for the terminal conditions and rates how close $x(T)$ is to the desired state $x_d$ (e.g. through $\frac 1 2\|\psi(T)-\psi_d\|^2$), $\tilde J(x,t)$ is a functional that accounts for other control objectives within $[0,T]$ (e.g. the wish to suppress the population of certain states), and the last term accounts for the limited laser resources. Our task now is to determine a control which minimizes $J(x,\Omega)$ subject to the constraint that $x$ fulfills the dynamic equations $\dot x_i=F_i(x,\Omega)$ with the initial condition $x(0)=x_0$. Within the framework of {\em optimal control}\/ \cite{rabitz:00,peirce:88,borzi.pra:02} this is accomplished by introducing Lagrange multipliers $\tilde x$ for the constraints, and turning the constrained minimization of \eqref{eq:optimal.cost} into an unconstrained one. For this purpose we define the Lagrangian function

\begin{equation}\label{eq:optimal.lagrange}
  L(x,\tilde x,\Omega)=\int_0^T dt\, \sum_i\tilde x_i\Bigl(\dot x_i-F_i(x,\Omega)\Bigr)
  +J(x,\Omega)\,.
\end{equation}

\noindent We next utilize that the Lagrange function admits a stationary point at the solution. Taking the functional derivatives of $L$ with respect to $x_i$, $\tilde x_i$, and $\Omega_k$ ($k$ labels the different components of the control fields) and performing integration by parts for the term $\tilde x_i\,\dot x_i$, we arrive at

\begin{equation}\label{eq:optimality}
  \dot x_i=F_i(x,\Omega)\,,\quad
  \dot{\tilde x}_i=-\frac{\partial F_i(x,\Omega)}{\partial x_i}-
  \frac{\partial\tilde J(x,t)}{\partial x_i}\,,\quad
  \Omega_k=\frac 1\gamma \sum_i\tilde x_i\,\frac{\partial F_i(x,\Omega)}{\partial\Omega_k}\,,
  \quad
\end{equation}

\noindent together with the intial $x_i(0)=x_0$ and terminal $\tilde x_i(T)=-\partial J_T(x,\Omega)/\partial x_i$ conditions. For the ``optimal control'' this set of equations has to be fulfilled simultaneously for $x$, $\tilde x$, and $\Omega$. In general, analytic solutions can be only found for highly simplified systems \cite{boscain:02}, whereas numerical calculation schemes have to be adopted for more realistic systems. A numerical algorithm for the solution of the optimality system \eqref{eq:optimality} was formulated in \citet{borzi.pra:02}. Suppose that we have an intial guess for the control fields $\Omega$. We can then solve the dynamic equations for $x$ subject to the initial conditions $x(0)=x_0$ {\em forwards}\/ in time, and the dynamic equations for $\tilde x$ subject to the terminal conditions $\tilde x_i(T)=-\partial J_T(x,\Omega)/\partial x_i$ {\em backwards}\/ in time. The last equation in \eqref{eq:optimality} then provides the search directions $d_{\Omega_k}=-(1/\gamma)\sum_i\tilde x_i\partial F_i(x,\Omega)/\partial\Omega_k$ for the new control fields $\Omega_k\to\Omega_k+\lambda d_{\Omega_k}$, where for sufficiently small $\lambda$ it is guaranteed that the new cost functional $J(x,\Omega)$ decreases \cite{borzi.pra:02}. Figure~\ref{fig:optimal.sketch} sketches the basic ingredients of the resulting algorithm.

\paragraph{Adiabatic passage.}

As a first example we shall revise the adiabatic passage scheme of sec.~\ref{sec:stirap} within the framework of optimal control. This can be done in absence of decoherence analytically \cite{boscain:02,kis:02b} or, as we shall do in the following, in presence of decoherence numerically \cite{borzi.pra:02}. Our starting point is given by the master equation \eqref{eq:lindblad} of Lindblad form, where the effective hamiltonian $H_{\rm eff}$ is given by \eqref{eq:stirap-ham}. Since we are aiming at solutions which minimize environment losses we neglect in eq.~\eqref{eq:lindblad} the in-scattering contributions and are left with the Schr\"odinger-like equation of motion $i\dot\psi=H_{\rm eff}\psi$ for the three-component wavefunction $\psi$. The effective hamiltonian reads

\begin{equation}
  H_{\rm eff}=\left(\Delta-i\frac \Gamma 2\right)\proj 2 2-\frac 12
  \sum_{\ell=p,s}\left(M_\ell^\dagger\,\Omega_\ell+M_\ell\,\Omega_\ell^*\right)\,,
\end{equation}

\noindent with the optical transition matrix elements $M_p=\proj 02$ and $M_s=\proj 12$ for the pump and Stokes pulse, respectively. We assume that at time zero the system is in state $\psi_0$. The objective of the control is expressed through

\begin{equation}\label{eq:optimal.cost2}
  J(\psi,\Omega)=\frac 1 2\left(1-|\psi_1(T)|^2\right)+\frac\gamma 2
  \int_0^T dt\,\sum_{\ell=p,s}|\Omega_\ell(t)|^2\,.
\end{equation}

\noindent The first term has its minimum zero for $|\psi_1(T)|=1$, i.e. when the system is finally in state $|1\rangle$, and the second term accounts for the limited laser resources and is needed to make the optimal-control problem well posed (we set $\gamma=10^{-8}$). For this system (as well as for its Lindblad generalization \cite{kim:95}) we can easily obtain the optimality system in complex form \cite{borzi.pra:02}

\begin{equation}\label{eq:optimality2}
  i\dot\psi=H_{\rm eff}\,\psi\,,\quad 
  i\dot{\tilde\psi}=H_{\rm eff}^\dagger\,\tilde\psi\,,\quad
  \Omega_\ell=-\frac 1{2\gamma}\left(
  \langle\psi|M_\ell^\dagger|\tilde\psi\rangle+\langle\tilde\psi|M_\ell^\dagger|\psi\rangle
  \right)\,,
\end{equation}

\noindent subject to the initial $\psi(0)=\psi_0$ and terminal $\tilde\psi_i(T)=-i\partial J_T(\psi)/\partial\psi_i=-i\delta_{i1}\psi_1(T)$ conditions. The last term in eq.~\eqref{eq:optimality2} is written in the usual bra and ket shorthand notation. Figure~\ref{fig:stirap-optimal1} shows results of prototypical optimal-control calculations and for different initial control fields $\Omega_{\rm trial}$. We start with pump and Stokes pulses of Gaussian form centered at time $30$ and with a full-width of half maximum of $15$, and use different pulse areas of $10$, $25$, and $50$. For the two fields of lowest area, the optimal control loop (fig.~\ref{fig:optimal.sketch}) comes up with control fields where in addition to the initial Gaussians components at the detuning frequency $\Delta$ are present. Reminiscent of the stimulated Raman adiabatic scheme of sec.~\ref{sec:stirap}, at this frequency the Stokes pulse is turned on prior to the pump one. In contrast to that, for the highest initial pulse areas the pump and Stokes fields keep their Gaussian shape (and no additional frequency components show up). The resulting control strategy is similar to that of the black regions in fig.~\ref{fig:stirap}b at zero time delay. Thus, different initial control fields $\Omega_{\rm trial}$ in the optimal  control algorithm lead to different control strategies. This is not particularly surprising since even for the relatively simple situation shown in fig.~\ref{fig:stirap}, where the pump and Stokes pulse are characterized by the two parameters of area and pulse delay, a huge number of successful control strategies (indicated by the black regions) is found. In the optimal control case shown in fig.~\ref{fig:stirap-optimal1}, where the control fields $\Omega$ are discretized at about $10\,000$ points in time, the control space is tremendously increased and corrispondingly an even much larger number of possible solutions can be expected. In absence of decoherence, i.e. for an isolated few-level system, it can be shown that the only allowed extrema of the Lagrange function \eqref{eq:optimal.lagrange} correspond to perfect control or no control \cite{rabitz:04}. In presence of decoherence things are modified. This is shown in fig.~\ref{fig:stirap-optimal1} and even more clearly fig.~\ref{fig:stirap-optimal2}, where one observes that the deviation of the final state $\psi(T)$ from the desired state $\psi_1$ differs for different initial conditions. In other words, starting at one point of the high-dimensional control space, the search algorithm of fig.~\ref{fig:optimal.sketch} proceeds  along the direction of the steepest descent and becomes trapped in a suboptimal local minimum. The way the search algorithm evolves can be studied in fig.~\ref{fig:stirap-optimal2}. For the small pulse areas (dotted and dashed lines) one observes that initially no population is transfered (and $J$ has its largest possible value of $0.5$), whereas for the largest pulse area (solid line) the initial guess for $\Omega_{\rm trial}$ already leads to a quite efficient transfer. Within the first few iterations of the optimal control loop $J$ does not decrease significantly. This is because the initial stepsize $\lambda$ is too small [panel (b)] to cause a significant decrease of $J$. After approximately ten iterations an adequate stepsize is obtained (through the increase of $\lambda\to 1.2\,\lambda$ after successful steps, see fig.~\ref{fig:optimal.sketch}), and $J$ decreases rapidly over approximately 100 iterations. This regime is followed by one with a much slower decrease of $J$, where the control approaches slowly that associated to the local minimum. This discussion allows us to pinpoint the respective advantages and disadvantages of the optimal control algorithm. If we start with a reasonable guess for $\Omega_{\rm trial}$ (whose solutions can be even completely away from the desired ones, such as for the smaller pulse areas), we obtain a strongly improved $\Omega$ whose solutions fulfill the objective of the control much better than that of $\Omega_{\rm trial}$. On the other hand, the solutions $\Omega$ of the control algorithm (fig.~\ref{fig:optimal.sketch}) are most probably associated to local minima of the control space rather than to the {\em global}\/ minimum.

\paragraph{Phonon-assisted dephasing.}

The adiabatic passage scheme which we have just discussed is an extreme example in the sense that there exist numerous solutions ---as indicated in fig.~\ref{fig:stirap}b---, and one expects that the search algorithm becomes quickly trapped in a suboptimal, local minimum. As another example we shall now discuss the coherent control of the two-level system in presence of phonon couplings. In sec.~\ref{sec:phonon-dephasing} we showed for the spin-boson model that in presence of an exciting laser pulse Rabi flopping occurs but is damped because of phonon decoherence. However, contrary to other decoherence channels in solids where the system's wavefunction acquires an uncontrollable phase through environment coupling, in the independent Boson model the loss of phase coherence is due to the coupling of the electron-hole state to an ensemble of harmonic oscillators which all evolve with a coherent time evolution but different phase. This results in destructive interference and dephasing, and thus spoils the direct applicability of coherent carrier control. On the other hand, the coherent nature of the state-vector evolution suggests that more refined control strategies might allow to suppress dephasing losses. To address the question whether such losses are inherent to the system under investigation, in the following we examine phonon-assisted dephasing within the optimal-control framework aiming at a most efficient control strategy to channel the system's wavefunction through a sequence of given states. We quantify the objective of the control through the cost function 

\begin{equation}\label{eq:cost.polaron}
  J(\bm u,\Omega)=\frac 1 2\left(\int_{-T}^T dt\,\beta(t)|\bm u(t)-\hat{\bm e}_3|^2+|\bm u(T)
  +\hat{\bm e}_3|^2+\gamma\int_{-T}^T dt\,|\Omega(t)|^2\right)\,,
\end{equation}

\noindent with $\beta$ a Gaussian centered at time zero with a narrow full-width of half maximum of $0.1\,\omega_c^{-1}$ and $\gamma=10^{-5}$ a small constant. In other words, we are seeking for solutions where $\bm u$ passes through the excited state $\hat{\bm e}_3$ at time zero and goes back to the ground state $-\hat{\bm e}_3$ at $T$. For the system dynamics we assume the equations of motion \eqref{eq:polaron.density} for the cumulants $\bm u$, $s_i$, and $\bm u_i$, subject to the initial conditions $\bm u(-T)=-\hat{\bm e}_3$, $s_i(-T)=0$, and $\bm u_i(-T)=0$. Again the method of Lagrange multipliers is used to minimize $J(\bm u,\Omega)$ subject to eq.~\eqref{eq:polaron.density}, and to obtain the adjoint equations \cite{hohenester.prl:04}

\begin{eqnarray}\label{eq:adjoint.polaron}
  \dot{\tilde{\bm u}}&=& \bm\Omega\times \tilde{\bm u}+\sum_i g_i
  \Bigl(\bigl(n_i+\frac 1 2\bigr)\,\hat{\bm e_3}\times\Re e(\tilde{\bm u}_i)+
  \frac 1 2\, \Im m\bigl((\tilde s_i-\bm u\,\tilde{\bm u}_i)\,\hat{\bm e}_3-
  u_3\,\tilde{\bm u}_i\bigr)\Bigr)\nonumber\\&&
  +\beta(t)(\bm u-\hat{\bm e}_3)\nonumber\\
  \dot{\tilde s}_i &=& -i\omega_i\tilde s_i+
  2\,g_i\,\biggl((\tilde{\bm u}\times \bm u)\,\hat{\bm e}_3+\Re e\sum_j
  (\tilde{\bm u}_j^*\times {\bm u}_j)\,\hat{\bm e}_3\biggr)\nonumber\\
  \dot{\tilde{\bm u}}_i &=& \bm\Omega\times\tilde{\bm u}_i-
  i\omega_i \tilde{\bm u}_i+2\,g_i\, \hat{\bm e}_3\times \tilde{\bm u}\,,
\end{eqnarray}

\noindent with terminal conditions $\tilde{\bm u}(T)=-\hat{\bm e}_3-\bm u(T)$, $\tilde s_i(T)=0$, and $\tilde{\bm u}_i(T)=0$. Equations~\eqref{eq:polaron.density} and \eqref{eq:adjoint.polaron} together with

\begin{equation}\label{eq:control.polaron}
  \Omega=\frac 1\gamma\left((\tilde{\bm u}\times \bm u)+
  \Re e\sum_i (\tilde{\bm u}_i^*\times \bm u_i)
  \right)\,(\hat{\bm e}_1-i\hat{\bm e}_2)
\end{equation}

\noindent form the optimality system. It is solved iteratively through the scheme depicted in fig.~\ref{fig:optimal.sketch}, with integration of eq.~\eqref{eq:polaron.density} forwards and eq.~\eqref{eq:adjoint.polaron} backwards in time, and computing an improved control by use of eq.~\eqref{eq:control.polaron}. Results of such optimal-control calculations are shown in fig.~\ref{fig:polaron-optimal}. Most remarkably, one can indeed obtain a control field for which $\bm u(t)$ passes through the desired states of $\hat{\bm e}_3$ at time zero and $-\hat{\bm e}_3$ at $T$. Thus, appropriate pulse shaping allows to fully control the two-level system even in presence of phonon couplings. We emphasize that, with the exception of the somewhat pathological quantum ``bang-bang'' control \cite{viola:98} where the system is constantly flipped to suppress decoherence, no such simple control strategy for suppression of environment losses is known in the literature. This result, which also prevails in presence of finite but low termperatures \cite{hohenester.prl:04}, clearly highlights the strength and flexibility of optimal control.

\paragraph{Genetic algorithms.}

In several cases of interest one is often neither able to solve the equations of motion nor knows the full hamiltonian characterizing the system. This holds in particular true for laser-induced reactions of molecules \cite{rabitz:00}, where the configuration landscape is highly complicated. \citet{judson:92} were the first to propose an evolutionary algorithm that allows the search for control fields even without any knowledge of the hamiltonian. For the sake of completeness, in the following we shall briefly outline the main ideas of this approach. Suppose that the laser pulse can be encoded in terms of a ``gene'', i.e. a vector $\Omega=(\Omega_1,\Omega_2,\dots \Omega_n)$ with typically $n\sim 10$--100 components. Within the evolutionary approach a population $N_{\rm pop}\cong 48$ of different genes $\Omega^\mu$ is considered. At the beginning the different components of each gene are chosen randomly. Next, we compute for each gene the objective function $J^\mu$. Within the evolutionary approach, the next population of genes is determined in biological terms according to the fitness $J^\mu$ of each individual \cite{judson:92,zeidler:01}. Following \citet{zeidler:01} this can be accomplished as follows: first, the individual with the best $J^\mu$ is included without change in the next generation ({\em elitism}); next, the $N_{\rm parent}\sim N_{\rm pop}/7$ individuals with the best $J^\mu$ are chosen as parents for the next generation; about $2\,N_{\rm parent}$ of the individuals of the next generation are determined by randomly choosing two parents, cutting their genes at one given point ({\em single-point crossover}) or two points ({\em two-point crossover}), and pasting the different pieces of the genes together ({\em recombination}); finally, the remaining individuals of the next generation are obtained by randomly taking one parent and randomly modifying its genetic information $\Omega_i^\mu$ ({\em mutation}). There exist numerous other implementations which differ in one or several points. However, the grand idea of all these approaches is to provide a sufficiently large gene pool and to let the individuals benefit from their respective advantages and peculiarities. This has the consequence that once an individual acquires a successful control strategy (either through recombination or mutation), it will distribute it to the next generation where it possibly becomes optimized through further mixing or mutation. For that it is compulsory to keep not only the fittest individual but to provide a larger gene pool. In this respect, mutation plays an important role as it determines the degree of modification from one generation to the next. For normalized control-field components $\Omega_i^\mu\in[0,1]$ mutation can be computed according to \cite{zeidler:01,press:02}

\begin{equation}\label{eq:genetic-algorithm}
  {\Omega_i^\mu}'=\Omega_i^\mu-\sigma\,\log\left(\tan\left(
  \frac\pi 2\,r\right)\right)\,,
\end{equation}

\noindent where $r\in(0,1)$ is a uniformly distributed random number (and ${\Omega_i^\mu}'\in[0,1]$ has to be asserted). $\sigma$ is the quantity that determines how much ${\Omega^\mu}'$ can deviate from $\Omega^\mu$. It has a similar role as the stepsize $\lambda$ in the optimal control scheme, and its value should be adapted during optimization. Here one can proceed as follows \cite{zeidler:01}: let $N_{\rm mut}$ be the number of mutated individuals and $N_{\rm succ}$ the number of successful mutations with ${J^\mu}'<J^\mu$; for $N_{\rm succ}<0.2\,N_{\rm perm}$ we conclude that the mutation rate is too high and set $\sigma\to 0.9\,\sigma$; otherwise we increase the rate according to $\sigma\to\sigma/0.9$. Quite generally, the genetic algorithm works formidably well for laser fields which can be by characterized by a few parameters (e.g. laser-pulse shaping experiments \cite{rabitz:00,zeidler:01}). For instance, parameterizing the pump and Stokes pulses of the adiabatic passage scheme by Gaussians (i.e. in terms of areas, detunings, time delay, and full-width of half maxima), a highly successful transfer scheme is found after a few generations. In comparison to the optimal control approach, the evolutionary one examines a larger portion of the control space (through its different individuals), and therefore chooses out of several local minima the lowest one. On the other hand, evolutionary approaches are usually much slower (no information about the steepest descent $\nabla_\Omega\,J(x,\Omega)$ is used) and have huge problems to find non-trivial pulse shapes, such as the one depicted in fig.~\ref{fig:polaron-optimal}a.

\subsection{Self-induced transparency}\label{sec:sit}

We conclude this section with an at first sight somewhat different topic, namely laser pulse propagation in a macroscopic sample of inhomogenously broadened quantum dots. Let the central frequency of the laser pulse be tuned to the exciton groundstate transition (see fig.~\ref{fig:absorption-strong}). We then describe the dot states in terms of generic two-level systems with different detunings $\Delta$. When the light pulse enters the dot region, it excites excitons and hereby suffers attenuation. It should, however, be emphasized that  inhomogeneous line broadening leads to losses which are substantially different from those induced by homogeneous broadening \cite{andreani:98}. Through the pulse propagation in the medium of inhomogenously broadened dots all of them are excited in {\em in phase},\/ where ---at variance with homogeneous broadening--- each dot has a coherent time evolution. However, the phase varies from dot to dot, thus leading to interference effects which in most cases prevent the observation of the coherent radiation-matter interaction. A striking exception is the phenomenon of {\em self-induced transparency}\/~\cite{mccall:67,mccall:69,gibbs:71,slusher:72,mandel:95,allen:75}, a highly nonlinear optical coherence phenomenon which directly exploits inhomogeneous level broadening. Light-matter coupling plays a crucial role in its theoretical analysis. Not only one has to consider the material response in the presence of the driving light pulse, but also the back-action of the macroscopic material polarization on the light propagation (through Maxwell's equations). For the inhomogenously broadened two-level systems we assume a time evolution according to the master equation \eqref{eq:lindblad} of Lindblad form, where the coherent part is given by the usual Bloch equations \eqref{eq:bloch-equations} for different detunings $\Delta$. For the light pulse we assume a geometry (fig.~\ref{fig:sit}) where the laser enters from the left-hand side into the sample of inhomogenously broadened dots. Denoting the pulse propagation direction with $z$ and assuming an electric-field profile with envelope $\bm{\mathcal{E}}_0(z,t)$ and a central frequency of $\omega_0$ \cite{panzarini.prb:02}, we describe the light propagation in the slowly varying envelope approximation \cite{mandel:95}

\begin{equation}\label{eq:maxwell-slowly}
  \left(\partial_z+\frac n c\partial_t\right)\bm{\mathcal{E}}_0(z,t)\cong
  -\frac{2\pi\omega_o}{nc}\,\Im m\bm{\mathcal{P}}(z,t)\,.
\end{equation}

\noindent Here, $\bm{\mathcal{P}}(z)=M_{0x}\hat{\bm e_\lambda}N\int g(\Delta)d\Delta\,\frac 12[u_1(z,\Delta)-i\,u_2(z,\Delta)]$ is the material polarization, with $M_{0x}$ the excitonic dipole moment (assumed to not depend on $\Delta$), $\hat{\bm e_\lambda}$ the exciton polarization, $N$ the dot density, $g(\Delta)$ the inhomogenous broadening, and $u_1(z,\Delta)$ and $u_2(z,\Delta)$ the real and imaginary part of the Bloch vector, respectively, at position $z$ and for a detuning $\Delta$. For a coherent time evolution there exists a remarkable theorem which asserts that the pulse area, defined through 

\begin{equation}\label{eq:sit-area}
  A(z)=M_{0x}\lim_{t\to\infty}\int_{-\infty}^t dt'\,\hat{\bm e}_\lambda^*\,,
  \bm{\mathcal{E}}(z,t')
\end{equation}

\noindent satisfies the equation \cite{mccall:69,mandel:95}

\begin{equation}\label{eq:sit.area-theorem}
  \frac{dA(z)}{dz}=-\frac \alpha 2\,\sin A(z)\,,\quad
  \alpha=\frac{2\pi^2N\omega_0 M_{0x}^2}{nc}\,g(0)\,.
\end{equation}

\noindent Here $\alpha$ provides a characteristic length scale. For a weak incident pulse one immediately observes from the linearized form of eq.~\eqref{eq:sit.area-theorem} that  $A$ decays according to $\exp -\alpha z/2$, as expected from Beer's law of linear absorption. Within the Bloch vector picture this decay is due to the small rotations of Bloch vectors out of their equilibrium positions and the resulting intensity loss of the light pulse. However, completely new features appear when $A \geq \pi$.  Most importantly, if $A$ is an integer of $\pi$ the pulse area suffers no attenuation in propagating along $z$. Indeed, such behavior is observed in fig.~\ref{fig:sit} which shows results of simulations for the more complete level scheme depicted in fig.~\ref{fig:level-scheme}c: for small field strengths, fig.~\ref{fig:sit}a, the pulse becomes attenuated quickly. However, if the pulse area exceeds a certain value, fig.~\ref{fig:sit}b, self modulation occurs and the pulse propagates without suffering significant losses. In a sense, this situation resembles a material control of the laser pulse. The latter acquires a $2\pi$ hyperbolic secant shape \cite{mccall:69,mandel:95}, which ---contrary to the situation depicted in fig.~\ref{fig:bloch.detuning} for a constant laser--- rotates all Bloch vectors from their initial state through a sequence of excited states back to the initial ones, irrespective of their detuning $\Delta$ (inset of fig.~\ref{fig:sit}b). Here, the leading edge of the pulse coherently drives the system in a predominantly inverted state; before decoherence takes place, the trailing edge brings then the population back to the ground state by means of stimulated emission, and an equilibrium condition is reached in which the pulse receives through induced emission of the system the same amount of energy transferred to the sample through induced absorption. Finally, at the highest pulse area, fig.~\ref{fig:sit}c, we observe pulse breakup \cite{mandel:95,mccall:67,mccall:69}. The inset shows a $4\pi$-rotation of the exciton states and an additional population of the biexciton ones. As apparent from the figure, this additional biexciton channel does not spoil the general pulse propagation properties (for details see ref.~\cite{hohenester.prb:02}). Self-induced transparency in semiconductor quantum dots has been demonstrated recently \cite{borri:02,schneider:03}. No pulse breakup was observed, a finding attributed to a possible dependence of the dipole moments $M_{0x}(L)$ on the quantum dot size $L$.

\section{Quantum computation}\label{sec:computation}

Quantum computation is a quantum control with unprecedented precision \cite{bouwmeester:00,bennett:00}. Its key elements are the quantum bit or {\em qubit},\/ which is a generic two-level system, and a register of such qubits (with a typical size ranging from a few tens to several hundreds). This register allows to store the quantum information, which is processed by means of unitary transformations ({\em quantum gates}) through an external control. Besides the single-qubit rotations ({\em unconditional gates}), one also requires two-qubit rotations ({\em conditional gates}) where the ``target qubit'' is only rotated when the ``control qubit'' is in an inverted state. Through the latter transformations it becomes possible to create entanglement, which is at the heart of quantum computation. In his seminal work, \citet{shor:94} showed that such quantum computation could ---if implemented successfully--- eventually outperform classical computation. Yet, the hardware requirements and the degree of controllability are tremendous (error-correction schemes permit only one error in approximately $10^4$ operations \cite{laflamme:96}), and it is completely unclear whether a quantum computer will be ever built. Despite this very unclear situation, recent years have seen huge efforts in identifying possible candidates for quantum computers and performing proof-of-principle experiments. Quite generally, any few-level system with sufficiently long-lived states, that allows for efficient readout and scalability, can serve as a possible candidate \cite{divincenzo:00}. Among the vast amount of work devoted to the implementation of quantum-information processing in physical systems, several have been concerned with optical and spin excitations in quantum dots. In the following we shall briefly review some of the key proposals and experiments.

There are a number of critical elements to be met in any implementation of a quantum computer, among which the most important ones are the identification of qubits with long decoherence times, of a coupling mechanism between different qubits (for performing conditional gates), and of a readout capability for the quantum information. In the field of quantum dots optical and spin excitations have been considered as qubits. With the limits discussed in the previous sections, long decoherence times, efficient control, and reliable readout schemes are available. The strategies for coupling qubits are motivated by related schemes in different fields of research, which either exploit some local nearest-neighbour interactions \cite{divincenzo:00b}, e.g. hyperfine interactions in nuclear magnetic resonance \cite{chuang:98}, or rely on a common ``bus'' which connects all qubits, e.g. phonon excitations of a linear chain of ions \cite{cirac:95}. For quantum dots \citet{barenco:95} were the first to propose the quantum-confined Stark effect as a means to couple optical excitations in different dots. This proposal was elaborated by \citet{troiani.prb:00} and \citet{biolatti:00,biolatti:02}, who proposed to use the Coulomb renormalizations of few-particle states for an efficient inter-qubit coupling. Let us briefly address the first proposal \cite{troiani.prb:00} at the example of the level scheme depicted in fig.~\ref{fig:level-scheme}c. We denote the groundstate with $|0\rangle|0\rangle$, where the first and second expression account, respectively, for the (missing) exciton with spin-up and spin-down orientation; within this qubit language, the single-exciton states correspond to $|1\rangle|0\rangle$ and $|0\rangle|1\rangle$, and the biexciton state to $|1\rangle|1\rangle$. Because of the polarization selection rules (sec.~\ref{sec:exciton.spin}) and the Coulomb renormalization $\Delta$ of the biexciton, all these states can be addressed individually by means of coherence spectroscopy. Indeed, optical control of these two exciton-based qubits was demonstrated \cite{chen:00,li:03,bianucci:04}. To allow within this framework for scalability, optical excitations in an array of quantum dots were proposed; enhancement of the Coulomb couplings between different dots could be either achieved by relying on the quantum-confined Stark effect \cite{biolatti:00,biolatti:02} ---in an electric field electron and hole wavefunctions become spatially separated, and in turn the dipole-dipole interaction between excitons in different dots is strongly enhanced---, or on intrinsic exciton-exciton couplings \cite{derinaldis:02,derinaldis:02b}. Other work has proposed F\"orster-type processes where optical excitations are near-field coupled \cite{quiroga:99,lovett:03,sangu:04}. 

Qubits based on optical excitations have the glaring shortcoming of a fast decoherence on the sub-nanosecond timescale. Much longer decoherence times are expected for spin excitations. \citet{loss:98} proposed a quantum computation scheme based on spin states of coupled single-electron doped quantum dots, with electrical gating as a means for the unconditional and conditional operations. A mixed approach was put forward by \citet{imamoglu:99}, where the quantum information is encoded in the spin degrees of freedom and coupling to the optical degrees is used for efficient and fast quantum gates. Within this proposal, the quantum computer is realized through single-electron charged quantum dots. The unconditional gates are performed through optical coupling of the different electron-spin states to the charged-exciton state (Voigt geometry, sec.~\ref{sec:exciton.spin}), and the conditional ones by means of cavity quantum electrodynamics where all quantum dots are located in a microcavity and coupled to a common cavity mode \cite{imamoglu:99,sherwin:99,feng:03}. Differently, \citet{piermarocchi:02b} proposed a coupling via virtual excitations of delocalized excitons as a genuine solid-state coupling mechanism between electron-spins in different quantum dots. There are a number of further proposals for quantum computation with spin memory and optical gating, where either the quantum-confined Stark effect \cite{pazy:03,calarco:03} or the enhanced flexibility of molecular states in artificial molecules \cite{troiani.prl:03,troiani:03} is used for switchable qubit-qubit interactions. In addition, some work has been concerned with strategies for sophisticated optical gating, e.g. based on pulse shaping \cite{chen:01,piermarocchi:02} or spin-flip Raman transitions \cite{chen:04}. As regarding stimulated Raman adiabatic passage (sec.~\ref{sec:stirap}) a slightly modified level scheme and a somewhat different control strategy is required for qubit rotations or entanglement creation by means of adiabatic population transfers \cite{unanyan:01,kis:02,zhang:03}. Corresponding quantum-dot implementations were proposed for unconditional and conditional gates \cite{troiani.prl:03}, and for storage qubits \cite{pazy:01}. Finally, in the context of  molecular systems the applicability of optimal control for quantum gates was shown to be feasible \cite{tesch:02}.

\section*{Acknowledgements}

Over the last years I had the opportunity to collaborate with many people who have influenced my way of thinking, and have substantially contributed to the results presented in this overview. I am indebted to all of them. My deepest gratitude goes to Giovanna Panzarini (1968---2001), who has introduced me to the field of quantum optics and has shared with me her physical intuition and her joy for the beauty of physics. Her memory will keep alive. Elisa Molinari is gratefully acknowledged for generous support, most helpful dicussions, and for providing a lively and stimualting atmosphere in the Modena group. I sincerely thank Filippo Troiani for his pioneering contributions in our collaboration on quantum-dot based quantum computation. Him, as well as Guido Goldoni, Costas Simserides, Claudia Sifel, Pekka Koskinen, Alfio Borz\'{\i}, Georg Stadler, and Jaro Fabian I wish to thank for many fruitful discussions and the pleasure of collaborating over the last years.

\begin{appendix}

\section{Rigid exciton and biexciton approximation}\label{sec:rigid}

\subsection{Excitons}

Consider the trial exciton wavefunction 

\begin{equation}\label{eq:exciton.trial}
  \Psi^x(\bm r_e,\bm r_h)=\Phi(\bm R)\,\phi_0(\bm\rho)\,,
\end{equation}

\noindent which consists of the groundstate exciton wavefunction $\phi_0(\bm\rho)$ of an ideal quantum well and an envelope function $\Phi(\bm R)$. In other words, we assume that in presence of a quantum confinement the electron and hole are Coulomb bound in the same way as they would be in an ideal quantum well, and only the center-of-mass motion is affected by the quantum confinement. In eq.~\eqref{eq:exciton.trial} the center-of-mass and relative coordinates are given by the usual expressions $\bm R=(m_e\bm r_e+m_h\bm r_h)/M$ and $\bm\rho=\bm r_e-\bm r_h$, respectively. We next insert the trial wavefunction \eqref{eq:exciton.trial} into the Schr\"odinger equation \eqref{eq:exciton-qd} and obtain

\begin{equation}\label{eq:schroedinger.trial}
  \biggl(\mathcal{H}+h+\sum_{i=e,h}U_i(\bm r_i)\biggr)\Phi(\bm R)\,\Phi_0(\bm\rho)=
  E\,\Phi(\bm R)\,\phi_0(\bm\rho)\,,
\end{equation}

\noindent with $\mathcal{H}$ and $h$ defined in eq.~\eqref{eq:exciton-bulk2}. The left-hand side can be simplified by using $h\phi_0(\bm\rho)=\epsilon_0\phi_0(\bm\rho)$. Multiplying eq.~\eqref{eq:schroedinger.trial} with $\delta(\bm R-\bm R_x)\,\phi_0(\bm\rho)$ and integrating over the entire phase space $\bm\tau$ finally gives

\begin{equation}\label{eq:schroedinger.trial2}
  \biggl(-\frac{\nabla_{R_x}^2}{2M}+\sum_{i=e,h}\int d\bm\tau\,
  \delta(\bm R-\bm R_x)\,U_i(\bm r_i)\,|\phi_0(\bm\rho)|^2\biggr)\Phi(\bm R_x)
  =\mathcal{E}\,\Phi(\bm R_x)\,.
\end{equation}

\noindent Comparing this expression with eq.~\eqref{eq:rigid-exciton} shows that the term on the left-hand side is identical to the averaged potential $\bar U(\bm R_x)$.

\subsection{Biexcitons}

A similar procedure can be applied for biexcitons. In analogy to eq.~\eqref{eq:exciton.trial} we make the ansatz

\begin{equation}\label{eq:biexciton.trial}
  \bar\Psi(\bar{\bm\tau})=
  \bar\Phi(\bm R)\,\bar\phi_0(\bar{\bm\tau})\,,
\end{equation}

\noindent with $\bar\phi_0$ the variational function \eqref{eq:kleinman} and $\bar\Phi(\bm R)$ the corresponding envelope function, which depends on the center-of-mass coordinate $\bm R=m_e(\bm r_e+\bm r_{e'})/M+m_h(\bm r_h+\bm r_{h'})/M$ with $M=2(m_e+m_h)$; finally $\bar{\bm\tau}$ denotes the set of variables $\bm r_{e}$, $\bm r_h$, $\bm r_{e'}$, and $\bm r_{h'}$. Suppose that the Hamiltonian can be decomposed into the parts

\begin{equation}\label{eq:schroedinger.trial3}
  H=-\frac{\nabla_R^2}{2M}+h+\sum_i U_i(\bm r_i)\,,
\end{equation}

\noindent where in analogy to excitons $h\bar\phi_0=\bar\epsilon_0\bar\phi_0$ gives the energy of the quantum-well biexciton \cite{kleinman:83} and $i$ runs over all electrons and holes. From the Schr\"odinger equation defined by eqs.~\eqref{eq:schroedinger.trial3} and \eqref{eq:biexciton.trial} we then obtain after multiplication with $\delta(\bm R-\bm R_b)\,\bar\phi_0$ and integration over the entire phase space $\bar{\bm\tau}$ the final result

\begin{equation}
 \biggl(-\frac{\nabla_{R_b}^2}{2M}+\sum_{i}\int d\bar{\bm\tau}\,
  \delta(\bm R-\bm R_b)\,U_i(\bm r_i)\,|\bar\phi_0(\bar{\bm\tau})|^2\biggr)\bar\Phi(\bm R_b)
  =\bar{\mathcal{E}}\,\bar\Phi(\bm R_b)\,,
\end{equation}

\noindent where the term on the left-hand side defines the effective confinement potential for biexcitons (see fig.~\ref{fig:confinement.weak}).

\subsection{Optical dipole elements}\label{sec:dipole.rigid}

Let us investigate the dependence of the dipole matrix elements \eqref{eq:dipole-exciton} on the confinement length $L$ for the exciton states under consideration. Because of the product-type exciton wavefunction \eqref{eq:exciton.trial} the optical dipole moment \eqref{eq:dipole-exciton} is given by the spatial average of the envelope part $\Phi(\bm R)$. We shall now show how this average depends on the confinement length $L$. Our starting point is given by the normalization condition $\int d\bm R\,|\Phi(\bm R)|^2=1$. We next introduce the dimensionless space variable $\bm\xi=\bm R/L$ which is of the order of one. Through $L^d\int d\bm\xi\,|\Phi(L\bm\xi)|=\int d\bm\xi\,|\tilde\Phi(\bm\xi)|^2=1$ we define the wavefunction $\tilde\Phi(\bm\xi)=L^{d/2}\Phi(L\bm\xi)$, where $d=2$ denotes the two-dimensional nature of the electron-hole states. Then,

\begin{equation}\label{eq:dipole-exciton.weak}
  M_{0x}=\mu_0\,\phi_0(0)\int d(L\bm\xi)\,L^{-d/2}\tilde\Phi(\bm\xi)=
  \mu_0\,\phi_0(0)L^{d/2}\int d\bm\xi\, \tilde\Phi(\bm\xi)
\end{equation}

\noindent is the dipole moment for excitonic transitions in the weak confinement regime. Equation~\eqref{eq:dipole-exciton.weak} is the result we were seeking for. The integral on the right-hand side of the last expression is of the order of unity. Thus, the oscillator strength for optical transitions scales with $|M_{0x}|^2\propto L^d$, i.e. it is proportional to the area $L^2$ of the confinement potential.

\section{Configuration interactions}\label{sec:CI}

\subsection{Second quantization}\label{sec:second-quantization}

Second quantization is a convenient tool for the description of few- and many-particle problems \cite{kadanoff:62,fetter:71,mahan:81,haug:96}. The central objects are the field-operators $\ppsi^\dagger(\bm r)$ and $\ppsi(\bm r)$ which, respectively, describe the creation and destruction of an electron at position $\bm r$. The field operators obey the usual anticommutation relations $\{\ppsi(\bm r),\ppsi^\dagger(\bm r')\}=\delta(\bm r-\bm r')$ and zero otherwise. Within the framework of second quantization one replaces all one- and two-particle operators $\mathcal{O}_1(\bm r)$ and $\mathcal{O}_2(\bm r_1,\bm r_2)$ by \cite{fetter:71,mahan:81}

\begin{eqnarray}\label{eq:second-quantization.a}
  \mathcal{O}_1(\bm r) &\longrightarrow &
  \int d\bm r\,\ppsi^\dagger(\bm r)\,\mathcal{O}_1(\bm r)\,\ppsi(\bm r)\\
  \label{eq:second-quantization.b}
  \mathcal{O}_2(\bm r,\bm r') &\longrightarrow &
  \int d\bm rd\bm r'\,\ppsi^\dagger(\bm r)\ppsi^\dagger(\bm r')\,
  \mathcal{O}_2(\bm r,\bm r')\,\ppsi(\bm r')\ppsi(\bm r)\,.
\end{eqnarray}

\noindent When a semiconductor is described in the envelope-function approximation electrons and holes have to be treated as independent particles. This can be accomplished by introducing the field operators $\ppsi_\lambda^e(\bm r)$ and $\ppsi_\lambda^h(\bm r)$ accounting for the electron and hole degrees of freedom, where $\lambda$ is the spin of the electron or hole (sec.~\ref{sec:exciton.spin}); below we shall denote the spin orientation orthogonal to $\lambda$ with $\bar\lambda$. We find it convenient to expand $\ppsi^{e,h}(\bm r)$ in the single-particle bases of eq.~\eqref{eq:sp.qd},

\begin{equation}
  \ppsi_\lambda^e(\bm r)=\sum_\mu\phi_\mu^e(\bm r)\,c_{\mu\lambda}\,,\qquad
  \ppsi_\lambda^h(\bm r)=\sum_\nu\phi_\nu^h(\bm r)\,d_{\nu\lambda}\,,
\end{equation}

\noindent where $c_{\mu\lambda}^\dagger$ creates an electron with spin orientation $\lambda$ in the single-particle state $\mu$, and $d_{\nu\lambda}^\dagger$ a hole with spin $\lambda$ in state $\nu$. With these field operators we can express the few-particle hamiltonian accounting for the propagation of electrons and holes in presence of the quantum confinement and mutual Coulomb interactions as \cite{mahan:81,haug:96}

\begin{eqnarray}\label{eq:h-second}
  H&=&\sum_{\mu\lambda}\epsilon_{\mu\lambda}^e\;c_{\mu\lambda}^\dagger c_{\mu\lambda}+
  \sum_{\nu\lambda}\epsilon_{\nu\lambda}^h\;d_{\nu\lambda}^\dagger c_{\nu\lambda}
  \nonumber\\
  &+&\frac 1 2\sum_{\mu\mu',\bar\mu\bar\mu'\atop\lambda\lambda'}
  V_{\mu'\mu,\bar\mu'\bar\mu}^{ee}\,
  c_{\mu'\lambda}^\dagger c_{\bar\mu'\lambda'}^\dagger c_{\bar\mu\lambda'}c_{\mu\lambda}+
  \frac 1 2\sum_{\nu\nu',\bar\nu\bar\nu'\atop\lambda\lambda'}
  V_{\nu'\nu,\bar\nu'\bar\nu}^{hh}\,
  d_{\nu'\lambda}^\dagger d_{\bar\nu'\lambda'}^\dagger d_{\bar\nu\lambda'}d_{\nu\lambda}
  \nonumber\\
  &-&\sum_{\mu'\mu,\nu'\nu\atop\lambda\lambda'}
  V_{\mu'\mu,\nu'\nu}^{eh}\,
  c_{\mu'\lambda}^\dagger d_{\nu'\lambda'}^\dagger d_{\nu\lambda}c_{\mu\lambda}\,,
\end{eqnarray}

\noindent where the terms in the first line account for the single-particle properties of electrons and holes, those in the second line for the mutual electron and hole Coulomb interactions, and those in the third line for the Coulomb attractions between electrons and holes. All Coulomb couplings in eq.~\eqref{eq:h-second} preserve the spin orientations of the particles. For simplicity we have neglected the electron-hole exchange interaction discussed in sec.~\ref{sec:exciton.spin} as well as Auger-type Coulomb processes \cite{axt:98,hohenester.prb:01}. The Coulomb matrix elements are given by

\begin{equation}\label{eq:coulomb-element}
  V_{\mu'\mu,\nu'\nu}^{ij}=\int d\bm rd\bm r'\,
  \frac{\phi_{\mu'\lambda }^{i\,*}(\bm r )\phi_{\mu\lambda }^i(\bm r)\,
        \phi_{\nu'\lambda'}^{j\,*}(\bm r')\phi_{\nu\lambda'}^j(\bm r')}
  {\kappa_s|\bm r-\bm r'|}\,.
\end{equation}

\noindent A word of caution is at place. It might be tempting to assume that eq.~\eqref{eq:h-second} can be obtained in a first-principles manner from eq.~\eqref{eq:dft}. This is not the case. While eq.~\eqref{eq:dft} has a rather precise meaning in the first-principles framework of density functional theory \cite{dreizler:90}, no comparably simple interpretation exists for the Coulomb terms of eqs.~(\ref{eq:h-second},\ref{eq:coulomb-element}). It turns out that dielectric screening in semiconductors is a highly complicated many-particle process \cite{hohenester.prb:01,onida:02} which, surprisingly enough, approximately results in the dielectric screening constant $\kappa_s$. Thus, eqs.~(\ref{eq:h-second},\ref{eq:coulomb-element}) should be understood as an effective rather than first-principles description.

\subsection{Direct diagonalization}

We will now show how the framework of second quantization can be used for the calculation of few-particle states in the strong confinement regime. Throughout we shall assume that the single-particle description of eq.~\eqref{eq:sp.qd} provides a good starting point and that Coulomb interactions only give rise to moderate renormalization effects. More precisely, for $\Delta\epsilon$ a typical single-particle level splitting and $V$ a typical Coulomb matrix element we assume that $V\ll\Delta\epsilon$, which allows to approximately describe the interacting few-particle system in terms of a limited basis of single-particle states --- typically around ten states for electrons and holes \cite{hartmann.prl:00,rinaldi.prb:00,findeis.prb:01}. We stress that exciton or biexciton states in the weak confinement regime, i.e. electron-hole complexes which are bound because of Coulomb correlations, could not be described within such an approach.

\subsubsection{Excitons}

Consider first the Coulomb correlated states for one electron-hole pair --- i.e. the exciton states in the strong-confinement regime. Although they could be easily calculated without invoking the framework of second quantization, this analysis will allow us to grasp the essential features of configuration interaction calculations. We first define the Hilbert space under consideration. In view of the above discussion and keeping in mind that we are aiming at a computational scheme, we restrict our basis to a limited number of single-particle states, e.g. the ten states of lowest energy for electrons and holes. Then, $|\mu,\nu\rangle=c_{\mu\lambda}^\dagger d_{\nu\bar\lambda}^\dagger\,|0\rangle$ provides a basis of approximately hundred states suited for the description of one electron and hole with opposite spin orientations (fig.~\ref{fig:exciton.strong}). We next expand the exciton in this basis,

\begin{equation}
  |x\rangle=\sum_{\mu\nu}\Psi_{\mu\nu}^x\,|\mu,\nu\rangle\,.
\end{equation}

\noindent The exciton eigenstates $\Psi_{\mu\nu}^x$ and energies $E_x$ are then obtained from the Schr\"odinger equation $H|x\rangle=E_x|x\rangle$, where $H$ is the many-body hamiltonian defined in eq.~\eqref{eq:h-second}. To this end, we multiply the Schr\"odinger equation from the left-hand side with $\langle\mu,\nu|$ and obtain after some straightforward calculation the eigenvalue equation

\begin{equation}\label{eq:exciton-eigenvalue}
  \sum_{\mu'\nu'}\biggl(
   (\epsilon_\mu^e+\epsilon_\nu^h)\delta_{\mu\mu'}\delta_{\nu\nu'}-V_{\mu\mu',\nu\nu'}^{eh}
   \biggr)\,\Psi_{\mu'\nu'}^x=E_x\Psi_{\mu\nu}^x\,.
\end{equation}

\noindent Here, the term in parentheses on the left-hand side is the hamiltonian matrix in the single-particle basis, and $E_x$ and $\Psi_{\mu\nu}^x$ can be obtained by its direct diagonalization.

\subsubsection{Biexcitons}

Things can be easily extended to biexcitons. Before presenting the details of the underlying analysis two points are worth mentioning. First, the proper anti-symmetrization of the electron-hole wavefunction is automatically guaranteed within the framework of second quantization. Second, the size of the Hilbert space for an $n$-body problem scales according to $\sim N^n$, where $N$ is the number of single-particle states under consideration. This number becomes exceedingly fast prohibitively large for computational approaches. One thus introduces a further cutoff adapted from the single-particle energies of the few-particle basis states. Consider the basis $|\mu,\nu;\mu',\nu'\rangle=c_{\mu\lambda}^\dagger d_{\nu\bar\lambda}^\dagger\,c_{\mu'\bar\lambda}^\dagger d_{\nu'\lambda}^\dagger\,|0\rangle$ for the description of a biexciton where the two electron-hole pairs have opposite spin orientations. Then,

\begin{equation}
  |b\rangle=\sum_{\mu\nu,\mu'\nu'}\bar\Psi_{\mu\nu,\mu'\nu'}^b\,
  |\mu,\nu;\mu',\nu'\rangle
\end{equation}

\noindent defines the biexciton state. The biexciton wavefunctions $\bar\Psi_{\mu\nu,\mu'\nu'}^b$ and energies $\bar E_b$ are obtained from Schr\"odinger's equation with the many-body hamiltonian \eqref{eq:h-second},

\begin{alignat}{4}\label{eq:biexciton-eigenvalue}
  \left(\epsilon_\mu^e+\epsilon_\nu^h+\epsilon_{\mu'}^e+\epsilon_{\nu'}^h\right)
  \bar\Psi_{\mu\nu,\mu'\nu'}^b\,
  +&\,\sum_{\bar\mu\bar\mu'}V_{\mu\bar\mu,\mu'\bar\mu'}^{ee}\,
  \bar\Psi_{\bar\mu\nu,\bar\mu'\nu'}^b\,
  &+&\sum_{\bar\nu\bar\nu'}V_{\nu\bar\nu,\nu'\bar\nu'}^{hh}\,
  \bar\Psi_{\mu\bar\nu,\mu'\bar\nu'}^b\nonumber\\
  -&\,\sum_{\bar\mu\bar\nu}V_{\mu\bar\mu,\nu\bar\nu}^{eh}\,
  \bar\Psi_{\bar\mu\bar\nu,\mu'\nu'}^b\,
  &-&\sum_{\bar\mu'\bar\nu'}V_{\mu'\bar\mu',\nu'\bar\nu'}^{eh}\,
  \bar\Psi_{\mu\nu,\bar\mu'\bar\nu'}^b\nonumber\\
  -&\,\sum_{\bar\mu\bar\nu'}V_{\mu\bar\mu,\nu'\bar\nu'}^{eh}\,
  \bar\Psi_{\bar\mu\nu,\mu'\bar\nu'}^b\,
  &-&\sum_{\bar\mu'\bar\nu}V_{\mu'\bar\mu',\nu\bar\nu}^{eh}\,
  \bar\Psi_{\mu\bar\nu,\bar\mu'\nu'}^b\nonumber\\
  =&\,\bar E_b\,\bar\Psi_{\mu\nu,\mu'\nu'}^b\,. 
\end{alignat}

\noindent Here, the terms in the first line account for the single-particle energies and the repulsive electron-electron and hole-hole interactions, and those in the second and third line for the various attractive Coulomb interactions between electrons and holes. Again, the biexciton eigenstates $\bar\Psi_{\mu\nu,\mu'\nu'}^b$ and energies $\bar E_b$ are obtained through direct diagonalization of the hamiltonian matrix. The same scheme can be further extended to other few-particle complexes, such as e.g. triexcitons or multi-charged excitons. It turns out to be advantageous to derive general rules for the construction of the hamiltonian matrix. The interested reader is refered to the literature \cite{mcweeny:92,brasken:00,corni:03}.

\section{Two-level system}\label{sec:two-level}

Two-level systems are conveniently described in terms of the Pauli matrices

\begin{equation}\label{eq:pauli2}
  \sigma_1=\left(\begin{array}{cc}0&1\\1&0\\ \end{array}\right)\,,\quad
  \sigma_2=\left(\begin{array}{cc}0&-i\\i&\phantom{-}0\\ \end{array}\right)\,,\quad
  \sigma_3=\left(\begin{array}{cc}1&\phantom{-}0\\0&-1\\ \end{array}\right)\,.
\end{equation}

\noindent They are hermitian $\sigma_i^\dagger=\sigma_i$, have trace zero $\tr\,\sigma_i=0$, and fulfill the important relation

\begin{equation}\label{eq:pauli3}
  \sigma_i\,\sigma_j=\delta_{ij}\,\one+i\epsilon_{ijk}\,\sigma_k\,.
\end{equation}

\noindent Here $\epsilon_{ijk}$ is the total anti-symmetric tensor, and we have used the Einstein summation convention. It immediately follows that $\sigma_i^2=\one$. For the commutation and anti-commutation relations we obtain

\begin{equation}\label{eq:pauli4}
  [\sigma_i,\sigma_j]=2i\epsilon_{ijk}\,\sigma_k\,,\quad
  \{\sigma_i,\sigma_j\}=2\delta_{ij}\,\one\,.
\end{equation}

\noindent We have now all important relations at hand. Let us first compute the expression $\exp(\lambda\,\bm a\bm\sigma)$, with $\bm a=a\hat{\bm e}$ an arbitrary real vector which has the norm $a=\|\bm a\|$ and the direction described by the unit vector $\hat{\bm e}=\bm a/a$. To this end, we expand the exponential into a power series and obtain

\begin{eqnarray}\label{eq:pauli.exponential}
  e^{\lambda\,\bm a\bm\sigma} &=& \one+\lambda(\bm a\bm\sigma)+\frac{\lambda^2}{2!}
  (\bm a\bm\sigma)^2+\frac{\lambda^3}{3!}(\bm a\bm\sigma)^3+
  \frac{\lambda^4}{4!}(\bm a\bm\sigma)^4+\dots\nonumber\\
  &=& \one+(\lambda a)\,\hat{\bm e}\bm\sigma+\frac{(\lambda a)^2}{2!}\one+
  \frac{(\lambda a)^3}{3!}\hat{\bm e}\bm\sigma+\frac{(\lambda a)^4}{4!}\one+\dots\nonumber\\
  &=& \cos a\lambda\,\one+\sin a\lambda\,\hat{\bm e}\bm\sigma\,.\phantom{\frac 1 2}
\end{eqnarray}

\noindent To arrive at the second line we have used $(\bm a\bm\sigma)^2=a^2\,\one$, which immediatly follows from eq.~\eqref{eq:pauli3}. This expression can be used, e.g for computing the time evolution operator of a two-level system. Things have to be slightly modified for the conditional time evolution in the unraveling \eqref{eq:conditional.dt} of the master equation. The conditional time evolution \eqref{eq:conditional.dt} of a two-level system driven by the resonant laser $\Omega$ and subject to spontaneous photon emissions is described by the effective hamiltonian

\begin{equation}
  H_{\rm eff}=-\frac 1 2\left(\Omega\,\sigma_1+i\Gamma\frac 12\left(\one+\sigma_3\right)
  \right)=-i\frac\Gamma 4\one-\Omega_{\rm eff}\,\hat{\bm e}\bm\sigma\,.
\end{equation}

\noindent Here $\Omega_{\rm eff}^2=\Omega^2-(\Gamma/2)^2$ and $\hat{\bm e}=(\cosh\theta,0,i\sinh\theta)$, where the angle $\theta$ is defined through $\tanh\theta=\Gamma/(2\Omega)$. By use of $\hat{\bm e}^2=1$ we obtain for the conditional time evolution operator $e^{-iH_{\rm eff}t}$ [in a similar manner to eq.~\eqref{eq:pauli.exponential}] the result

\begin{equation}
  U_{\rm eff}(t)=e^{-iH_{\rm eff}t}=e^{-\frac\Gamma 4 t}\left(
    \cos\frac{\Omega_{\rm eff}t}2\,\one+
    i\sin\frac{\Omega_{\rm eff}t}2\,\hat{\bm e}\bm\sigma
  \right)\,.
\end{equation}

\noindent For the initial density operator $\proj 00=(\one-\sigma_3)/2$ the probability that within $[0,t]$ the system has not emitted a photon is

\begin{equation}
  P_0(t)=\frac 1 2\tr\left( U_{\rm eff}(t)\,(\one-\sigma_3)\,U_{\rm eff}^\dagger(t)\right)\,.
\end{equation}

\noindent In the evaluation of the above expression we only have to consider the products of those terms which give $\one$ (because those with $\sigma$ vanish when performing the trace).  Then,

\begin{equation}
  P_0(t)=e^{-\frac\Gamma 2 t}\left(
  \cos^2\frac{\Omega_{\rm eff}t}2+\sin^2\frac{\Omega_{\rm eff}t}2-
  i\cos\frac{\Omega_{\rm eff}t}2\,\sin\frac{\Omega_{\rm eff}t}2\,
  (\hat{\bm e}-\hat{\bm e}^*)\,\hat{\bm e}_3
  \right)\,,
\end{equation}

\noindent which after a few minor manipulations finally gives eq.~\eqref{eq:prob.no-photon}.

\paragraph{Bloch equations.} 

Consider the hamiltonian \eqref{eq:hamiltonian.two-level} of a two-level system subject to the driving laser $e^{-i\omega_0 t}\,\Omega$. In the interaction representation according to $\omega_0\,\proj 11$ we can remove the fast time dependence of $e^{-i\omega_0 t}\,\Omega$ ({\em rotating frame}\/ \cite{scully:97,hohenester:02}), and obtain

\begin{eqnarray}
  H&=& \frac 1 2\biggl(\Delta\,\sigma_3-\Omega^*\,\proj 0 1 - \Omega\,\proj 1 0\biggr)
  \nonumber\\
  &=& \frac 12\,\Delta\,\sigma_3-\frac 1 4\biggl( 
  (\Re e\,\Omega-i\Im m\,\Omega)(\sigma_1-i\sigma_2)+
  (\Re e\,\Omega+i\Im m\,\Omega)(\sigma_1+i\sigma_2)\biggr)\nonumber\\
  &=&\frac 1 2\biggl(\Delta\,\sigma_3-\Re e\,\Omega\,\sigma_1+
  \Im m\,\Omega\,\sigma_2\biggr)\,.
\end{eqnarray}

\noindent Here we have used $\proj 01=(\sigma_1-i\sigma_2)/2$ and $\proj 10=(\sigma_1+i\sigma_2)/2$, and have decomposed $\Omega$ into its real and imaginary part. $\Delta=E_0-\omega_0$ is the detuning of the two-level system with respect to the laser frequency $\omega_0$. Inserting this hamiltonian together with the density operator \eqref{eq:density.pauli} into the Liouville von-Neumann equation gives 

\begin{equation}
  \dot{\bm u}\bm\sigma=
  -i\left[\frac 1 2\,\bm\Omega\bm\sigma,\bm u\bm\sigma\right]=
  (\bm\Omega\times\bm u)\,\bm\sigma\,,
\end{equation}

\noindent where we have used eq.~\eqref{eq:pauli4} to arrive at the last term. We next multiply this equation with $\bm\sigma$ and take the trace, to finally arrive at the coherent part \eqref{eq:bloch-equations} of the optical Bloch equations. For the incoherent part we express the Lindblad operators according to $L=a_0\one+\bm a\bm\sigma$, with the complex coefficients $a_0=a_0'+ia_0''$ and $\bm a=\bm a'+i\bm a''$. When this operator is inserted into the master equation \eqref{eq:lindblad} of Lindblad form, we obtain after some lengthy but straightforward calculation the incoherent part of the Bloch equations \cite{hohenester.prb:03}

\begin{equation}\label{eq:pauli.lindblad}
  \dot{\bm u}\cong 2\biggl( 2\,(\bm a'\times \bm a'')-(a_0'\bm a''-a_0''\bm a')\times\bm u-
  |\bm a|^2\bm u+(\bm a'\bm u)\,\bm a'+(\bm a''\bm u)\,\bm a''\biggr)\,.
\end{equation}

\noindent For the Lindblad operator $L=\sqrt\Gamma\proj 01=\sqrt\Gamma\,(\sigma_1-i\sigma_2)/2$, corresponding to $a_0=0$ and $\bm a=\sqrt\Gamma\,(\hat{\bm e}_1-i\hat{\bm e_2})/2$, we finally arrive at eq.~\eqref{eq:bloch.t}, with the longitudinal and transverse scattering times $T_1=1/\Gamma$ and $T_2=2/\Gamma$, respectively.

\section{Independent boson model}\label{sec:independent-boson}

In this appendix we show how to evaluate the polarization fluctuations $G(t)=\langle\sigma_-(0)\sigma_+(t)\rangle$ for the spin-boson hamiltonian $H=E_0\,\proj 1 1+H_0+V$ of eq.~\eqref{eq:spin-boson}, with $H_0=\sum_i\omega_i\,a_i^\dagger a_i^{\phantom\dagger}$ and $V$ the dot-phonon coupling. $E_0\,\proj 1 1$ commutes with both $H_0$ and $V$, and correspondingly $e^{iHt}=e^{iE_0t\,\proj 11}\,e^{i(H_0+V)t}$. Inserting this expression into $G(t)$ allows to evaluate all expressions involving the system operators $\proj 11$ and $\sigma_\pm$ explicitly, and we obtain

\begin{equation}\label{eq:spin-boson.G}
  G(t)=e^{iE_0 t}\,\left< e^{i(H_0+V)t}\,e^{-iH_0t}\right>\,.
\end{equation}

\noindent Here we have assumed that the expectation value $\langle .\rangle$ is for the system in the ground state and for a thermal distribution of phonons, and we have used that $e^{-i(H_0+V)t}|0\rangle=e^{-iH_0t}|0\rangle$ which follows upon expanding the exponential in its power series and using that $V|0\rangle=0$. We next introduce the displacement operator 

\begin{equation}\label{eq:displacement}
  D(\xi)=e^{\xi a^\dagger -\xi^* a}\,,\quad
  D^\dagger(\xi)=D^{-1}(\xi)=D(-\xi)
\end{equation}

\noindent of the harmonic oscillator \cite{walls:95,barnett:97}. It has the important properties 

\begin{eqnarray}\label{eq:displacement.property}
  D^\dagger(\xi)\,a\,D(\xi)&=&a+\xi\nonumber\\
  D^\dagger(\xi)\,a^\dagger\,D(\xi)&=&a^\dagger+\xi^*\nonumber\\
  D^\dagger(\xi)\,f(a,a^\dagger)\,D(\xi)&=&f(a+\xi,a^\dagger+\xi^*)\,,
\end{eqnarray}

\noindent with $f(a,a^\dagger)$ an arbitrary function of the field operators $a$ and $a^\dagger$. The last expression can be easily proven by inserting $D(\xi)D^\dagger(\xi)=\one$ in the power series of $f(a,a^\dagger)$. Because in eq.~\eqref{eq:spin-boson.G} the different oscillators propagate independently of each other, in the following it suffices to consider only one phonon mode. Then, eq.~\eqref{eq:displacement.property} can be used to simplify the time evolution operator according to

\begin{equation}
  e^{i(H_0+V)t}=e^{-i\xi^2\omega t}\,D^\dagger(\xi)\,e^{iH_0t}\,D(\xi)\,,
\end{equation}

\noindent with $\xi=g/\omega$. Accordingly we can express $G(t)$ for a single phonon mode through

\begin{equation}\label{eq:spin-boson.G2}
  G(t)=e^{i\bar E_0t}\,\left< D^\dagger(\xi)\,e^{iH_0t}\,D(\xi)\,e^{-iH_0t}\right>=
       e^{i\bar E_0t}\,\left< D^\dagger(\xi)\,D(\xi\,e^{i\omega t})\right>\,,
\end{equation}

\noindent where $\bar E_0=E_0-\xi^2\omega$ and the last expression has been derived by evaluating $D(\xi)$ in the interaction representation according to $H_0$. We can now use the relation $D(\xi)\,D(\xi')=e^{-i\Im m(\xi^*\xi')}\,D(\xi+\xi')$ for the displacement operators \cite{walls:95,barnett:97} to simplify expression \eqref{eq:spin-boson.G2} to

\begin{equation}\label{eq:spin-boson.G3}
  G(t)=e^{i(\bar E_0t+\xi^2\sin\omega t)}\,
  \left< D\left(\xi\left[e^{i\omega t}-1\right]\right)\right>\,.
\end{equation}

\noindent In the remainder we discuss how this expression can be evaluated for a thermal phonon distribution. To this end we use the factorization $D(\xi)=e^{-|\xi|^2}\,e^{\xi a^\dagger}e^{-\xi^*a}$ of the displacement operator. For a Bose-Einstein distribution $n(\omega)$ of the phonons it can be shown that \cite{mahan:81,breuer:02,barnett:97}

\begin{equation}
  \left<(a^\dagger)^\ell a^\ell\right>=  \ell!\,[n(\omega)]^\ell\,.
\end{equation}

\noindent To compute $\langle D(\xi)\rangle$ we expand the exponentials $e^{\xi a^\dagger}$ and $e^{-\xi^*a}$ in power series, and use that only terms with an equal number of creation and annihilation operators give a non-vanishing contribution. Then,

\begin{equation}
  \langle D(\xi)\rangle=\exp\Bigl(-|\xi|^2\,
  \left(n(\omega)+\mbox{$\frac 1 2$}\right)\Bigr)\,.
\end{equation}

\noindent The final result \eqref{eq:polaron.fluc} is obtained by using $|e^{i\omega t}-1|^2=2(1-\cos\omega t)$, $n(\omega)+\frac 1 2=\frac 1 2\coth(\beta\omega/2)$, and introducing an appropriate summation over all phonon modes.

\end{appendix}

\newpage

\begin{figure*}
\caption{
Schematic sketch of the (a) weak and (b) strong confinement regime. In the {\em weak confinement regime}\/ the carriers are usually confined at monolayer fluctuations in the width of a narrow quantum well, which form terraces of typical size $100\times 100\times 5$ nm$^3$ \cite{zrenner:94,hess:94,gammon:96,gammon:96b,matsuda:03}. In the {\em strong confinement regime}\/ the carriers are confined within pyramidal or lens-shaped islands of lower-bandgap material, usually formed in strained layer epitaxy, with typical spatial extensions of $10\times 10\times 5$ nm$^3$ \cite{marzin:94,grundmann:95,leon:95,bimberg:98}.
}\label{fig:confinement}
\end{figure*}
\efloatseparator
 
\begin{figure*}
\caption{
Schematic representation of optical spectroscopy (shaded boxes) and quantum control (dashed lines): an external field acts upon the system and promotes it from the ground to an excited state; the excitation decays through environment coupling, e.g. photo emission, and a measurement is performed indirectly on the environment, e.g. photo detection. In case of quantum control the perturbation is tailored such that a given objective, e.g. the wish to channel the system from one state to another, is fulfilled in the best way. This is usually accomplished by starting with some initial guess for the external field, and to improve it by exploiting the outcome of the measurement. The arrows in the figure indicate the flow of information.}\label{fig:scope}
\end{figure*}
\efloatseparator
 
\begin{figure*}
\caption{Schematic representation of the Bloch vector $\bm u$. The $z$ component accounts for the population inversion and gives the probability for finding the system in either the upper or lower state. The $x$ and $y$ components account for {\em quantum coherence},\/ i.e. the phase relation between the upper and lower state, which is responsible for quantum-interference effects. For the {\em coherent}\/ time evolution of an isolated quantum system $\bm u$ stays at the surface of the Bloch sphere. For an incoherent time evolution in presence of environment couplings $\bm u$ dips into the Bloch sphere: the system {\em decoheres}.
}\label{fig:bloch}
\end{figure*}
\efloatseparator
 
\begin{figure*}
\caption{Trajectories of the Bloch vector $\bm u$ for a $2\pi$-pulse and for: (a) an isolated two-level system [$\bm\Omega=-\Omega\,\hat{\bm e}_1$ is the vector defined in eq.~\eqref{eq:bloch.rabi}]; (b) a two-level system in presence of phonon-assisted dephasing (see sec.~\ref{sec:phonon-dephasing}) and a Gaussian pulse envelope; because of decoherence the length of $\bm u$ decreases \cite{foerstner:03a,foerstner:03b}; (c) same as (b) but for an optimal-control pulse envelope (for details see sec.~\ref{sec:optimal-control}).}
\label{fig:schematics.control}
\end{figure*}
\efloatseparator
 
\begin{figure*}
\caption{Confinement potential along $x$ for the center-of-mass motion of excitons (solid line) and biexcitons (dashed line). The insets report the probability distributions for finding an (e) electron or (h) hole at a given distance from the center-of-mass coordinate $\bm R$; (e',h') same for biexcitons --- the reduced probability at the center of (h') is attributed to the repulsive part $\chi$ of the trial wavefunction. In the calculations we use material parameters representative for GaAs and assume an interface-fluctuation confinement of
rectangular shape with dimensions $100\times 70$ nm$^2$, and monolayer fluctuations of a 5 nm thick quantum well \cite{hohenester.apl:04}.}\label{fig:confinement.weak}
\end{figure*}
\efloatseparator
 
\begin{figure*}
\caption{Contour map showing the square modulus of the wavefunction $\Phi_x(\bm R)$ for the center-of-mass motion of (a) the $s$-type groundstate, (b,c) the $p$-type first excited states with nodes along $x$ and $y$, and (d) the third excited state with two nodes along $x$. We use material parameters of GaAs and the confinement potential depicted in fig.~\ref{fig:confinement.weak}.
}\label{fig:exciton.weak}
\end{figure*}
\efloatseparator
 
\begin{figure*}
\caption{Schematic representation of the two possible groundstates of bright excitons in the strong confinement regime. The solid lines indicate the single-particle states of lowest energy, and the dotted lines the first excited states. Excited exciton states can be obtained by promoting the electron or hole to excited single-particle states. The black triangles in the left and right panel indicate the different spin orientations of the electron and hole, as discussed in more detail in sec.~\ref{sec:exciton.spin}.
}\label{fig:exciton.strong}
\end{figure*}
\efloatseparator
 
\begin{figure*}
\caption{Schematic sketch of the configuration-interaction calculation of the Coulomb-correlated electron-hole states in the restricted single-particle basis of the ground states $0$ and the first excited states $-1$ (left) and $+1$ (right); the numbers correspond to the angular momenta of electrons and holes. The figures show the allowed Coulomb transitions between different electron and hole states, indicated by the filled and open triangles.
}\label{fig:CI-exciton}
\end{figure*}
\efloatseparator
 
\begin{figure*}
\caption{Contour plot of the square modulus of (a) the exciton and (b) the biexciton groundstate. The confinement potential is depicted in fig.~\ref{fig:confinement.weak} and computational details are presented in appendix \ref{sec:rigid}. Because of the larger spatial extension of the biexciton ---see insets (e',h') of fig.~\ref{fig:confinement.weak}--- the center-of-mass motion of the biexciton becomes more confined \cite{hohenester.apl:04}.
}\label{fig:biexciton.weak}
\end{figure*}
\efloatseparator
 
\begin{figure*}
\caption{Schematic sketch of the biexciton ground state which consists of two electron-hole pairs with opposite spin orientations. Because of non-compensating Coulomb interactions and/or Coulomb correlation effects, the energy of the biexciton is modified by a small amount $\Delta$. In the figure $X$ and $X\!X$ refer to the exciton and biexciton ground state, respectively.}
\end{figure*}
\efloatseparator
 
\begin{figure*}
\caption{This figure schematically sketches the creation of charged or multi-charged excitons. A quantum dot is placed inside a field-effect structure. By applying an external gate voltage it becomes possible to transfer electrons one by one from the nearby $n$-type reservoir to the dot. When the sample is optically excited an additional electron-hole pair is created --- i.e. a charged or multi-charged exciton is formed.}\label{fig:charged-exciton}
\end{figure*}
\efloatseparator
 
\begin{figure*}
\caption{Absorption spectra for quantum dots in the weak-confinement regime and for (a) an inhomogenously broadened ensemble of quantum dots and (b) a single dot of dimension $100\times 70$ nm$^2$. For the inhomogenous broadening we assume a Gaussian distribution of the confinement lengths $L_1$ and $L_2$ which is centered at 70 nm and has a full-width of half-maximum of 60 nm, and for the homogeneous lifetime broadening $\gamma=10$ $\mu$eV. We use material parameters representative for GaAs. Photon energy zero is given by the exciton energy of the two-dimensional quantum well.
}
\label{fig:absorption-weak}
\end{figure*}
\efloatseparator
 
\begin{figure*}
\caption{Absorption spectra for quantum dots in the strong confinement regime and for (a) an inhomogenously broadened ensemble of quantum dots and (b) a single dot. We use prototypical material and dot parameters for In$_x$Ga$_{1-x}$As/GaAs dots \cite{sosnowskii:98,findeis.prb:01}. We assume a 2:1 ratio of single-particle splittings between electrons and holes and a 20\% stronger hole confinement associated to the heavier hole mass and possible piezoelectric fields. The absorption spectra are computed within a full configuration-interaction approach for the respective six electron and hole single-particle states of lowest energy.
}
\label{fig:absorption-strong}
\end{figure*}
\efloatseparator
 
\begin{figure*}
\caption{
Luminescence spectra for multi excitons (left panel) and multi-charged excitons (right panel) as computed from a full configuration-interaction approach with a basis of approximately 10000 states  \cite{hohenester.pss-b:00}. We use material parameters for GaAs and single-particle level-splittings of 20 meV for electrons and 3.5 meV for holes. The insets report the electron-hole configuration of the groundstate before photon emission (for multi excitons we use the same configuration for electrons and holes). For clarity we use a relatively large peak broadening.
}
\label{fig:luminescence}
\end{figure*}
\efloatseparator
 
\begin{figure*}
\caption{(a--d) Real-space map of the square modulus of the wavefunctions for  the exciton (a) ground state, (b) first and (c) third excited  state, and (d) the biexciton ground state. The
dashed lines indicate the boundaries of the assumed interface fluctuation. (a'--d') Near-field spectra for a spatial resolution of 25 nm and (a''--d'') 50 nm, as computed according to
eqs.~\eqref{eq:absorption.near} and \eqref{eq:absorption.near2}. The full-width of half maximum $\sigma_{\mathcal{E}}$ is indicated in the second and third row.
}\label{fig:snom}
\end{figure*}
\efloatseparator
 
\begin{figure*}
\caption{Trajectories of the Bloch vector $\bm u$ for a laser pulse with $\Omega\,T=2\pi$ and for detunings $\Delta$ of (a) zero, (b) $-\Omega/2$, and (c) $-\Omega$. The light and dark arrows indicate the final position of the Bloch vector and the driving field $\bm\Omega$, respectively. Only on resonance $\bm u$ returns to its initial positions, whereas off resonance $\Omega_{\rm eff}>\Omega$ and the Bloch  vector is rotated further. }
\label{fig:bloch.detuning}
\end{figure*}
\efloatseparator
 
\begin{figure*}
\caption{Memory function $\int_0^\infty\omega^3 d\omega\,\sin[(1-\omega)t]/(1-\omega)$ of eq.~\eqref{eq:memory2} which describes the temporal buildup of photon scatterings. The asymptotic limit is reached on a timescale of $1/E_0$, where $E_0$ is the energy difference of ground and excited state. For a typical value of $E_0=1$ eV the corresponding time is approximately 0.66 fs. The large negative values at early times are attributed to the somewhat unphysical assumption made in \eqref{eq:liouville2} that system and reservoir are initially completely decoupled. }
\label{fig:gamma}
\end{figure*}
\efloatseparator
 
\begin{figure*}
\caption{
Time evolution of the Bloch vector as computed from eq.~\eqref{eq:bloch-damped} for a constant resonant laser with $\Omega=0.2$ meV and for a finite upper-state lifetime $1/\Gamma=40$ ps. The solid and dashed lines in panel (a) show $u_2(t)$ and $u_3(t)$, respectively, and figure (b) shows the corresponding trajectory of the Bloch vector $\bm u$.}
\label{fig:bloch-damped}
\end{figure*}
\efloatseparator
 
\begin{figure*}
\caption{(a) Probability distribution $P_0(t)$ of eq.~\eqref{eq:prob.no-photon} that a two-level system initially in the ground state and subject to a driving laser field and spontaneous photon emissions, has up to time $t$ not emitted a photon. Times are measured in units of $1/\Omega_{\rm eff}$ and the angles are $\theta=0.1$ (solid line), $\theta=0.2$ (dashed line), and $\theta=0.5$ (dotted line). (b) Probability distribution $-\dot P_0(t)$ that the second photon is emitted at time $t$. (c) Two-photon probability distribution that any other photon is detected at $t$, as computed from the quantum regression theorem \eqref{eq:regression2}.
}
\label{fig:photon-count}
\end{figure*}
\efloatseparator
 
\begin{figure*}
\caption{
(a) Polarization fluctuations and (b) their Fourier transforms, which are proportional to absorption, as computed within the spin-boson model of eq.~\eqref{eq:polaron.fluc} for temperatures of 0.1 (solid lines), 1 (dashed lines), and 10 (dotted lines). For material and dot parameters representative for GaAs ---i.e. a mass density $\rho=5.27$ g\,cm$^{-3}$, a longitudinal sound velocity $c_\ell=5110$ m/s, deformation potentials $D_e=-14.6$ eV and $D_h=-4.8$ eV for electrons and holes, respectively, and a carrier localization length $L=5$ nm \cite{krummheuer:02,foerstner:03a}---, time, energy, and temperature are measured in units of 1 ps, 0.7 meV, and 7.8 K, respectively. We assume a dot-phonon coupling strength $\alpha_p=0.033$ (for details see text). In panel (b) energy zero is given by the renormalized energy $\bar E_0$ of the two-level system.
}
\label{fig:polaron}
\end{figure*}
\efloatseparator
 
\begin{figure*}
\caption{Rabi flopping in presence of phonon-assisted dephasing at temperatures of (a) $T=1$ and (b) $T=10$ and for a Gaussian laser pulse with a full-width of half maximum of 5 (for units see figure caption \ref{fig:polaron}). The solid and dashed lines show $u_3$ and $\|\bm u\|$, respectively. The dark curves show results of calculations including $\bm u$, $s_i$, and $\bm u_i$ as dynamic variables, and the gray ones [which are indistinguishable in panel (a)] those of calculations which additionally include $\langle\!\langle a_i a_j\bm\sigma\rangle\!\rangle$, $\langle\!\langle a_i^\dagger a_j^\dagger\bm\sigma\rangle\!\rangle$, and $\langle\!\langle a_i^\dagger a_j\bm\sigma\rangle\!\rangle$. The insets report the trajectories of the Bloch vector.
}\label{fig:rabi-polaron}
\end{figure*}
\efloatseparator
 
\begin{figure*}
\caption{
Prototypical dot-level schemes. (a) Two-level system with $|0\rangle$ and $|1\rangle$ the ground and excited state; $\Omega$ denotes the Rabi frequency in presence of a light field. (b) $\Lambda$-type scheme, e.g. carrier states in coupled dots \cite{hohenester.apl:00}: $|0\rangle$ and $|1\rangle$ are long-lived states, whereas $|2\rangle$ is a short-lived state which is optically coupled to both $|1\rangle$ and $|2\rangle$ (for details see sec.~\ref{sec:stirap}); the wiggled line indicates spontaneous photon emission. (c) Exciton states in a single dot: $|0\rangle$ is the vacuum state; $|\blacktriangle\rangle$ and $|\blacktriangledown\rangle$ are the spin-degenerate single-exciton groundstates, and $|X\!X\rangle$ is the biexciton groundstate; optical selection rules for light polarizations $\hat{\bm e}_\lambda$ or $\hat{\bm e}_{\bar\lambda}$ (either circular or linear) apply as indicated in the figure.
}\label{fig:level-scheme}
\end{figure*}
\efloatseparator
 
\begin{figure*}
\caption{Simulations of coherent population transfer in coupled dots: (a) transients of the populations $\rho_{00}$, $\rho_{11}$, and $\rho_{22}$ [for level scheme see fig.~\ref{fig:level-scheme}b]; (b) contour plot of final population $\rho_{11}$ as a function of time delay between Stokes and pump pulse and of pulse area $A_s=A_p$; white corresponds to values below $0.1$, black to values above $0.9$; the dashed line gives the contour of $\rho_{11}\ge 0.999$ and the cross indicates the values used in panel (a). In our simulations we use the same Gaussian envelopes for the Stokes and pump pulses (with the time delay given in the figure) and a full-width of half maximum of 20 ps. Parameters are chosen according to ref.~\cite{hohenester.apl:00}.}\label{fig:stirap}
\end{figure*}
\efloatseparator
 
\begin{figure*}
\caption{Schematic sketch of the numerical algorithm for optimal control \cite{borzi.pra:02}. One starts with a guess $\Omega_{\rm trial}$ for the control fields and sets the initial stepsize $\lambda$ to, e.g. $\gamma/10$. Then the equations of motion for $x$ are solved forwards in time and the cost functional $J_{\rm trial}$ is computed. In the first iteration $\Omega_{\rm trial}$ is accepted. Upon acceptance the equations of motion for the dual variables $\tilde x$ are solved backwards in time, one sets $J=J_{\rm trial}$ and $\Omega=\Omega_{\rm trial}$, and finally the new search directions $d_\Omega$ are computed. Finally a new $\Omega_{\rm trial}$ is computed and the loop is started again. In the ensuing iterations $\Omega_{\rm trial}$ is only accepted if $J_{\rm trial}<J$, and otherwise the stepsize $\lambda$ is decreased and the loop restarted (i.e., linesearch with Armijo-backtracking). Through the procedure of increasing and decreasing $\lambda$ it is guaranteed that the algorithm always finds an appropriate stepsize. Finally, the algorithm comes to an end after a certain number of iterations or when $x$ has come close enough to the desired state $x_d$.
}\label{fig:optimal.sketch}
\end{figure*}
\efloatseparator
 
\begin{figure*}
\caption{Results of optimal control calculations for the adiabatic population transfer. We assume that the system is initially in state $|1\rangle$. The objective of the optimal control is to maximize the population of $|1\rangle$ at time $T=60$. As an initial guess for the pump and Stokes laser pulses we assume Gaussians centered at time 30 with a full-width of half maximum of 15 and areas of $10$, $25$, and $50$. We use $\Delta=20$ and $\Gamma=0.1$. Panel (a) shows the population transients [see fig.~\ref{fig:stirap}a] for the initial areas of $10$ (dotted lines), $25$ (dashed lines), and $50$ (solid lines). Panel (b) shows the corresponding optimal control fields; the solid and dashed lines, respectively, correspond to frequency components centered around zero and $\Delta$. The zero-frequency components are attributed in order of increasing magnitude to the initial areas of $10$, $25$, and $50$.
}\label{fig:stirap-optimal1}
\end{figure*}
\efloatseparator
 
\begin{figure*}
\caption{Details of the optimal control calculations shown in fig.~\ref{fig:stirap-optimal1} for initial pulse areas of $10$ (dotted lines), $25$ (dashed lines), and $50$ (solid lines). Panel (a) shows the decrease of the cost functional $J(\psi,\Omega)$ of eq.~\eqref{eq:optimal.cost2} as a function of the number of iterations, panel (b) the stepsize $\lambda$ chosen in the optimal control algorithm, and panel (c) the derivative $\|d_\Omega \|$ (which should vanish at the extremum).
}\label{fig:stirap-optimal2}
\end{figure*}
\efloatseparator
 
\begin{figure*}
\caption{Results of our calculations with a Gaussian $2\pi$ (dashed lines) and optimal-control (solid lines) laser pulse and for zero temperature and an electron-phonon coupling of $\alpha_p=0.1$. Panel (a) shows $|\Omega(t)|$ and panel (b) the time evolution of $u_3(t)$, and the insets the trajectories of the Bloch vector $\bm u(t)$. For the Gaussian $2\pi$-pulse Rabi flopping occurs but is damped due to electron-phonon interactions. For the optimal control decoherence losses are completely suppressed, and the the system passes through the desired states of $\hat{\bm e}_3$ at time zero and $-\hat{\bm e}_3$ at $T$.
}\label{fig:polaron-optimal}
\end{figure*}
\efloatseparator
 
\begin{figure*}
\caption{Results of our simulations for $\mathcal{E}_0(z,t)$ of pulse propagation in a sample of inhomogenously broadened quantum dots and for different pulse areas; we assume a setup where the pulse enters from a dot-free region (negative $z$-values) into the dot region. Length is measured in units of $z_0=1/\alpha$, time in units of $t_0=nz_0/c$, and energy in units of $E_0=1/t_0$, with $z_0\sim 250$ $\mu$m, $t_0\sim 3$ ps, and $E_0\sim 0.2$ meV for typical InGaAs dot samples. The insets report contour plots of the time evolution of the exciton and biexciton population at position $z=5$ \cite{panzarini.prb:02,hohenester:02,hohenester.prb:02}. }\label{fig:sit}
\end{figure*}
\efloatseparator

\end{document}